\documentclass[useAMS,usenatbib,usegraphicx]{mn2e}
\usepackage{amsmath}

\def \bea {\begin{eqnarray}}
\def \ena {\end{eqnarray}}                  
\def \bee {\begin{equation}}
\def \ene {\end{equation}}

\def    \bQ  {{\bf Q}}

\def    \bG     {{\bf \Gamma}}
\def \mc {\mbox{ cos }}
\def \ms {\mbox{ sin }}
\def \mcs {\mbox{cos}^{2}}
\def \mss {\mbox{sin}^{2}}
\def \ma {\hat{{\bf a}}}
\def \me {\hat{\bf e}}
\def \mx {\hat{\bf x}}
\def \my {\hat{\bf y}}
\def \mz {\hat{\bf z}}
\def \mJ {\frac{J}{J_{th}}}
\def \mJs {\frac{J^{2}}{J_{th}^{2}}}
\title{Radiative Torque Alignment: Essential Physical Processes} 

\author[Thiem Hoang \& A. Lazarian]{Thiem Hoang
       $^1$\thanks{E-mail: hoang@astro.wisc.edu}, and A. Lazarian
       $^2$\thanks{E-mail:lazarian@astro.wisc.edu} \\Department of Astronomy, University of Wisconsin, Madison, 
       WI 53706, USA}

\begin{document}

\maketitle
\begin{abstract}

We study the physical processes that affect the alignment of grains subject to
radiative torques (RATs). To describe the action of RATs, we use the 
analytical model (AMO) of RATs introduced in Paper I.  We focus our discussion on the alignment by anisotropic
radiation flux with
respect to magnetic field, which defines the axis of grain Larmor precession. Such an alignment
does not invoke paramagnetic dissipation (i.e. Davis-Greenstein mechanism), but,
nevertheless, grains tend to be aligned  with long axes perpendicular to the magnetic
field. When we account for thermal fluctuations within grain
material, we show that for grains, which are
characterized by a triaxial ellipsoid of inertia,
 the zero-$J$ attractor point obtained in our earlier 
study develops into a low-$J$ attractor point. The value of angular momentum at the
low-J attractor point is the order of the thermal angular momentum corresponding to the
 grain temperature. We show that, for situations when the direction of radiative flux was nearly
 perpendicular to a magnetic field, the alignment of grains with long axes
parallel to the magnetic field (i.e. ``wrong alignment'') reported in Paper I,
 disappears in the presence of thermal fluctuations. Thus, all grains are aligned with their long axes perpendicular to the magnetic field.
To gain insight into the origin and 
stability of the low-$J$
attractor points, we study the dynamics of the crossovers
in the presence of both RATs and thermal fluctuations. 
We study effects
of stochastic gaseous
bombardment and show that gaseous bombardment can drive grains from low-$J$ to high-$J$ attractor points in cases when the high-$J$ attractor points are present. As the alignment of grain
axes with respect to angular momentum is higher for higher values of $J$, 
counter-intuitively, gaseous bombardment can increase the degree of grain
 alignment in respect to the magnetic field. We also study the effects of torques induced 
by H$_2$ formation and show that they can change the value of angular
momentum at high-$J$ attractor point, but marginally affect the value of angular momentum at low-$J$ attractor 
points. We compare the AMO results with those obtained using the
direct numerical calculations of RATs acting upon irregular grains and validate the use of the AMO for realistic situations of RAT alignment.

\end{abstract}

\section{Introduction}

Polarization from absorption and emission by aligned grains is widely believed to trace magnetic field topology (see Hildebrand et al. 2000, Hildebrand 2002; Aitken et al. 2002; Hough et al. 1989). The reliability of the interpretation of polarization maps in terms of magnetic fields depends crucially on the understanding of grain alignment theory.

When, nearly 60 years ago, the polarization of starlight was discovered (Hall 1949;
Hilner 1949), it was immediately
explained  by absorption by elongated dust grains, which are aligned with
respect to the interstellar magnetic field. Since then, the problem of grain alignment
has been addressed by many authors (see a recent review by Lazarian 2007 and
references therein). As a result, substantial progress has
been made towards understanding how these tiny particles can become aligned.

The original theory of grain alignment formulated by
Davis \& Greenstein (1951) is based on
the paramagnetic dissipation of energy for a grain rotating in an {\it external} magnetic field. However, the paramagnetic relaxation time-scale for
a typical interstellar field strength is long compared to the gas damping
time. Moreover, this mechanism has been found to be more efficient for small grains (see Lazarian 1997a; Roberge \& Lazarian 1999), contrasting with observation data, which have testified  that small grains are not aligned (Kim \& Martin 1995;  Andersson \& Potter 2007).

Study of the paramagnetic alignment mechanism was
 given new life by Purcell (1979). In his classical paper, Purcell suggested three processes that can spin
a grain up to suprathermal rates (i.e., much higher than thermal value): $H_{2}$ formation, photoemission,
 and the variation of the accommodation coefficient over the grain surface. $H_{2}$
 formation was identified as the most powerful of these
three processes. In terms of alignment, fast rotation is advantageous, as
fast rotating
 grains are immune to randomization by gas bombardment. As a result, paramagnetic
 dissipation can give rise to good alignment of angular momentum with the magnetic
 field. An obvious limitation of the Purcell mechanism is that the
spin-up process is most efficient, provided that a sufficient fraction of hydrogen is in
 atomic form. Therefore, this mechanism would fail in dark molecular clouds
 where most of the hydrogen is in molecular form. In fact, all spin-up processes
 fail in molecular clouds (see Lazarian, Goodman \& Myers
 1997; Roberge \& Lazarian 1999), while observations have testified that the grains were aligned there (see Ward-Thompson et al. 2000).

Nevertheless, combining the ideas of suprathermal rotation with magnetic inclusions,
in the spirit of the classical Jones \& Spitzer (1967) paper, researchers hoped
to explain the observational data. For instance, a model by
Mathis (1986) provided a good fit to the observed 
Serkowski polarization curve (Serkowski 1975). Problems concerning the Purcell (1979) alignment mechanism
became obvious, however, when Lazarian \& Draine (1999a) reported the effect of
thermal trapping. This effect stems from the coupling of vibrational and
rotational degrees of freedom via the internal relaxation (e.g. Barnett relaxation; Purcell 1979). Thermally trapped grains undergo fast
flips that average out un-compensated Purcell torques. Later, Lazarian \&
Draine (1999b) reported a new effect which they termed nuclear relaxation\footnote{The Barnett relaxation arises from the unpaired electron spins in
  the grain. The nuclear relaxation is resulted from unpaired nuclear
  spins. The mechanisms are different and should not be mixed up.};  they showed this
to be $10^6$ times stronger than the Barnett
relaxation. After this, it became clear that most of the interstellar grains cannot be 
spun up by Purcell torques.   

The mechanical alignment mechanism, which is based on the relative motion of gas and dust, was
pioneered by Gold (1952a). However, this
mechanism, as well as its more sophisticated cousins (Lazarian 1995, 1997; 
Lazarian \& Efroimsky 1996; Lazarian, Efroimsky \& Ozik 1996), 
requires supersonic motion of gas relative
to dust (see Purcell 1969; Lazarian 1994; Roberge, Hanany \& Messinger 1995; Lazarian 1997b), and this is only available in special
circumstances. In his original paper, Gold (1952b) proposed collisions between clouds
as a way to drive supersonic motion. Soon after this, Davis (1955) pointed out that such a process can
only align grains in a tiny fraction of the interstellar medium (ISM). Other, more promising means of obtaining supersonic
drift are ambipolar diffusion (Roberge \& Hanany 1990, Roberge, Hanany \& Messinger 1995) and MHD turbulence (Lazarian
1994; Lazarian \& Yan 2002; Yan \& Lazarian 2003; Yan, Lazarian \& Draine 2004). Nevertheless, while 
applicable for particular 
environments, the mechanical mechanisms are 
unable to explain the ubiquitous alignment of dust in a diffuse medium
and molecular clouds. A more promising mechanism is the mechanical
alignment of helical grains, first mentioned in Lazarian (1995) and Lazarian,
Goodman \& Myers (1997). We briefly described this in Lazarian (2007) and Lazarian \& Hoang (2007a). The consequences of this have to be evaluated, but these are beyond the scope of the present
paper.

Alignment by radiative torques (RATs) has recently
become a favored mechanism to explain grain alignment. This mechanism was initially
proposed by Dolginov \& Mytraphanov (1976), but was not sufficiently appreciated at the time of its introduction. 
Draine \& Weingartner (1996, 1997, hereafter DW96 and DW97) reinvigorated the study of the RAT mechanism by showing that RATs induced by anisotropic radiation upon rather arbitrarily shaped grains can spin-up and directly align them with the magnetic field. These papers refocused attention on the RAT mechanism and made it a promising candidate to
change the grain alignment paradigm. It is also very noble that Bruce Draine 
has kindly modified the publicly available DDSCAT code (Draine \& Flatau 1994) to
include RATs. This has enabled other researchers to access a useful tool for studying the RAT alignment.

The RAT mechanism seems to be able to address major puzzles presented by 
observations. For instance, the observation of polarized  emission emanating
from starless cores (see Ward-Thompson 2000, 2002)
initially seemed completely unexplainable.\footnote{These findings are also contrasting with observational claims based on visible and near-infrared radiation (Goodman et al. 1995). The difference in results was explained in Lazarian (2003)} Indeed, all mechanisms seemed
to fail in such cores, which are presumably close to thermodynamic equilibrium
(see Lazarian, Goodman \& Myers 1997). The RATs seem to be too weak as well
(DW96). However,
Cho \& Lazarian (2005, hereafter CL05) found that the efficiency of
RATs increases fast with grain size (see also Lazarian \& Hoang 2007a), and thus large grains can still be aligned in dark clouds. They found that grains as large as $0.6 \mu m$ can be 
aligned in dark clouds by radiation attenuated by the column density with 
 $A_{V} \approx 10$. This was for much higher extinction than expected 
in the absence of embedded stars (see DW96), and was claimed to be observed
with optical and near infrared polarimetry (Acre et al. 1998). 

The pre-stellar cores studied in Ward-Thompson et al. (2000) correspond to $A_V = 50-60$; the shielding column, assuming uniformity, is approximately half of these 
values. However, Crutcher et al. (2004) pointed out that polarization
data do not sample the innermost core regions\footnote{The peaks
$A_{V}$ of 150 were claimed for the clouds in Pagani et al. (2004).
According to Crutcher (2007, private communication) these peaks are likely
not to produce polarized dust emission.}, and this provides an explanation
for polarization ``holes'' (see Matthews \& Wilson 2000, 2002; Matthews et al. 2001; Lai et al. 2002). The
reported decrease in the percentage polarization with the optical depth also
agrees well with the findings in CL05.

The approach of CL05 was further 
elaborated in the studies by Pelkonen et al (2007) and Bethell et al. (2007),
in which the 
synthetic maps obtained via MHD simulations were analyzed.  In particular,
Bethell et al. (2007) obtained the polarization maps were for
a simulated cloud with peaks of extinction $A_{V}$ as high as $25$. The
study showed non-trivial nature of radiative transfer in a fractal turbulent
cloud, and confirmed the  
the ability of aligned grains to trace a magnetic
field in dark clouds and proved a decrease of percentage polarization
at the highest $A_{V}$. Note that the studies above were
in contrast to the previous studies, which used rather  arbitrary criteria (e.g. $A_{V}=3$) for the alignment to shut down, or even 
more unrealistic assumption that all grains were perfectly aligned.

Similarly, the change of the degree of polarization with optical depth observed
by Whittet et al. (2001) was also explained by RATs in Lazarian (2003, 2007). Detailed modeling of this effect is provided in Whittet et al. (2007) and Hoang \& Lazarian (in preparation).

The encouraging 
correspondence of the theoretical expectations\footnote{Although being
a step forward compared to the earlier naive predictions of polarization
which were mostly detached from the grain alignment theory, 
CL05 and the follow up studies (e.g. Pelkonen et al. 2007; Bethell et al. 2007; Cho \& Lazarian 2007) are also not exact, as they are based
on the alignment efficiencies inferred from idealized numerical studies, rather
than on the exact RAT alignment theory.} and the observed polarization
arising from aligned grains makes it essential to understand
the reason why RATs align grains.
 In our first paper (Lazarian \& Hoang 
2007a, hereafter Paper I), we subjected to scrutiny the properties of RATs.
Using a simple analytical model (AMO) of a helical grain we studied the properties
of RATs and the alignment driven by such RATs. The analytical results were found to 
be in good correspondence with numerical calculations for irregular grains obtained with DDSCAT. Evoking
the generic properties of the RAT components, we explained
the RAT alignment of grains in both the absence and presence of magnetic fields. 
Intentionally, for the sake of simplicity, in Paper I
 we studied a simplified dynamical model to demonstrate the effect of the RAT alignment. This model 
disregards the wobbling of grain axes with respect to the angular momentum that arises
 from thermal fluctuations (Lazarian 1994; Lazarian \& Roberge 1997), and
thermal flipping of grains (Lazarian \& Draine 1999a, b). The simplifications
 allowed us to provide a vivid illustration of the RAT alignment, however, there is still a question
concerning the modification of the alignment by thermal fluctuations, as well as the action
of additional (e.g. random) torques.

The first study to combine the RAT alignment and the physics
of thermal fluctuations and flipping was carried 
out by Weingartner \& Draine (2003; hereafter WD03). They
studied the alignment by RATs produced by a monochromatic
radiation field and for one particular radiation direction $\psi=70^{\circ}$, taking into 
account thermal fluctuations and thermal flipping. They observed the appearance of new
attractor points at low angular momentum, but the underlying physics of this
effect remained unclear.

WD03 also posed a question
concerning what is happening when the entire spectrum of
the interstellar radiation field (the ISRF) is accounted for, and when radiation arrives from other
directions. Their study does not consider the effect of gaseous bombardment and effects of $H_{2}$ formation. We address these and other issues in the present below.

The structure of the paper is as follows. In \S 2 we present current theoretical
understandings of the RAT alignment and formulate our theoretical predictions. We briefly describe
thermal wobbling and our method of averaging RATs in \S~3. In \S 4 we derive the
analytical expressions for RATs for the AMO averaged over thermal fluctuations, and study
the influence of thermal fluctuations to the grain alignment by RATs. In \S 5 we study the alignment for irregular grains with RATs from
DDSCAT. We provide an explanation to the appearance of the low attractor point based on the AMO and discuss the stability of low and high J attractor points in \S 6. We study the physics of
 crossovers induced by RATs in \S 7. We consider fast alignment in the presence of thermal fluctuations in \S 8. In \S 9 and \S 10 show  the influences of random collisions and $H_{2}$
formation torques in the framework of RAT alignment for both the AMO and irregular
grains. An extended discussion is presented in \S
11. Finally, we summarize our results in \S 12. 

\section{Theoretical considerations}

In Paper I we provided analytical calculations for RATs for a grain model that
consists of a perfectly reflecting oblate
spheroid with a ``massless'' mirror attached at its pole (see the
upper panel in Fig. \ref{f1}). 
This helical grain demonstrates RATs that are similar to those of irregular
grains obtained by DDSCAT (see Paper I). 

The basic properties of RATs were the subject of a detailed analysis in Paper I. We established that the projection of RATs onto
${\me}_3$ axis (i.e. $Q_{e3}$; see the upper panel in Fig. \ref{f1})
determines the precession of the grain axis about the
radiation direction. This component is also present for
non-helical (e.g. simple spheroid) grains, and neither case induces grain alignment. It is the other two RAT components, $Q_{e1}$ and $Q_{e2}$, that are responsible for grain
alignment. In Paper I we explain why the RAT alignment tends to occur with the long axis of the grain either perpendicular to the magnetic field or perpendicular to the radiation direction (the choice of which, for a given external magnetic field, is
 determined by the ratio of the rate of radiative precession about the radiation direction and Larmor precession).

In Paper I we found an important parameter that affects the grain alignment, the ratio $Q_{e1}^{max}/Q_{e2}^{max}$, where $Q_{e1}^{max}$ and $Q_{e2}^{max}$ are amplitudes of $Q_{e1}$ and $Q_{e2}$, respectively. Different grain shapes illuminated by different radiation fields can have different ratios $Q_{e1}^{max}/Q_{e2}^{max}$, and thus the resulting alignment is different.

We carried out the study of the RAT alignment in Paper I assuming that the
grain always rotates about its principal inertia axis ${\bf a}_1$, corresponding
to the maximal moment of inertia (hereafter called the maximal inertia axis). 
However, this assumption is only valid for fast rotating grains. Such grains, are subjected to 
fast internal relaxation (Purcell 1979; Lazarian \& Efroimsky 1999; Lazarian \& Draine 1999b)
that couples the angular momentum ${\bf J}$ and ${\bf a}_{1}$. As the grain slows
down to thermal angular velocity, the coupling becomes weaker. As a result, ${\bf a}_{1}$ wobbles about 
${\bf J}$ (Lazarian
1994, Lazarian \& Roberge 1997). Similar to
DW97, in Paper I we disregarded this effect.

Some implications of the thermal wobbling are self-evident. For instance, in Paper I we found that
for a narrow range of angles, when the radiation beam is close to being perpendicular
to magnetic field, the alignment becomes ``wrong'' (i.e. it occurs with the
maximal inertia axis {\it perpendicular} to the magnetic field) \footnote{``Wrong'' alignment without specifying
the conditions for it was also
reported in DW97.}. Such a "wrong" alignment  is in contrast to what is widely observed in the ISM. However, we found that the "wrong" alignment in a diffuse medium corresponds to
low angular momentum $J$. Therefore, we predicted in Paper I
that thermal wobbling of grains would destroy the ``wrong'' alignment.

Qualitatively, in Paper I, we show that, in most cases, a high fraction of grains are aligned at
a zero- $J$ attractor point, assuming that ${\bf a}_{1}$ is always parallel to ${\bf
  J}$. In reality, thermal fluctuations induce the wobbling of ${\bf a}_{1}$ with respect to ${\bf J}$, resulting in the modification of RATs. This, in turn, changes the alignment of ${\bf J}$ with respect to ${\bf B}$. Hence, we
expect grains to be trapped at attractor points with thermal angular momentum J
(hereafter called low-$J$ attractor points).

One direct consequence of being trapped at the low-$J$ attractor point is the decrease of the degree of the alignment of grain axes with  respect to the magnetic field. The degree of
alignment is defined by the Rayleigh reduction factor (Greenberg 1968)
\bea
R=1.5 \langle\mcs\beta\rangle-0.5,\label{eq1} 
\ena
where $\beta$ is the angle between the maximal inertia axis ${\bf a}_{1}$ and
the magnetic 
field ${\bf B}$, $\langle...\rangle$ denotes the averaging over an ensemble of grains. 
Because the process of internal alignment between ${\bf a}_{1}$ and ${\bf J}$ occurs
much faster than the alignment of ${\bf J}$ with respect to ${\bf B}$, $R$ could 
be approximately separated (see Lazarian 1994) as
\bea 
R \sim \langle R(\theta)\rangle \langle R(\xi)\rangle.\label{eq2}
\ena 
Here $\langle R(\theta)\rangle$ and $\langle R(\xi)\rangle$ are given by 
\bea
\langle R(\theta)\rangle=1.5\langle \mcs\theta \rangle-0.5,\label{eq31}\\ 
\langle R(\xi)\rangle=1.5\langle\mcs\xi\rangle-0.5,\label{eq4}
\ena 
where $\theta$ is the angle between ${\bf a}_{1}$ and ${\bf J}$, and $\xi$ is the
angle of ${\bf J}$ and ${\bf B}$.

When thermal fluctuations are absent, we have a perfect alignment of ${\bf
  a}_{1}$ 
with ${\bf J}$, i.e. $\langle R(\theta)\rangle=1$. Therefore the Rayleigh reduction factor
  depends only on the degree of alignment of ${\bf J}$ with ${\bf B}$. In the
  presence of thermal fluctuations, and assuming that the relaxation process obeys the
  normal distribution, $\langle \mcs\theta \rangle$ is given by
\bea
\langle \mcs\theta\rangle=Z \int_{0}^{\pi}\mcs\theta\ms\theta e^{-E(\theta)/kT_{d}}d\theta,\label{r3}
\ena
where $T_{d}$ is the dust temperature, $Z=\int_{0}^{\pi}\ms\theta
  e^{-E(\theta)/kT_{d}}d\theta $ is the normalization factor, and $E(\theta)=J^{2}(1+(h-1)\mss\theta)/2I_{1}kT_{d}$ is the kinetic energy of the spheroidal
  grain. Clearly, as the value of angular momentum $J$ becomes comparable with the thermal value, $J_{th}=\sqrt{2 I_{1} k
  T_{d}}$ where $I_{1}$ is the maximal inertia moment of the grain, the aforementioned thermal
wobbling should decrease the alignment of ${\bf a}_{1}$ with ${\bf B}$.

On the basis of our findings in Paper I, we can qualitatively address some
questions posed in WD03.
What will be the effect of random collisions on grain alignment? How do suprathermal torques, for instance, torques
arising from H$_2$ formation (Purcell 1979) influence the alignment by RATs? 

We expect that grains trapped at low $J$ attractor points, can be
significantly affected by random collisions of gas atoms. When the phase trajectory map of grains has attractor points at high- $J$, 
we expect that random collisions can depopulate grains aligned at the lower attractor point,
and stochastically direct grains to the high-$J$ attractor point. As
the grains with high $J$ are immune to randomization and internal
alignment for high $J$ is close to perfect (Purcell 1979), counter-intuitively, random gaseous bombardment 
can {\it increase} the degree of alignment. 
However, the removal of grains from the low-$J$ attractor points is expected to make grain alignment time dependent, althoughh the corresponding time may be long compared with other processes in the system.
With regards to H$_2$ un-compensated torques (Purcell 1979; Lazarian 1995),
because they are fixed in the grain body coordinate system, their effect depends on the flipping rate of the grains.

Obviously, at the high attractor point corresponding to $J\gg J_{th}$, the flipping rate is very low, 
therefore, $H_{2}$ torques act together with RATs and change the angular momentum depending upon the resurfacing process. 
In contrast, the grains flip fast in the process of heading to low attractor points. Assuming that the correlation
time-scale is shorter than the alignment time, then grains are rapidly driven to low angular momentum at which 
the grain flips very fast; thus H$_2$ torques will
be averaged out to zero (see equation \ref{eq41a}), and could result in an additional
randomization at low attractor point.

Below we test our expectations with numerics.
\section{RATs and Effect of Thermal Wobbling}
In this section we briefly discuss the general definitions of RAT components, then present methods to average the torques over free motion and thermal wobbling. We also analyze the role of the third component $Q_{e3}$ of RATs.

\subsection{RATs: spin-up, alignment and precession}
 Similar to Paper I, in order to easily compare our results to those in
earlier works, wherever possible, we preserve the notations adopted in DW97. 

\begin{figure}
\includegraphics[width=0.49\textwidth]{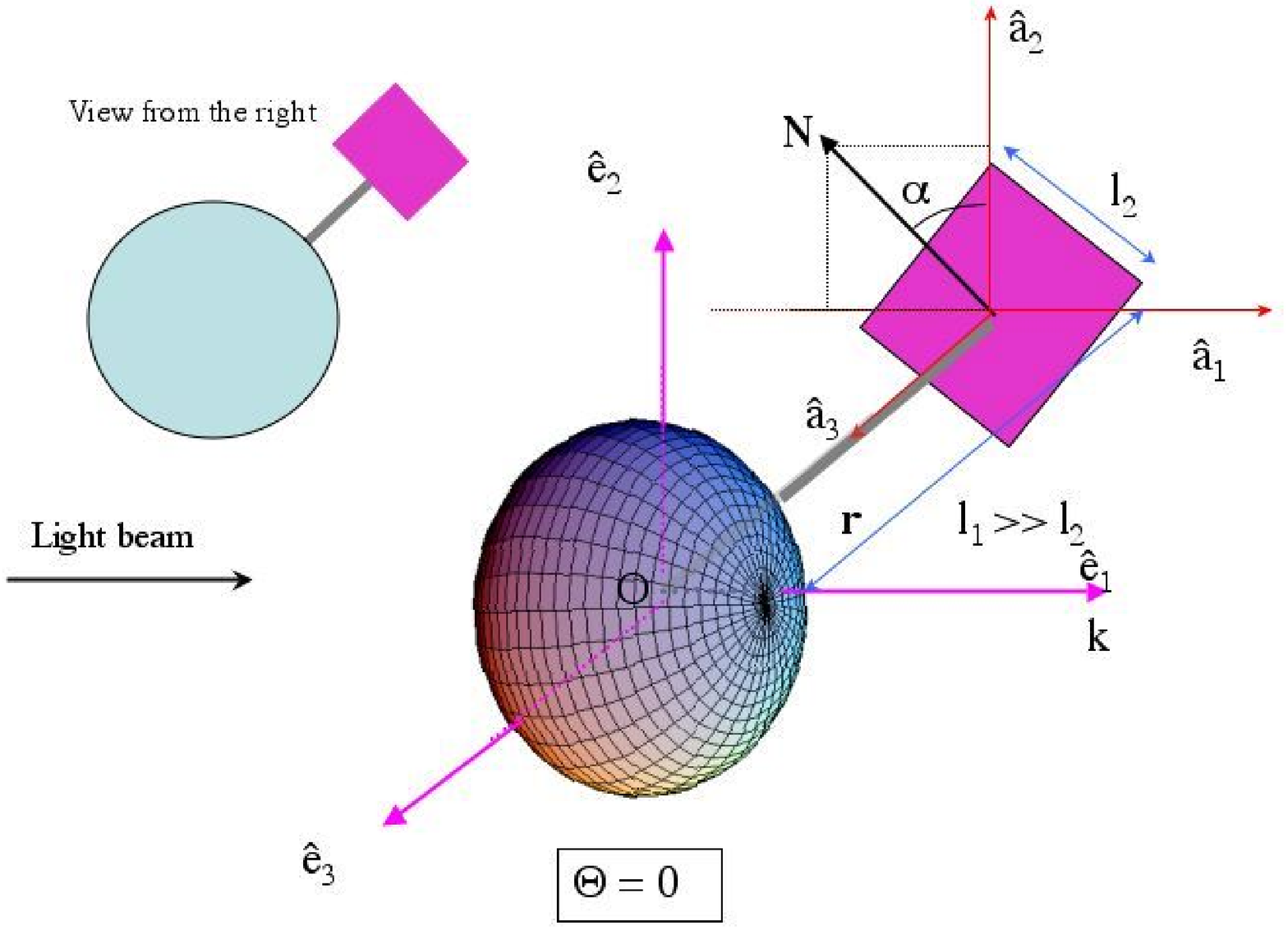}
\includegraphics[width=0.49\textwidth]{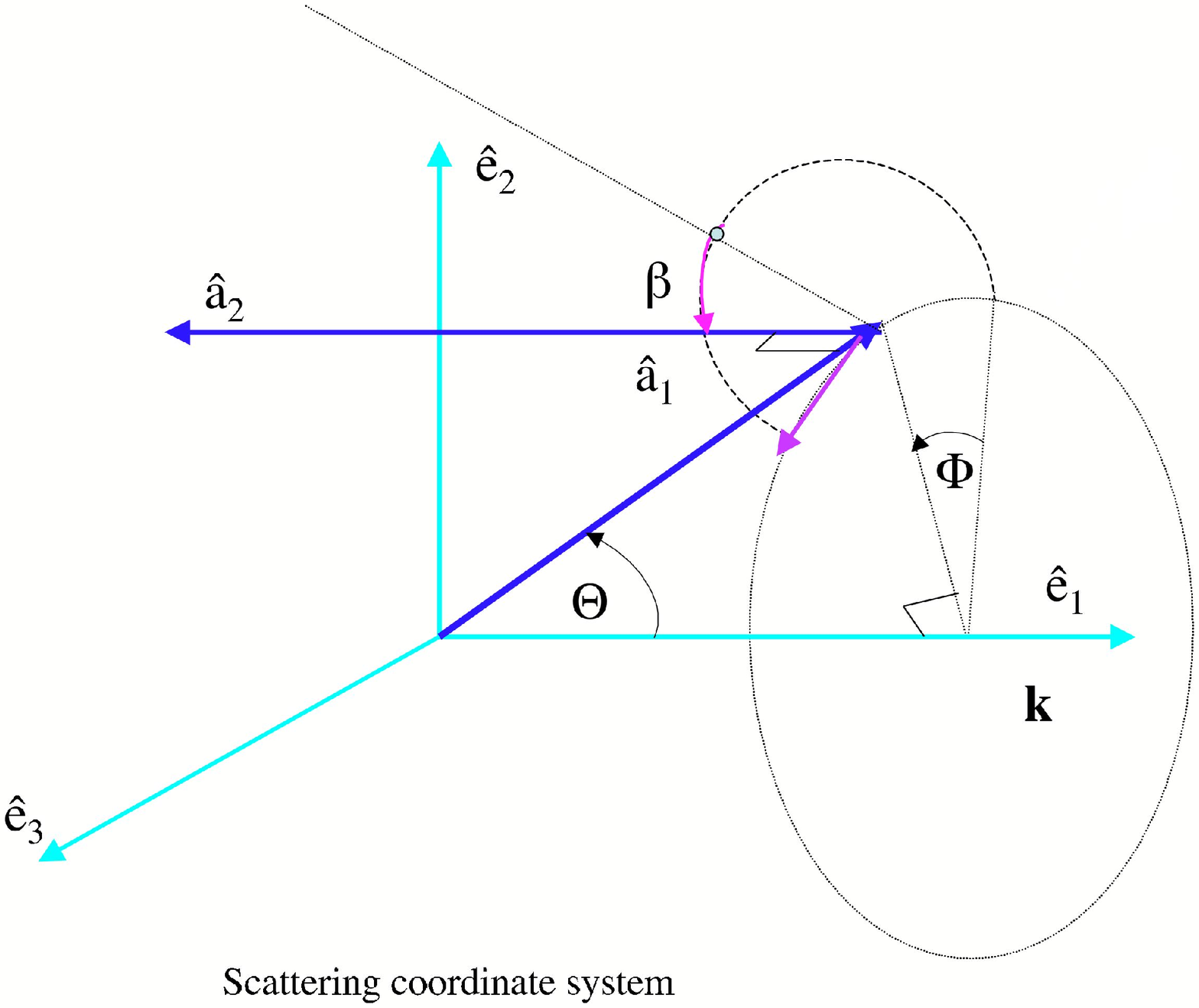} 
\caption{{\it Upper panel:} Geometry for AMO consisting of a perfectly reflecting spheroid and a
  weightless mirror. {\it Lower panel:} The orientation of a grain, described by three principal axes 
  $\hat{a}_{1},\hat{a}_{2}, \hat{a}_{3}$, in the laboratory coordinate system (scattering reference
system)
  $\hat{e}_{1},\hat{e}_{2}, \hat{e}_{3}$ is
  defined by three angles $\Theta, \beta, \Phi$. The
  direction of incident photon beam ${\bf k}$ is along $\hat{e}_{1}$ .} 
\label{f1}
\end{figure}

As in Paper I, we consider only the anisotropic component of the radiation field, the radiative torque is then defined by
\bea 
\bG_{rad}=\frac{\overline{u}_{rad}a_{eff}^{2}\overline{\lambda}}{2}\gamma{\bQ}_{\Gamma}(\Theta, \beta, \Phi),\label{eq7}
\ena 
where $\gamma$ is the degree of radiation anisotropy, $a_{eff}$ is the
effective size of the grain (see DW96; Paper I), and $\overline{\lambda},
\overline{u}_{rad}$ are the mean wave length and mean energy density of the radiation
field. The RAT efficiency vector ${\bf Q}_{\Gamma}$ is represented in the lab system $\me_{1}, \me_{2},$ and $\me_{3}$
\begin{align}
{\bf Q}_{\Gamma}(\Theta, \beta, \Phi)&=Q_{e1}(\Theta, \beta, \Phi)\me_{1}+Q_{e2}(\Theta, \beta, \Phi)\me_{2}+Q_{e3}(\Theta, \beta, \Phi)\me_{3},\label{eq7}
\end{align}
where $\Theta, \beta$ and $\Phi$ are angles describing the orientation of ${\bf a}_{1}$ in the lab system (see the lower panel in Fig. \ref{f1}). It was shown in Paper I that components $Q_{e1}(\Theta, \beta, \Phi=0)$ and $Q_{e2}(\Theta, \beta, \Phi=0)$ have universal properties and play a major role in the process of grain alignment, whereas the component $Q_{e3}(\Theta, \beta, \Phi=0)$ does not affect either spin-up or alignment, provided that $\ma_{1}$ is coupled with ${\bf J}$.

\begin{figure}
\includegraphics[width=0.49\textwidth]{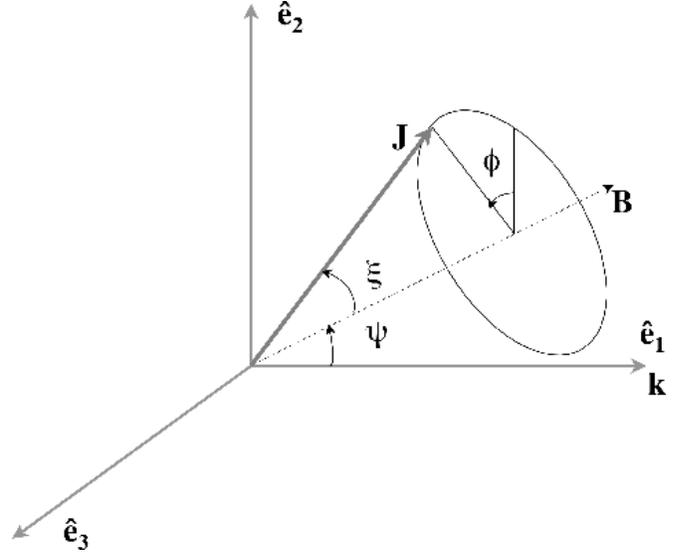} 
\caption{Alignment coordinate system in which $\psi$ 
  is the angle between the magnetic field {\bf B} and the radiation
  direction {\bf k}, $\xi$ is
  the angle between
  the angular momentum vector ${\bf J}$ and {\bf B}, $\phi $ is the Larmor precession angle.} 
\label{ff1}
\end{figure}

Here, we only deal with the alignment of angular momentum with respect to the radiation direction or magnetic field (i.e. external alignment), therefore it is convenient to consider RATs in the spherical coordinate system $J, \xi, \phi$ (see Fig. \ref{ff1}).

In this coordinate system, the RAT can be written as
\begin{align} 
\bG_{rad}=\frac{\gamma \overline{u}_{rad}a_{eff}^{2}\overline{\lambda}}{2}[F(\psi,\phi,\xi)\hat{\bf \xi}
  +G(\psi,\phi,\xi) \hat{\bf \phi}+H(\psi,\phi,\xi)\hat{\bf J}],\label{eq8}
\end{align}
where $F$, which is the torque component parallel to $\hat{\bf \xi}$, acts to change the orientation of ${\bf J}$ in respect to ${\bf B}$; $H$, the component parallel to $\hat{\bf J}$, is to spin grains up and $G$ induces the precession of ${\bf J}$ about the magnetic field or radiation. These are given by 
\begin{align} 
F(\psi,\phi,\xi)&=Q_{e1}(\xi, \psi, \phi)(-\mbox{sin }\psi \mbox{cos }\xi \mbox{cos }\phi-\mbox{sin  }\xi \mbox{cos }\psi)\nonumber\\
&+Q_{e2}(\xi, \psi, \phi)(\mbox{cos }\psi \mbox{cos }\xi \mbox{cos }\phi-\mbox{sin }\xi
\mbox{sin }\psi)\nonumber\\ 
&+Q_{e3}(\xi, \psi, \phi)\mbox{cos }\xi \mbox{sin }\phi,\label{eq9}\\ 
G(\psi,\phi,\xi)&=Q_{e1}(\xi, \psi, \phi)\mbox{sin }\psi \mbox{sin }\phi-Q_{e2}(\xi, \psi, \phi)\mbox{cos }\psi \mbox{sin }\phi\nonumber\\
&+Q_{e3}(\xi, \psi, \phi)\mbox{cos }\phi,\label{eq10}\\ 
H(\psi,\phi,\xi)&=Q_{e1}(\xi, \psi, \phi)(\mbox{cos }\psi \mbox{cos }\xi -\mbox{sin }\psi \mbox{sin  }\xi \mbox{cos }\phi)\nonumber\\
&+Q_{e2}(\xi, \psi, \phi)(\mbox{sin}\psi \mbox{cos }\xi + \mbox{cos }\psi\mbox{sin }\xi\mbox{cos  }\phi) \nonumber\\
&+Q_{e3}(\xi, \psi, \phi)\mbox{sin }\xi \mbox{sin }\phi, \label{eq11} 
\end{align}
where $Q_{e1}(\xi, \psi, \phi), Q_{e2}(\xi, \psi, \phi), Q_{e3}(\xi, \psi, \phi)$, as functions of $\xi, \psi$ and $\phi$, are components of the RAT efficiency vector in the
lab coordinate system (see DW97; Paper I). To obtain $Q_{e1}(\xi, \psi, \phi),
Q_{e2}(\xi, \psi, \phi)$ and $Q_{e3}(\xi, \psi, \phi)$ from ${\bf
  Q}_{\Gamma}(\Theta, \beta, \Phi)$, which is calculated using the AMO or provided by DDSCAT for irregular
grains, we need to use the
relations between $\xi, \psi, \phi$ and $\Theta, \beta, \Phi$ (see WD03 and
Appendix A).

\subsection{Free-torque motion}
\begin{figure}
\includegraphics[width=0.49\textwidth]{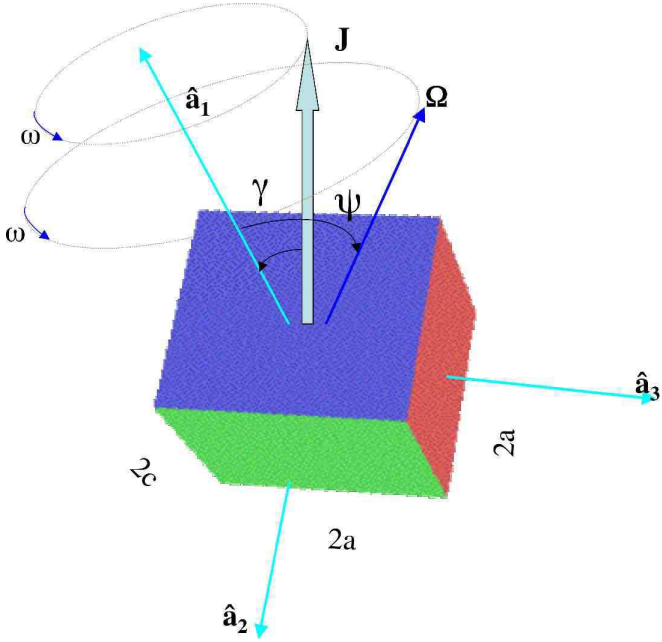}
\caption{In body frame, the motion of a brick consists of the nutation of ${\bf \Omega}$ about $\ma_{1}$ and the precession of ${\bf J}$ around $\ma_{1}$.} 
\label{f1b}
\end{figure}

In the absence of external torques, a grain rotates around its principal axes. This motion is called free-torque motion. For a symmetric grain (e.g. the spheroidal body of AMO or brick; Paper I, Spitzer \& McGlynn 1979), the free-torque motion consists of the nutation of angular velocity ${\bf \omega}$ around ${\bf J}$ at a constant angle $\gamma$ and the rotation around the maximal inertia axis ${\bf a}_{1}$ (see Fig. \ref{f1b}) with a period $P_{\tau}$. 

In general, an irregular grain can be characterized by an ellipsoid with moments of inertia
$I_{1}, I_{2}$ and $I_{3}$ around three principal  axes ${\bf a}_{1}, {\bf a}_{2},
{\bf a}_{3}$, respectively. For a freely rotating grain, its angular momentum is
conserved (i.e., ${\bf J}$ is
fixed in space), while the angular velocity ${\bf \omega}$ nutates around
${\bf J}$. We can call the wobbling associated with the irregularity in the grain shape {\it irregular wobbling}, to avoid confusion with thermal wobbling (also thermal fluctuations) induced by the Barnett effect (Purcell 1979) and nuclear relaxation (Lazarian \& Draine 1999b).

Obviously, the time-scales of the free-torque motion (e.g. rotation period $P_{\tau}$) are much shorter than other time-scales (e.g., internal relaxations and the gas damping time; see Lazarian 2007). As a result, it is always feasible to average RATs over the free-torque motion. A description of the  free-torque motion for an asymmetric top in terms of Euler angles $\gamma, \alpha$ and $\zeta$ (see Fig. \ref{fap2}) can be found in classical textbooks (e.g. Landau \& Lifshitz (1976; see also WD03). For an AMO with a spheroidal body, as in Fig. \ref{f1}, the Euler angle $\gamma$ is constant, while $\alpha $ and $\zeta$ are functions of time (see Spitzer \& McGlynn 1979). For irregular grains, we can get the Euler angles by numerically solving the equations of motion (see Appendix \ref{apen3}).

\subsection{Averaging over free torque motion and role of $Q_{e3}$}

As mentioned earlier, the rotation period $P_{\tau}$ about the maximal inertia axis and the precession time are
both much shorter than the gas damping time. Therefore, we can average RATs over
these processes. 

\begin{figure}
\includegraphics[width=0.49\textwidth]{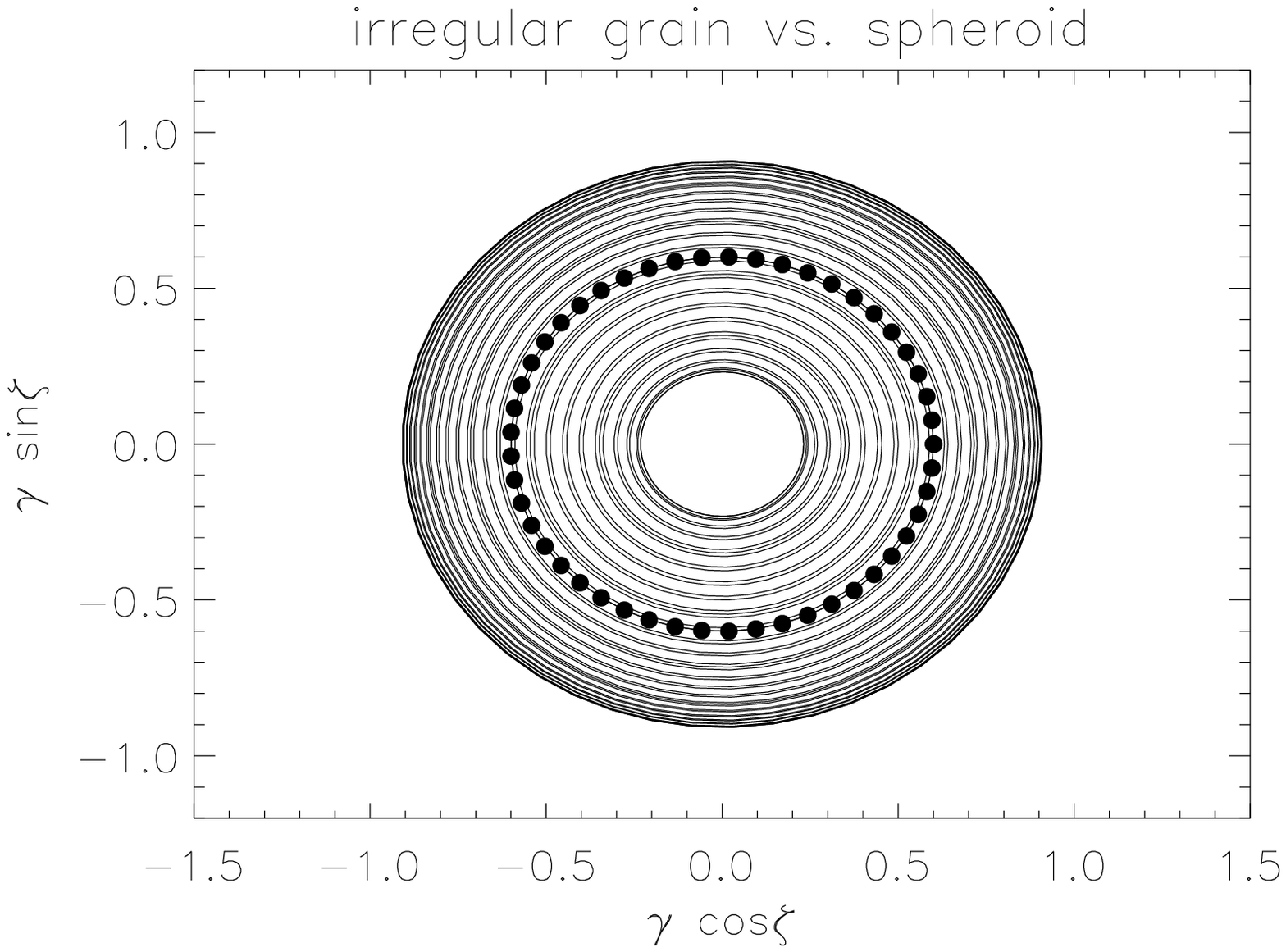}
\includegraphics[width=0.49\textwidth]{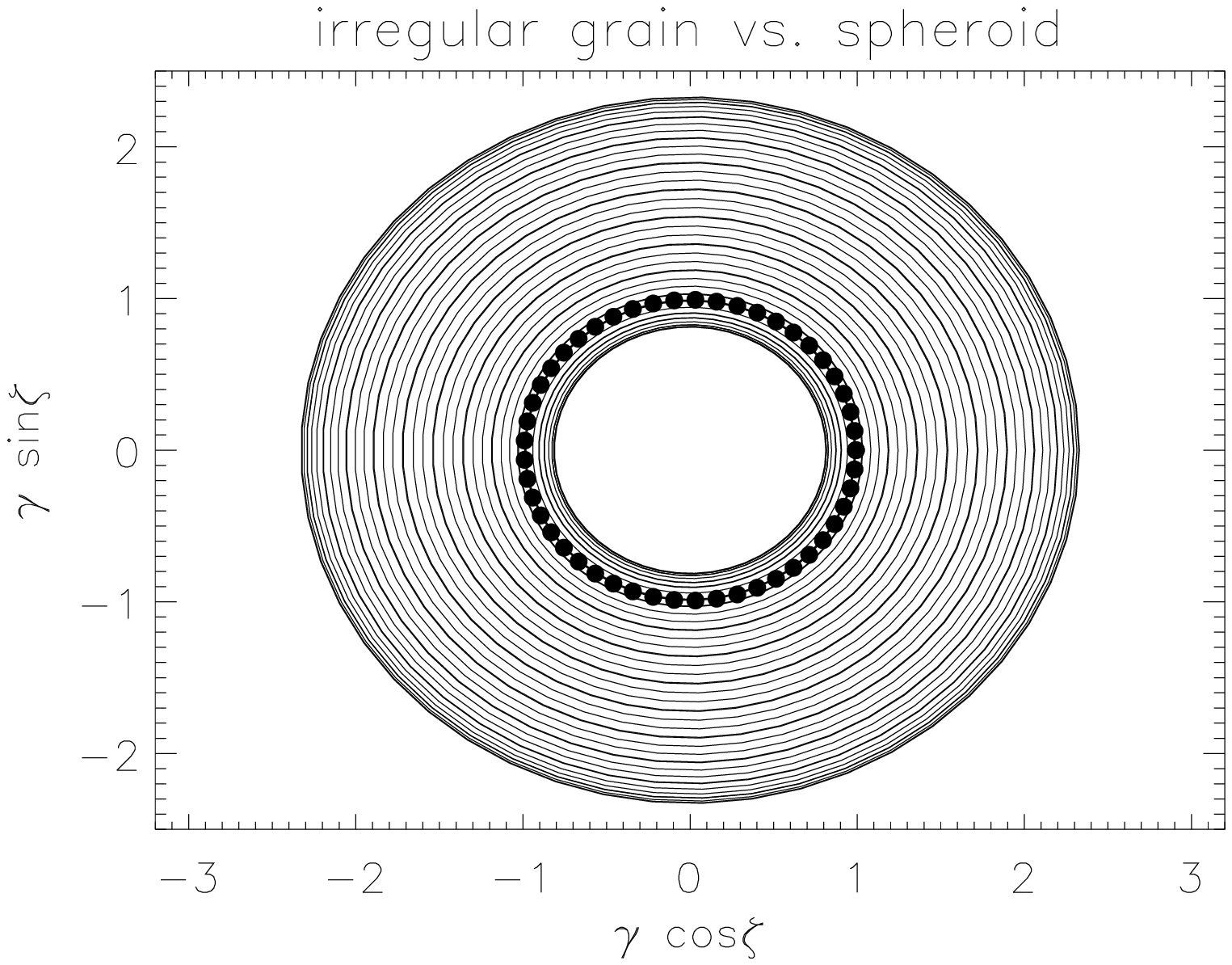}
\caption{The evolution of the tip of the maximal inertia axis ${\bf a}_{1}$ around the
  angular momentum for two initial angles $\gamma_{0}=\pi/10$ ({\it upper panel}) and
  $\gamma_{0}=\pi/4$ ({\it lower panel}) of ${\bf a}_{1}$ and ${\bf J}$. Here $\zeta$ is the nutation angle of $\ma_{1}$ about ${\bf J}$ and $\gamma$ is the azimuthal angle between $\ma_{1}$ and ${\bf J}$. Filled dots and torus show the evolution of ${\bf a}_{1}$  around ${\bf J}$ for the spheroid and the irregular grain respectively.} 
\label{f2}
\end{figure}

In Figure \ref{f2} we show the phase trajectory of the tip of ${\bf a}_{1}$ about ${\bf J}$ for a spheroid and a triaxial ellipsoid for two initial angles $\gamma_{0}=\pi/10$ and $\gamma_{0}=\pi/4$. It is shown that in the case of spheroid, the trajectory is a circle corresponding to the precession of ${\bf a}_{1}$ about ${\bf J}$ at a constant angle $\gamma$, but the evolution of the tip of vector ${\bf a}_{1}$ produces a torus shape for the irregular grain. As a result, the average over free-torque motion for spheroid is concerned only the average over a circle, i.e. over the precession, whereas it is required to average RATs over the torus area (see Fig. \ref{f2}).

The averaging algorithm over free-torque motion for triaxial ellipsoid  has been realized in \S 3 in WD03. However, they averaged RATs over a single period of rotation $P_{\tau}$. 
 
We feel that a different
averaging over a longer time is appropriate. This is well
motivated due to a many orders of magnitude difference of the rotation time
and the time-scale of internal relaxation (see Table \ref{tab1}). 

\begin{figure}
\includegraphics[width=0.49\textwidth]{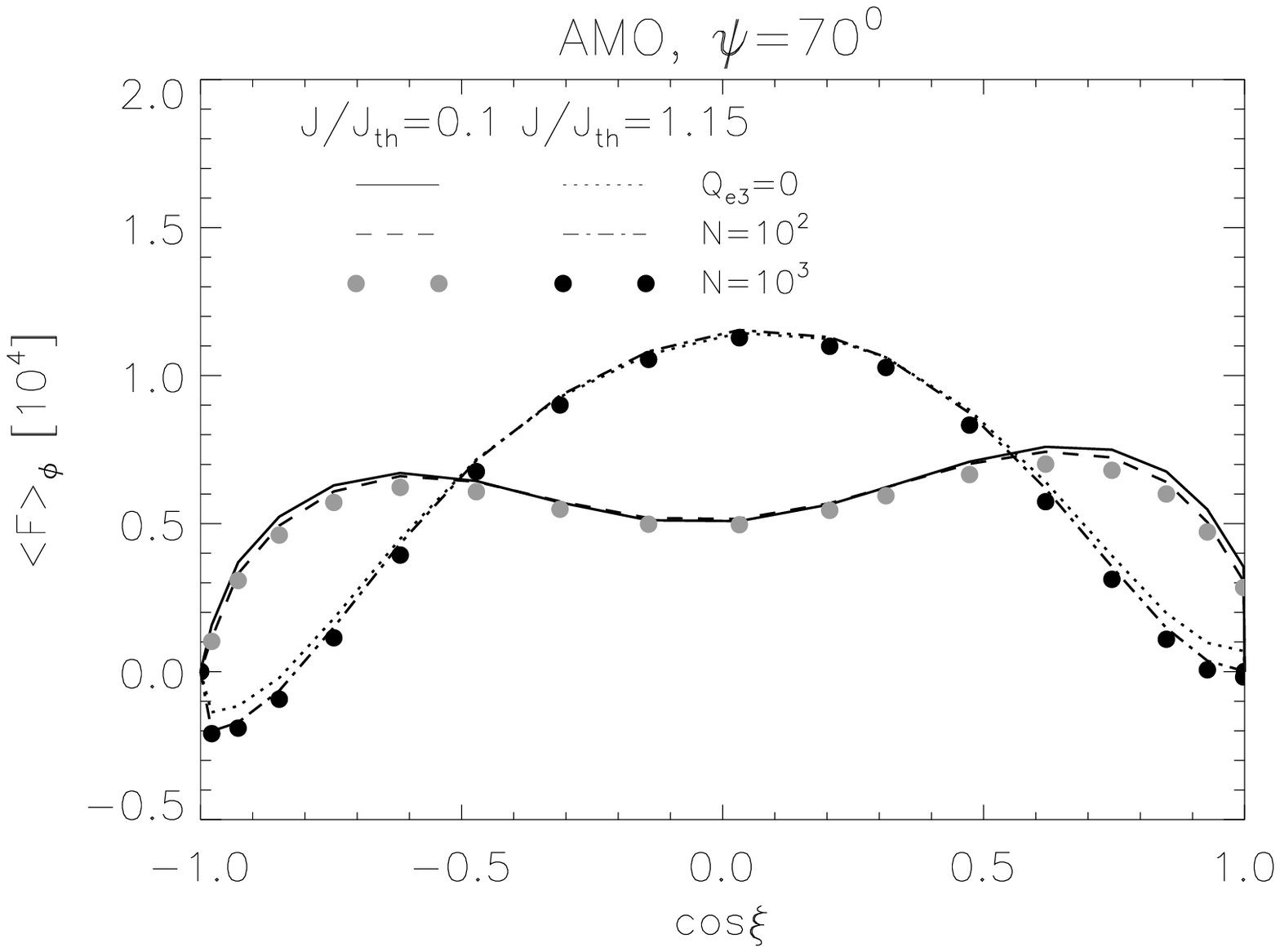}
\includegraphics[width=0.49\textwidth]{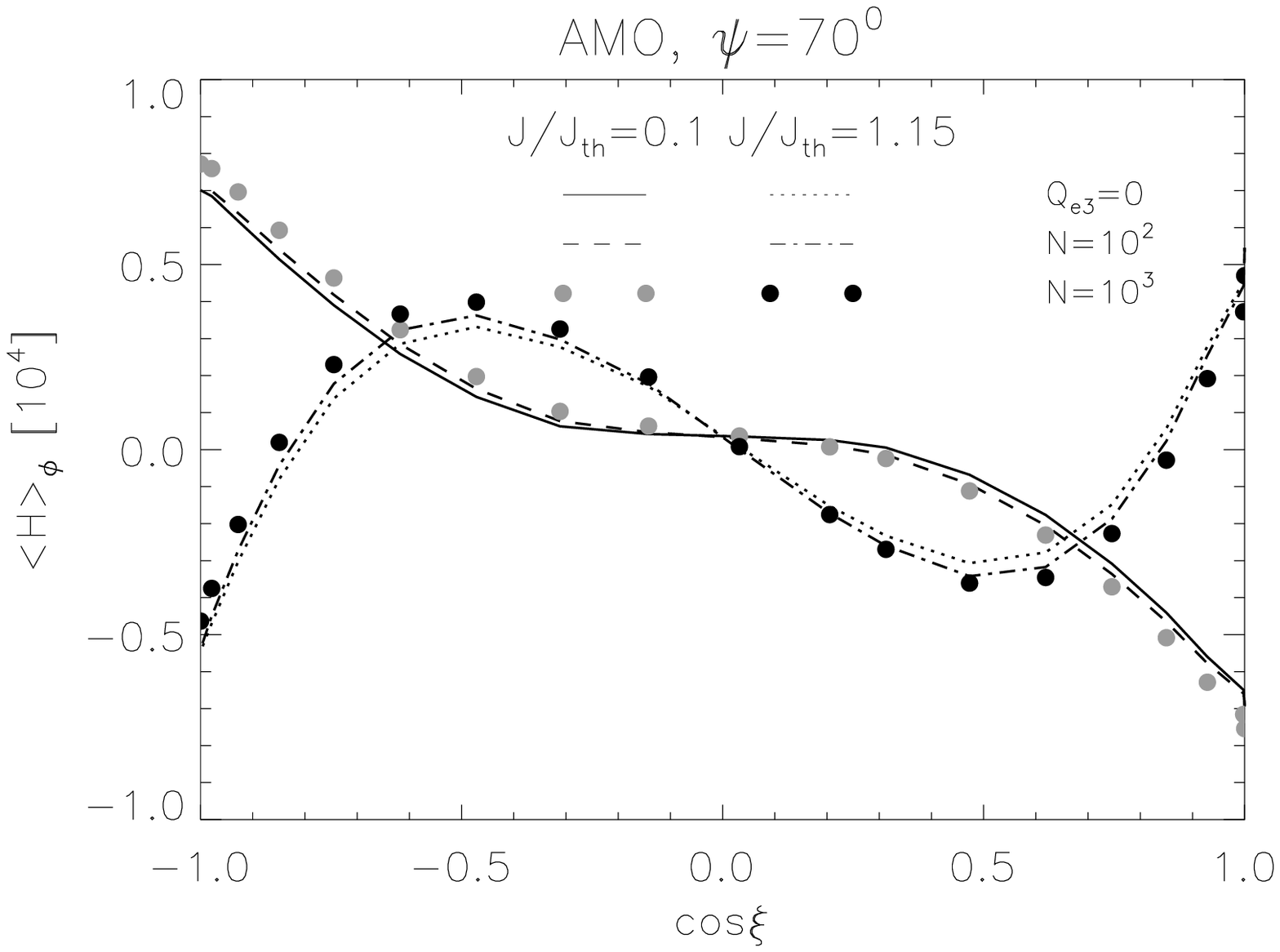}
\caption{$\langle F\rangle_{\phi}$ and $\langle H\rangle_{\phi}$ are shown for different
time-steps $N$ of averaging  and for the case $Q_{e3}=0$ corresponding to two values of
  angular momentum. It can be seen that the
  torques obtained with $N=10^{3}$ steps do not depend on $Q_{e3}$.}
\label{f1a}
\end{figure}

We compare the results obtained by averaging RATs from an AMO (see equations (\ref{eq21})-(\ref{eq23})) with the body being a triaxial ellipsoid over thermal fluctuations for time step $N=10^{2}$ and $N=10^{2}$ with results for $Q_{e3}=0$ and $N=10^{2}$ in Fig. \ref{f1a}. It can be seen that the
torque components $\langle F\rangle_{\phi}, \langle H\rangle_{\phi}$ for the first and the last cases are almost the same, but their averaged torques have a small difference with those obtained in the second case. Therefore, we can expect that the contribution of $Q_{e3}$ to the
spinning and aligning torques is negligible when the averaging of RATs is
performed with a sufficient accuracy (i.e. with sufficient high time step). In our study for the AMO, $Q_{e3}$
is disregarded for both the alignment and spin-up process.

\subsection{Thermal fluctuations and thermal flipping} 

Thermal fluctuations (i.e. thermal wobbling) and flipping arise from the coupling of rotational
and vibrational degrees of freedom induced by internal relaxations. The effect was first discussed in Lazarian (1994). At that time,
the strongest internal relaxation was believed to be associated with the
Barnett effect (Purcell 1979). 

The Barnett relaxation can be easily understood.  
Indeed, a freely rotating paramagnetic 
grain acquires a magnetic moment that is parallel to
angular velocity ${\bf \Omega}$ due to the Barnett effect (first discussed in this context in Dolginov \&
Mytrophanov 1976). The 
Barnett effect is the phenomenon of transferring the macroscopic angular momentum of
a rotating body to 
electrons through flipping electronic spins. To some degree, the 
magnetization of the rotating
body is analogous with the body at rest, which is magnetized by a rotating external
magnetic 
field. The Barnett equivalent magnetic field is ${\bf H}_{Be}={\bf \Omega}/\gamma$ where
$\gamma=e/2mc$ is the magnetogyric ratio of electrons (Purcell 1979).  

Because ${\bf \Omega}$ 
may not coincide with the maximal inertia axis ${\bf a}_{1}$, it precesses continuously in
the grain coordinate system (see Fig. \ref{f1b}). Therefore, the Barnett equivalent 
field can be decomposed into the constant component parallel to the precession axis and the rotating
component which is perpendicular to it\footnote{A more accurate description of the process that account for the finite spin-lattice relaxation is provided in Lazarian \& Draine 1999b}. Apparently, the rotating 
component will induce a dissipation of rotational energy. As a result, we
have an alignment of angular velocity and maximal inertia axis with angular
momentum. 

Lazarian \& Draine (1999a) found a new 
effect related to nuclear spins, which they termed nuclear relaxation. Similar to the rearranging of electronic spins, nuclear spins also
become oriented by angular momentum transferred from the body. Although the 
nuclear magnetization arising from grain rotation is mostly negligible compared to that arising from electrons (see Purcell
1979), the nuclear relaxation was shown to 
be much more efficient  than Barnett relaxation for $10^{-5} < a_{eff} <10^{-4}$ cm grains. This can be easily understood. Spin flipping is a mechanical effect that depends on the angular momentum of the species rather than on their magneto-magnetogyric ratio $\gamma$. Thus, a rotating grain will have the same portion of nuclear and electron spins flipped. The magnetic field that induces such a flip is different for electrons and nuclei (i.e. it is inversely proportional to $\gamma$). The dissipation is proportional to the "equivalent" field squared (i.e. to $B_{eq}^2\sim \frac{1}{\gamma^{2}}$). It does not depend on the value of the nuclear magnetic moment, as the lag between magnetization and $\Omega$ increases with the decrease of the magnetic moments. In other words, the coupling between nuclear spins is less efficient than the 
coupling between electron spins; hence, there is a substantial lag in the nuclear
spin alignment when ${\bf \Omega}$ precesses around ${\bf a_{1}}$.
All in all, although the  magnetic moment arising from nuclear spins of a rotating body is negligible, the corresponding relaxation (i.e. nuclear relaxation) is approximately $10^{6}$ times higher than the Barnett one.

Thermal fluctuations
within the grain body (see Purcell 1979) coupled  via internal relaxation with 
the macroscopic rotation of the grain  can affect the internal alignment, and result in random 
deviations of the major axis ${\bf a_{1}}$ in respect to ${\bf J}$. 
Following the fluctuation-dissipation theorem (Landau \& Lifshitz 1976), the thermal equilibrium distribution of $\ma_{1}$ deviations can be
established. The average of a torque $A$ over thermal fluctuations for a spheroid with $I_{1}>I_{2}=I_{3}$ is defined by (Lazarian \& Roberge 1997)
\begin{align}
\langle A \rangle&=\frac{\int_{0}^{\pi/2} d\gamma  A(\gamma,J)\ms\gamma\mbox{exp}[-E/kT_{d}]}{\int_{0}^{\pi/2}
    d\gamma \ms\gamma \mbox{exp}[-E/kT_{d}]},\label{eq12}
\end{align}
where $h=\frac{I_{1}}{I_{2}}$, $\gamma$ is the deviation angle of ${\bf a}_{1}$ and ${\bf J}$, and the kinetic energy $E$ of the grain is 
\bea 
E(\gamma)=\frac{J^{2}}{2I_{1}}[1+(h-1)\mbox{sin}^{2}\gamma].\label{eq13}
\ena 
For irregular grains with $I_{1}>I_{2}>I_{3}$, the corresponding average is given in Appendix \ref{apen3}.

Using the RAT expressions obtained for the AMO (see Paper I), one can  explicitly  integrate equation (\ref{eq12}) to obtain RATs induced by thermal fluctuations. However, for irregular grains, we need to numerically average RATs according to
equation (D1). The resulting averaged RAT components are plugged into
equations (\ref{eq9}-\ref{eq11}) to obtain $\langle F(\xi, \phi, \psi,
J)\rangle,\langle H(\xi, \phi, \psi, J)\rangle, \langle G(\xi, \phi, \psi,
J)\rangle$, which are required to solve the equations of motion.

The internal relaxation can induce the axis ${\bf a}_{1}$ to flip over with respect to ${\bf J}$. This phenomenon is called thermal
flipping (Lazarian \& Draine 1999a). The probability of  thermal flipping has been obtained in Lazarian \& Draine (1999a), 
\bea 
t_{tf}^{-1}=t_{Bn}^{-1} \mbox{exp}[-0.5(\frac{J^{2}}{J_{th}^{2}}-1)],\label{eq14}
\ena 
where $J_{th}=\sqrt{2 I_{1} k T_{d}}$ is the thermal angular momentum corresponding to the dust temperature, $t_{Bn}=(t_{Bar}^{-1}+t_{nucl}^{-1})^{-1}$ is the internal relaxation time
for Barnett and nuclear relaxation processes.
Because $t_{nucl}$ is usually much shorter than $t_{Bar}$ for grains from $10^{-5}$ to $10^{-4}$ cm, the nuclear relaxation 
dominates the process of thermal flipping for typical {\it aligned} interstellar grains (see Kim \& Martin 1995).

In our analysis, we assume that thermal fluctuation and flipping time-scales
are shorter than the
Larmor precession time. Fortunately, for the
magnetic field 
of the ISM and for astronomical silicate material, the Larmor precession time-scale is much longer than that of thermal
fluctuations and thermal flipping. Therefore, averaging over the later motion is 
appropriate.

\begin{figure}
\includegraphics[width=0.5\textwidth]{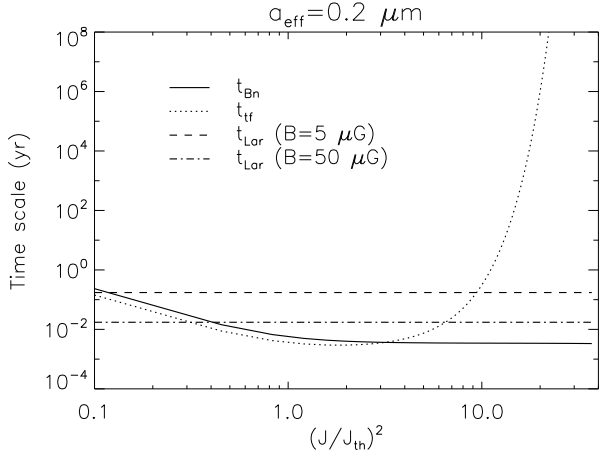}
\caption{Larmor precession, internal relaxation and thermal flipping time-scales for $B=5 
  \mu G$ and $B=50 \mu G$ for a grain size of 
  $a_{eff}=0.2 \mu m$. When magnetic field increases, the Larmor precession 
  time-scale decreases and becomes comparable to the Barnet relaxation time-scale for
  small J.} 
\label{f3}
\end{figure}

For circumstances in which magnetic field is stronger, the Larmor 
precession time-scale may be comparable with the thermal flipping time-scale. For instance, Fig.
\ref{f3} shows that for $B=50 \mu G$ and for ordinary paramagnetic grains, $t_{Lar} \sim t_{tf}$ for $J <10 J_{th}$. For this case, in order to treat properly the grain dynamics, it is necessary to follow both the thermal fluctuation and Larmor precession.

\subsection{Equations of motion} 
The motion of a grain subjected to a net torque is  completely determined 
by three variables: the angle  $\xi$  between the angular momentum vector {\bf J} and 
the magnetic field direction {\bf B}, the precession angle $\phi $ of ${\bf J}$ around {\bf B} and the value of the angular momentum $J$ (see the lower panel of
Fig. 1). The equations of motion, if we {\it disregard paramagnetic dissipation} for these variables are
\begin{align} 
\frac{d\phi}{dt}&= \frac{\gamma
  u_{\mbox{rad}}a_{\mbox{eff}}^{2}\overline{\lambda}}{2J
  \mbox{sin }\xi}\langle G(\xi, \phi, \psi,
J)\rangle-\Omega_{B},\label{eq15}\\ 
\frac{d\xi}{dt}&=\frac{\gamma u_{\mbox{rad}}a_{\mbox{eff}}^{2}\overline{\lambda}}{2J}\langle F(\xi, \phi, \psi,
J)\rangle,\label{eq16}\\ 
\frac{dJ}{dt}&=\frac{1}{2}\gamma u_{\mbox{rad}}a_{\mbox{eff}}^{2}\overline{\lambda} \langle H(\xi, \phi, \psi,
J)\rangle-\frac{J}{t_{gas}},\label{eq17}
\end{align} 
where $\langle F(\xi,\phi,\psi, J)\rangle, \langle G(\xi,\phi,\psi, J)\rangle$, 
and $\langle H(\xi,\phi,\psi, J)\rangle$ are defined by equations (\ref{eq9})-(\ref{eq11}), which are already averaged over thermal fluctuations (see WD03), $t_{gas}$ is the gas damping time-scale (see Table \ref{tab1}) and $\Omega_{B}$ is the Larmor precession 
frequency of the angular momentum about the magnetic
field. 
                    
For the ISM, the Larmor precession time is always shorter than the gas damping; thus, we can average equations (\ref{eq16}) -(\ref{eq17}) over a precession period. As a result, equations (\ref{eq15})-(\ref{eq17}) can be reduced to a couple equation for $\xi$ and $J$ only in which the spinning and aligning torque $\langle F\rangle$ and $\langle H\rangle$ are replaced by $\langle F\rangle_{\phi}$ and $\langle H\rangle_{\phi}$, which denote the quantities obtained from averaging corresponding RATs over the Larmor precession angle $\phi$ from 0 to $2\pi$.\footnote{For the sake of simplicity, hereafter, we denote $\langle F\rangle =\langle F(\xi,\phi,\psi, J)\rangle$, $\langle H\rangle =\langle H(\xi,\phi,\psi, J)\rangle$, and $\langle F\rangle_{\phi} =\langle F(\xi,\phi,\psi, J)\rangle_{\phi}$, $\langle H\rangle_{\phi} =\langle H(\xi,\phi,\psi, J)\rangle_{\phi}$.}

A stationary point in a phase trajectory map is determined by $\xi_{s}$, $J_{s}$, which
are solutions of the equations of motion (see DW97) 
\begin{align}
\frac{d\xi}{dt}&=0,\\ 
\frac{dJ}{dt}&=0.\label{eq18} 
\end{align}
The above stationary point is an attractor point if 
\begin{align}
\left. \frac{1}{\langle H\rangle_{\phi} }\frac{d \langle F\rangle_{\phi}}{d\xi}\right|_{\xi_{s}, J_{s}}<0,\label{eq18_a}
\end{align}
and is a repellor point otherwise (see DW97).

From equations (\ref{eq16}) and (\ref{eq17}), we get $\xi_{s}$, $J_{s}$ that satisfy  
\begin{align}
\langle F(\xi_{s}, \psi, J)\rangle_{\phi}&=0,\label{eq19}\\
J_{s}&=t_{gas} \frac{1}{2}\gamma 
u_{\mbox{rad}}a_{\mbox{eff}}^{2}\overline{\lambda} \langle H(\xi_{s}, \psi, J)\rangle_{\phi}.\label{eq20}
\end{align} 
\section{RAT alignment for AMO}\label{amo1}
We first consider the role of thermal fluctuations in grain alignment based on an AMO consisting of a reflecting spheroid ($I_{2}=I_{3}$, hereafter {\it spheroidal} AMO) and a mirror as in Paper I. Then, in order to see the correspondence of the AMO with irregular grains in terms of dynamics, we  replace the spheroid by an ellipsoid with the principal moments of inertia $I_{1}>I_{2}>I_{3}$ (hereafter {\it ellipsoidal} AMO).

\subsection{RATs: general expressions}
In terms of RATs, an AMO is formally only applicable for $\lambda \ll
a_{eff}$ as only in this case, is geometric optics approach, used to derive the analytical
formulae in Paper I, appropriate. However, in Paper I we proved that the
functional forms of RATs for the AMO and irregular grains for $\lambda \ge a_{eff}$
are similar. By choosing the appropriate ratio $Q_{e1}^{max}/Q_{e2}^{max}$ it is possible to see that the dynamics of the AMO is similar to that of irregular grains, if their torque ratio $Q_{e1}^{max}/Q_{e2}^{max}$ is the same. Therefore, the AMO can act as a proxy for actual grains, and the
advantage of this is that it allows an analytical insight into grain alignment.

For an AMO in which the mirror is tilted by an angle $\alpha$ with the axis ${\bf
  a}_{2}$, the RAT components are given by
(see Paper I)
\begin{align}
Q_{e1}(\Theta, \beta, 0)&=-\frac{4l_{1}}{\lambda}(|n_{1}\mc\Theta-n_{2}\ms\Theta\mc\beta|[n_{1}n_{2}\mc^{2}\Theta\nonumber\\
&+\frac{n_{1}^{2}}{2}\mc\beta\ms2\Theta
-\frac{n_{2}^{2}}{2}\mc\beta\ms2\Theta\nonumber\\
&-n_{1}n_{2}\ms^{2}\Theta\mc^{2}\beta]),\label{eq21}\\
Q_{e2}(\Theta, \beta, 0)&=\frac{4l_{1}}{\lambda}(|n_{1}\mc\Theta-n_{2}\ms\Theta
\mc\beta|[n_{1}^{2}\mc\beta\mc^{2}\Theta\nonumber\\
&-\frac{n_{1}n_{2}}{2}\mc^{2}\beta\ms2\Theta-\frac{n_{1}n_{2}}{2}\ms2\Theta\nonumber\\
&+n_{2}^{2}\mc\beta\ms^{2}\Theta]),\label{eq22}\\
Q_{e3}(\Theta, \beta, 0)&=\frac{4l_{1}}{\lambda}(|n_{1} \mc\Theta-n_{2} \ms\Theta \mc\beta|
n_{1}\ms\beta \nonumber\\
&\times [n_{1}\mc\Theta-n_{2}\mc\beta\ms\Theta])\nonumber\\
&+(\frac{b}{l_{2}})^{2}\frac{2 e a}{\lambda}(s^{2}-1)K(\Theta)\ms 2\Theta,\label{eq23}
\end{align}
where $\Theta$ is the angle between the axis of major inertia ${\bf a}_{1}$ and the radiation direction ${\bf k}$ (see Fig. \ref{f1}{\it lower}); $l_{1}$ is the distance from the square mirror of side $l_{2}$ to the center mass, $\lambda$ is the wavelength, $n_{1}=-\ms\alpha, n_{2}=\mc\alpha$ are components of the normal vector
of the mirror in the grain coordinate system (see Fig.~\ref{f1})\footnote{Note that RATs in equations (\ref{eq21}-\ref{eq23}) have opposite signs compared with those in Paper I because in Paper I we have defined $n_{1}=\ms\alpha$ (i.e. we incorporated the minus sign of RATs into $n_{1}$)}, $a, b$ are minor and major semi-axes of the spheroid, $s=a/b<1$ and $e$ is the eccentricity of the spheroid; $K(\Theta)$ is the fitting function (see also Paper I).
As in Paper I, we treat the AMO with $\alpha=45^{\circ}$ as our default model.

RATs at a precession angle $\Phi$ (see Fig. \ref{f1}{\it lower}) are given by (see DW97)
\begin{align}
Q_{e1}(\Theta, \beta, \Phi)&=Q_{e1}(\Theta, \beta, 0),\label{eq24}\\ 
Q_{e2}(\Theta, \beta, \Phi)&=Q_{e2}(\Theta, \beta,
0)\mbox{cos}\Phi-Q_{e3}(\Theta, \beta, 0)\mbox{sin}\Phi,\label{eq25} 
\\  
Q_{e3}(\Theta, \beta, \Phi)&=Q_{e2}(\Theta, \beta,
0)\mbox{sin}\Phi+Q_{e3}(\Theta, \beta, 0)\mbox{cos}\Phi.\label{eq26}  
\end{align} 

\subsection{Alignment with respect to {\bf k}}\label{kalign}
First we consider the alignment in respect to the direction of radiation ${\bf k}$, which also correspond to the situation when the direction of light ${\bf k}$ coincides with that of magnetic field ${\bf B}$ (i.e. $\psi=0^{\circ}$).
\subsubsection{Analytical averaging RATs for one component $Q_{e1}$}
As discussed earlier in \S 3.4, $Q_{e3}$ does not affect RAT alignment,
apart from inducing the precession of angular momentum ${\bf J}$ about ${\bf k}$. Thus, the alignment problem
only involves $Q_{e1}$ and $Q_{e2}$. To clarify the role of the torque components, we first consider the alignment driven by the component $Q_{e1}$ in the presence of thermal fluctuations.

For the default AMO (i.e. $\alpha=45^{\circ}$), the contribution arising from the change in the cross-section is insignificant (see Fig. 6 in Paper I). Thus, we can ignore the factor $A_{\perp}=|n_{1}\mc\Theta-n_{2}\ms\Theta\mc\beta|$ present in equations (\ref{eq21})-(\ref{eq23}). As a result, with an accuracy of $5\%$, equation (\ref{eq21}) can be rewritten as
\begin{align}
Q_{e1}\approx Q_{e1}^{max} [\frac{(3\mcs\Theta-1)}{2}+\mc\Theta\mc\beta+\mss\Theta\mc2\beta],\label{eq27a}
\end{align}
where $Q_{e1}^{max}=\frac{-4l_{1}}{\lambda}n_{1}n_{2}$.

When there is incomplete alignment of $\ma_{1}$ and ${\bf J}$, we have
\begin{align}
\mc\Theta=\mc\xi\mc\gamma-\ms\xi\ms\gamma\ms\eta,\label{eq212}
\end{align}
where $\eta$ is the precession angle of ${\bf a}_{1}$ around ${\bf J}$,
$\gamma$ is the angle of ${\bf a}_{1} $ and ${\bf J}$, and ${\xi}$ is the
angle between ${\bf J}$ and ${\bf B}$. In addition, $\beta$ is a complicated function of Euler angles.

Substituting equation (\ref{eq212}) into (\ref{eq27a}) and averaging the
resulting expression over the precession angle $\eta$, the second and third term involved $\beta$ are averaged to zero. Hence, we obtain
\begin{align}
Q_{e1}&\approx Q_{e1}^{max}[3(\mcs\gamma\mcs\xi +\frac{\mss\gamma\mss\xi}{2})-1].\label{eq213}
\end{align}
Using equations ({\ref{eq12}) and (\ref{eq13}) for the expression (\ref{eq213}) we obtain
  \begin{align}
\langle Q_{e1}\rangle&\approx Q_{e1}^{max}\frac{\int_{0}^{\pi} d\gamma (\mcs\gamma\mcs\xi
  +\frac{\mss\gamma\mss\xi}{2}-1)}{\int_{0}^{\pi} d\gamma \ms\gamma
  e^{-\frac{J^{2}}{J_{th}^2}[1+(h-1)\mss\gamma]}}\nonumber\\
&\times \ms\gamma
  e^{-\frac{J^{2}}{J_{th}^2}[1+(h-1)\mss\gamma]},\label{eq214}
\end{align}
where $J_{th}=\sqrt{2I_{1}kT_{d}}$.
The aligning and spin-up torque are respectively given by (see
Eqs. \ref{eq11}-\ref{eq13})
\begin{align}
\langle F\rangle_{\phi}&=-\langle Q_{e1}\rangle \ms\xi,\label{eq215}\\
\langle H\rangle_{\phi}&=\langle Q_{e1}\rangle \mc\xi.\label{eq216}
\end{align}
Hence,
\begin{align}
\langle F
\rangle_{\phi}&\approx -\ms\xi Q_{e1}^{max} \frac{\sqrt{\pi}}{4}\frac{e^{(-1+h)\mJs}(1+3\mc2\xi)[e^{h\mJs}\sqrt{-1+h}\mJ}{\mJs(-1+h)e^{-h\mJs}Erfi(\sqrt{-1+h\mJ})}\nonumber\\
&+\ms\xi Q_{e1}^{max}\frac{e^{\mJs}(3+2(-1+h)\mJs)]\sqrt{\pi}Erfi(\sqrt{-1+h}\mJ)}{\mJs(-1+h)e^{-h\mJs}Erfi(\sqrt{-1+h\mJ})}\label{eq28a},\\
\langle H \rangle_{\phi}&\approx \mc\xi Q_{e1}^{max} \frac{\sqrt{\pi}}{4}\frac{e^{(-1+h)\mJs}(1+3\mc2\xi)[e^{h\mJs}\sqrt{-1+h}\mJ}{\mJs(-1+h)e^{-h\mJs}Erfi(\sqrt{-1+h\mJ})}\nonumber\\
&-\mc\xi Q_{e1}^{max}\frac{e^{\mJs}(3+2(-1+h)\mJs)]\sqrt{\pi}Erfi(\sqrt{-1+h}\mJ)}{\mJs(-1+h)e^{-h\mJs}Erfi(\sqrt{-1+h\mJ})}.\label{eq29a}
\end{align}

Equation (\ref{eq28a}) reveals that $\langle F\rangle_{\phi}=0$ for $\xi=0, \pi$ and $\mc2\xi=-1/3$, regardless of angular momentum $J$.  In addition, zero points $\mc\xi=0$ and $\mc2\xi=-1/3$ of $\langle H\rangle_{\phi}$ do not depend on $J$ either. This indicates that thermal fluctuations within the {\it spheroidal} grains do not alter the value of angular momentum (i.e. $J =0$) at low-$J$ attractor points produced by RATs when $J\rightarrow J_{th}$. Therefore, we expect that the resulting alignment is not significantly affected by thermal fluctuations.
\begin{figure}
\includegraphics[width=0.49\textwidth]{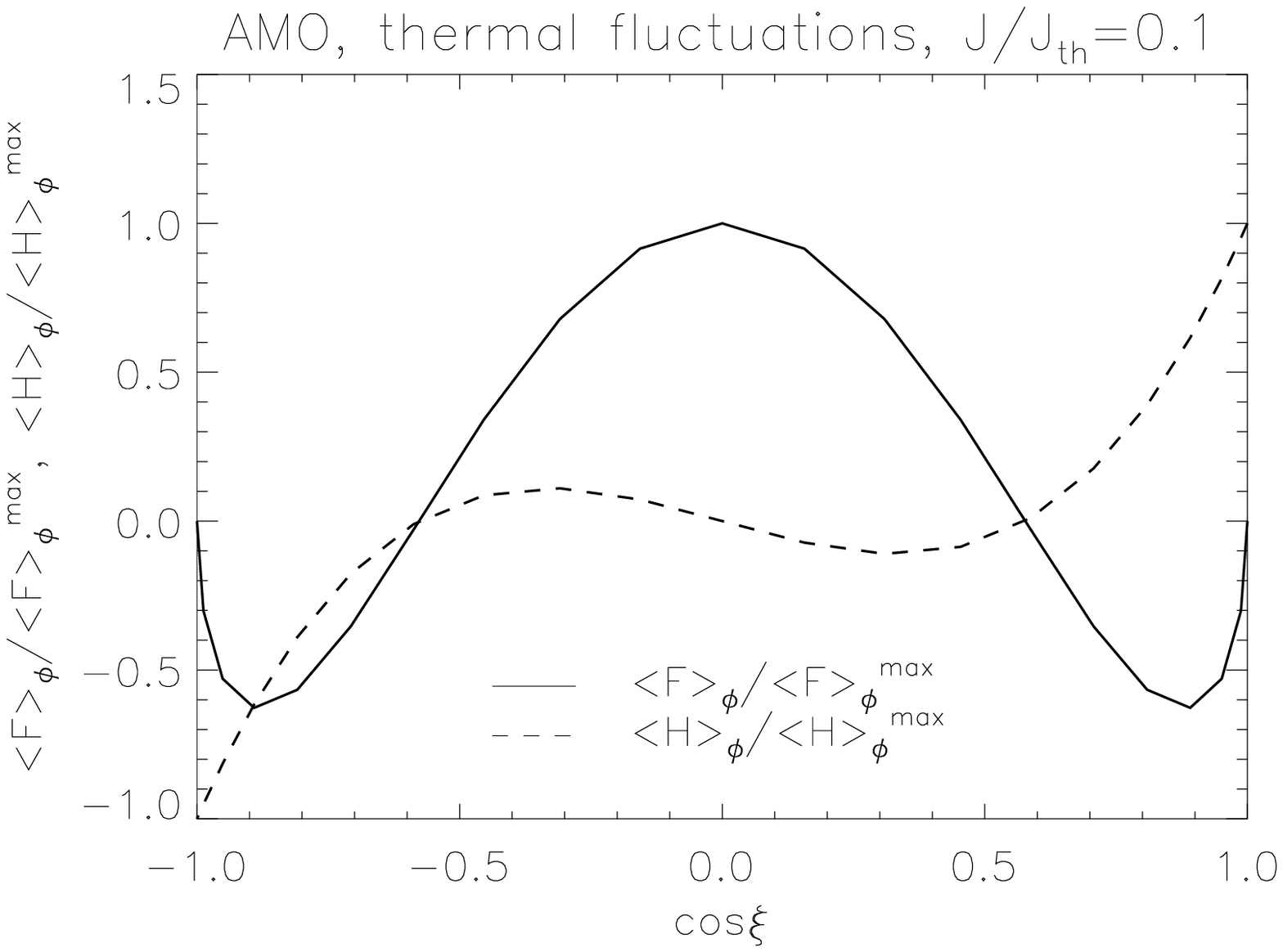} 
\includegraphics[width=0.49\textwidth]{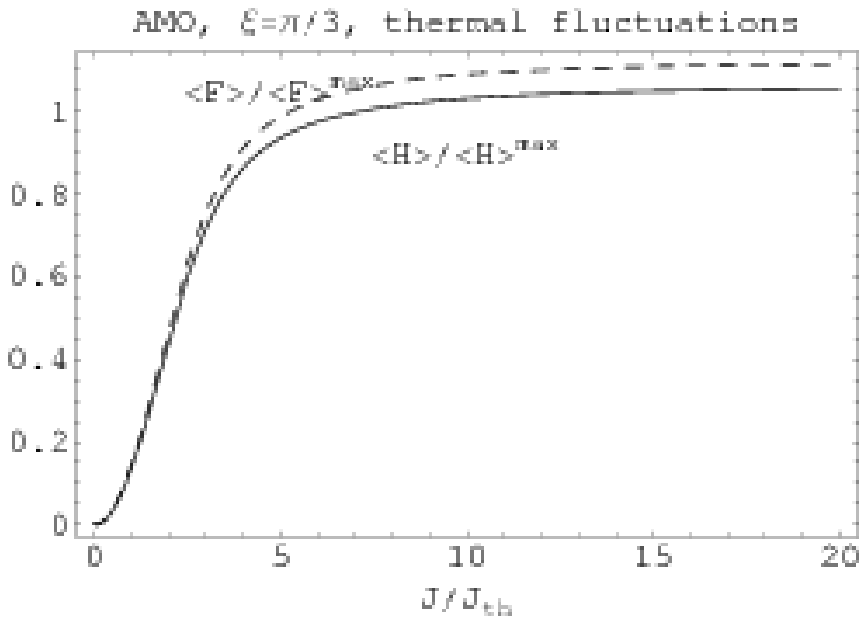} 
\caption{{\it Spheroidal} AMO: aligning $\langle F\rangle_{\phi}$ and spinning $\langle
  H\rangle_{\phi}$ torques  averaged over thermal fluctuations for the case $\psi=0^{\circ}$ and $Q_{e2}=0$. {\it Upper
  panel} shows $\langle F\rangle_{\phi}$ and $\langle H\rangle_{\phi}$
  as functions of $\xi$ for $J=0.1 J_{th}$. {\it Lower panel} shows that $\langle
  F\rangle_{\phi}$ and $\langle H\rangle_{\phi}$ for $\xi=\pi/3$, decrease rapidly with
  $J/J_{th}$ decreasing and get saturated as $J/J_{th} \gg 1$.}
\label{f41}
\end{figure}

The above features of RATs can be seen in Fig. \ref{f41}, which shows RAT components  $\langle F\rangle_{\phi}$
(solid line) and $\langle H\rangle_{\phi}$ (dashed line) as
functions of $\xi$ at a value of  angular momentum $J=0.1 J_{th}$. It follows that there will be
four stationary points in the phase map, corresponding to $\langle F\rangle_{\phi} =0$ at $\xi=0,
\pi/3, 2\pi/3, \pi$. In addition, the stationary point $\xi=0$ is a high-$J$
attractor point as $\left. \frac{d\langle F\rangle_{\phi}}{\langle H\rangle_{\phi}d\xi}\right|_{\xi=0}<0$ and $\langle H\rangle_{\phi}(\xi=0)>0$ (see the upper panel in Fig. \ref{f41}). The lower panel of Fig. \ref{f41} shows
$\langle F\rangle_{\phi}$ and $\langle H\rangle_{\phi}$ as functions of $J/J_{th}$ for $\xi=\pi/3$.
There, it can be seen that for $J\gg J_{th}$,   $\langle F\rangle_{\phi}$ and $\langle H\rangle_{\phi}$ are
saturated as a result of the perfect coupling of ${\bf a}_{1}$ with ${\bf J}$,
which makes RATs  independent on the angular momentum. As $J/J_{th}$ decreases,
$\langle F\rangle_{\phi}$ and $\langle H\rangle_{\phi}$ decrease steeply (see Fig. \ref{f41}).

\subsubsection{Averaged RATs for both components $Q_{e1}, Q_{e2}$}
The analytical averaging over thermal fluctuations for $Q_{e2}$ is more complicated because of its dependence upon $\Phi$, which is a
function of Euler angles $\alpha, \gamma$ and $\xi, \phi, \psi$ (see equation \ref{eq25} and Appendix A). Therefore, we use numerical averaging
for RATs, rather than deriving analytical expressions for them.
 The resulting torques $\langle F\rangle_{\phi},
\langle H\rangle_{\phi}$ are used to solve the equations of motion (\ref{eq16})-(\ref{eq17}). 

 Fig. \ref{f4} shows spin-up and alignment torques for $J =10^{-3} J_{th}, 0.9 J_{th}$ and $10 J_{th}$ corresponding to cases in which thermal fluctuations are dominant, important and negligible, respectively. It can be seen that for $J \gg J_{th}$, $\langle F\rangle_{\phi}=0$ for  $\mc\xi=\mp
1$. It turns out that for $J \gg J_{th}$, the AMO creates only two stationary points corresponding to
perfect alignment of ${\bf J}$ with ${\bf B}$. When $J = 10^{-3} J_{th}$, there are
new stationary points at $\mc\xi=\pm0.9$ corresponding to $\langle F\rangle_{\phi}=0$. This means that thermal
fluctuations can produce  new stationary points. At the same time, the magnitude of spinning torque $\langle H\rangle_{\phi}$ decreases as $J$ decreases (i.e. when thermal fluctuations increase; see Fig. \ref{f4}{\it lower}). As a result, the alignment of grains is expected to be similar to the case without thermal fluctuations.

\begin{figure}
\includegraphics[width=0.49\textwidth]{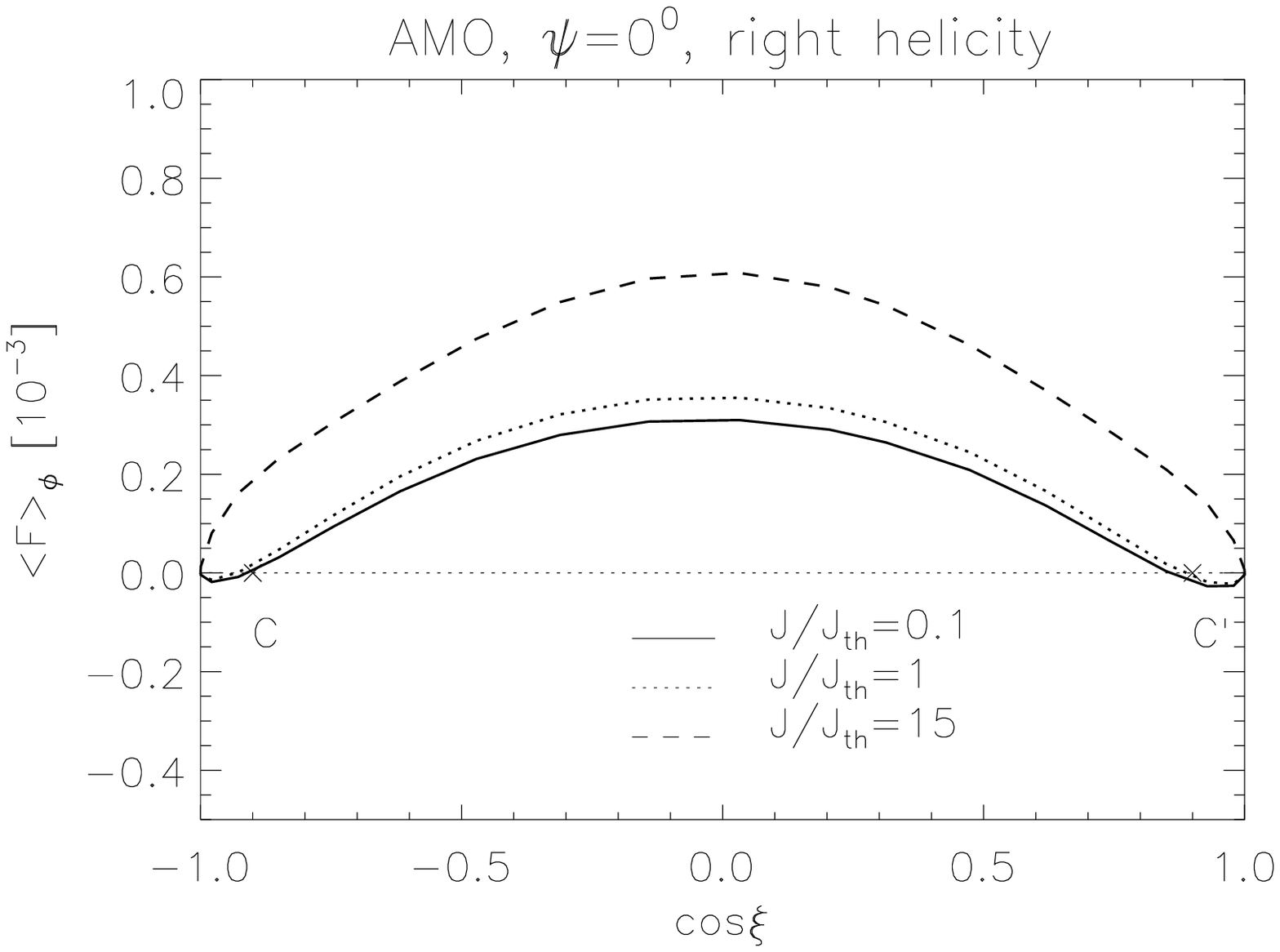} 
\includegraphics[width=0.49\textwidth]{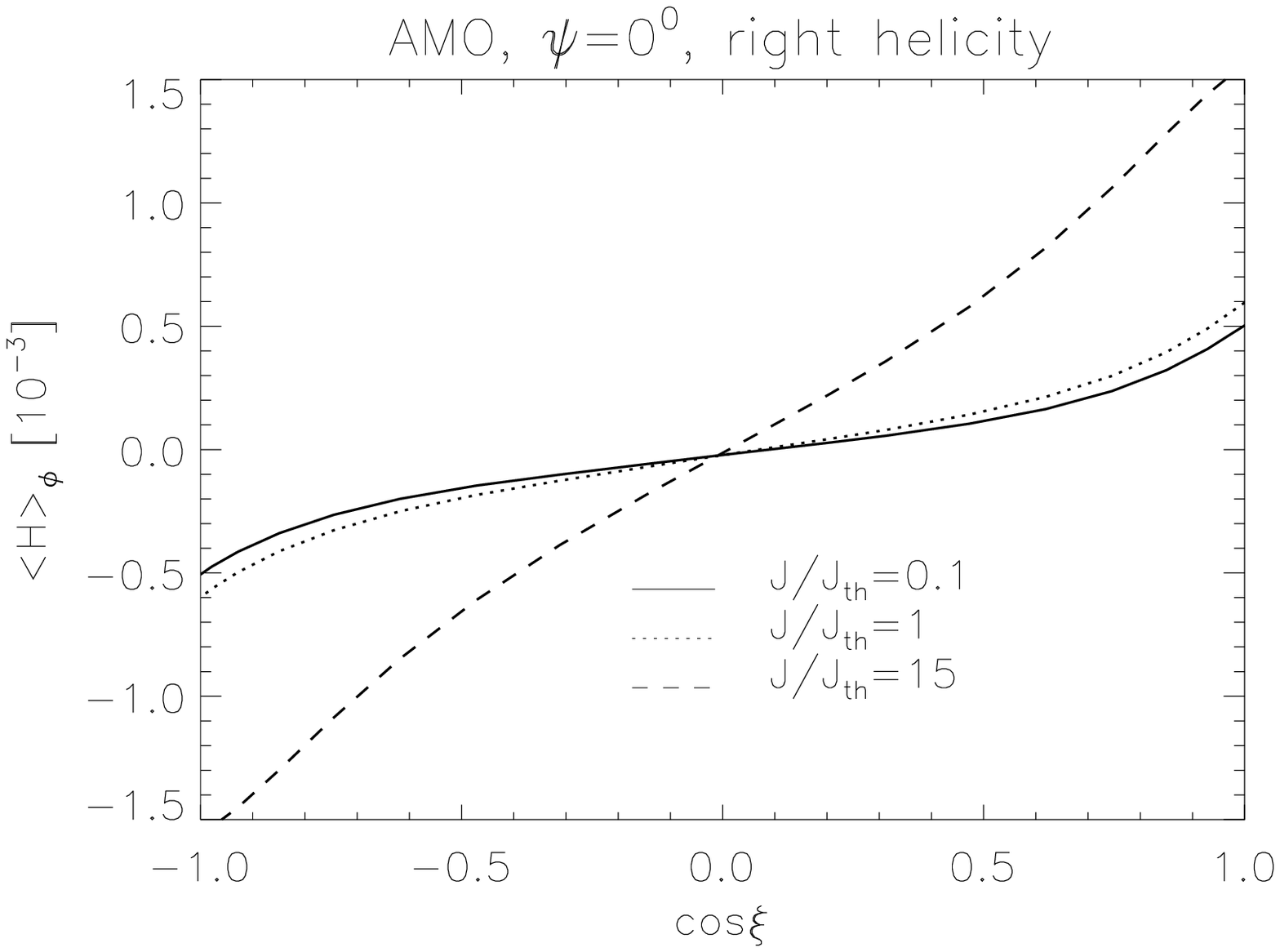} 
\caption{{\it Spheroidal} AMO: Aligning ({\it upper panel}) and spinning ({\it lower panel}) torques  averaged over thermal fluctuations for different $J/J_{th}$. {\it Upper panel} shows that as $J =10^{-3}
  J_{th}$, one new zero point of $\langle F\rangle_{\phi}$ appears at
  $\mc\xi=-0.9$. Both $\langle F\rangle_{\phi}$ and $\langle H\rangle_{\phi}$
  exhibit rapid decrease with $J/J_{th}$ decreasing.}
\label{f4}
\end{figure}

\subsubsection{Trajectory maps}
Let us consider the dynamics of the {\it spheroidal} AMO driven by RATs averaged over thermal fluctuations, and test the predictions using the analytical results above.

We solve the equations of motion for grains having the same initial
angular momentum and uniform orientation distribution with respect to the interstellar
magnetic field.  Parameters necessary for calculations for the ISM conditions are given in
Table \ref{tab1}. Phase trajectory maps are plotted in coordinates ($\mc\xi$,$J/I_{1}\omega_{T}$)
where $\xi$ is the angle between ${\bf J}$ and ${\bf B}$, and $\omega_{T}=\sqrt{kT_{d}/I_{1}}$. The upper and lower parts of which correspond to ${\bf a}_{1}$ initially parallel and anti-parallel to ${\bf J}$,
respectively. For a grain size $a_{eff}=0.2 \mu m$ and shape 1 (see Fig. \ref{sh14}), the ISRF can produce high stationary points with $J_{high} \sim 200 I_{1}\omega_{T}$ for $\psi=0^{\circ}$. To represent high and low attractor points together, we consider the ISRF with $u_{rad}=u_{ISRF}/10$ where $u_{ISRF}=8.64 \times 10^{-13} erg/cm^{3}$ is the energy density of the ISRF. Thus, the phase trajectory maps are shown with $J_{max}=20 I_{1}\omega_{T}$ in the present paper.

 Further in the paper, a stationary point on the phase trajectory maps is marked
by a circle is an attractor point, which is the point to which adjacent
trajectories tend to converge, and a stationary point marked by a cross denotes a repellor point (i.e., the
trajectories are repulsed while approaching it; see also Paper I).\footnote{In some trajectory maps, we do not label all existing repellor points.} In general, a phase trajectory map may have low-$J$
and  high-$J$ attractor points or only low-$J$ attractor points (see more in Paper
I). To see clearly the modification induced by thermal fluctuations on grain dynamics, we frequently show side by
side the trajectory maps for the case without thermal fluctuations as in Paper I, and with thermal
fluctuations. Phase trajectory maps for the {\it spheroidal} AMO are shown in Figs. \ref{f5a}, \ref{f5} and \ref{f6} in which the upper
and lower panel correspond to the cases without and with thermal fluctuations.

 \begin{table}
\caption{Physical parameters for diffuse ISM }
\begin{displaymath}
\begin{array}{rrrr} \hline\hline\\
\multicolumn{1}{c}{\it Definitions} & \multicolumn{1}{c}{\it Values} \\[1mm]
\hline\\
{\rm Gas~ density}& {\rm n=30~ cm^{-3}}\\[1mm]
{\rm Gas~temperature}& {\rm T_{gas}=100~K}\\[1mm]
{\rm Gas~damping~time}& {\rm t_{gas}=4.6\times
  10^{12}(\frac{\hat{\rho}}{\hat{n}\hat{T}_{gas}^{1/2}}) a_{-5}~ s}\\[1mm]
{\rm Dust~temperature}& {\rm T_{d}=20~K}\\[1mm]
{\rm Anisotropy~degree}&{\rm \gamma=0.1}\\[1mm]
{\rm Mean~wavelength}&{\rm \overline{\lambda}=1.2~ \mu m}\\[1mm]
{\rm Mean~density~of~the ISRF}&{\rm {u}_{ISRF}=8.64 \times 10^{-13}~ erg~ cm^{-3}}\\[1mm]
{\rm Effective~ grain~ size}&{\rm a_{eff}=0.2 \mu m}\\[1mm]
\\[1mm]
\hline\hline\\
\end{array}
\end{displaymath}
Here $\hat{T}_{gas}=T_{gas}/100~ K,~\hat{n}=n/30~ g~ cm^{-3}$, and
$a_{-5}=a_{eff}/10^{-5}~ cm $.~ $\hat{\rho}=\rho/3~ g~ cm^{-3}$ where $\rho=3~ g~ cm^{-3}$ is the mass density of
the grain.
\label{tab1}
\end{table}

\begin{figure}
\includegraphics[width=0.49\textwidth]{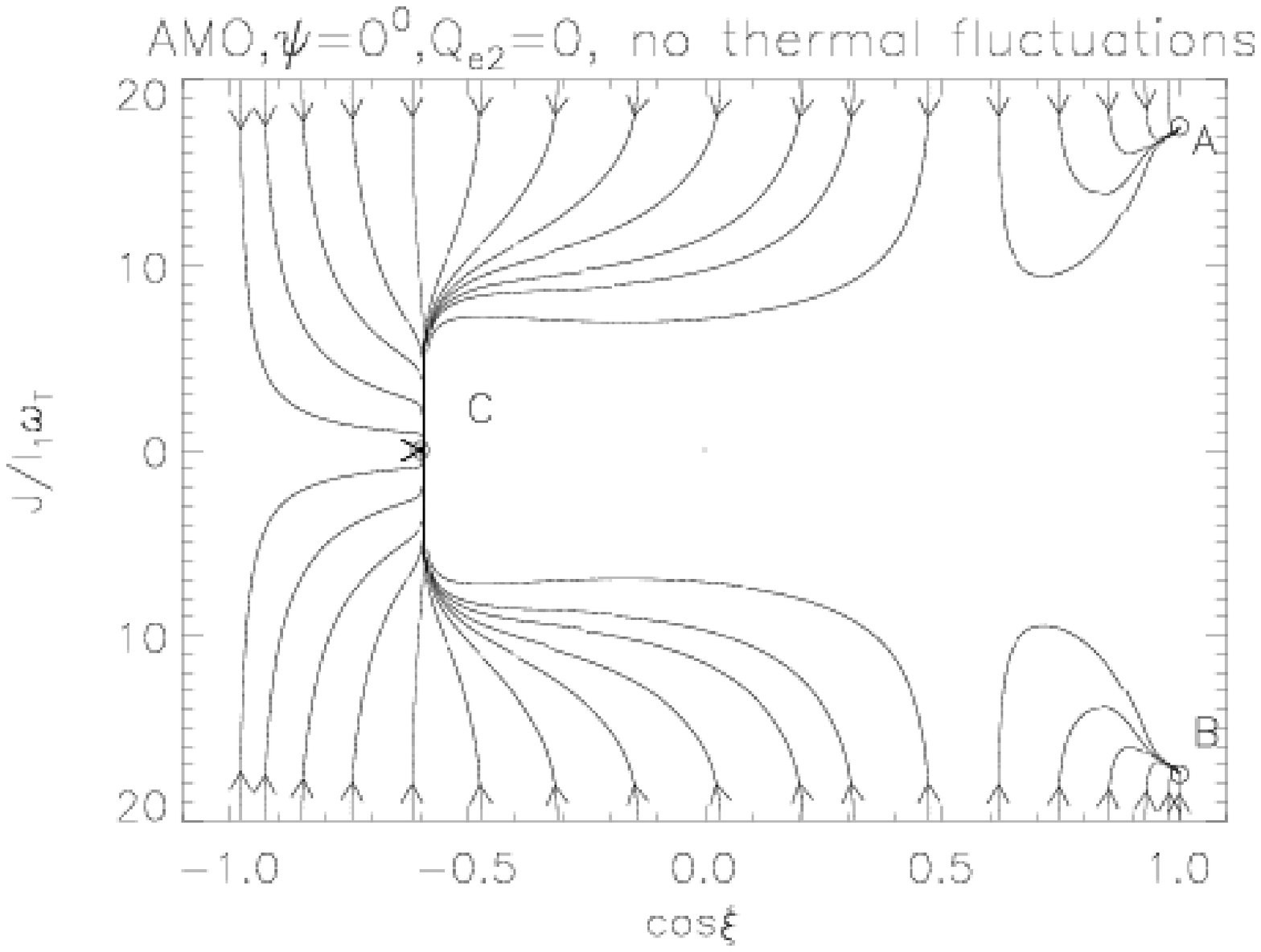} 
\includegraphics[width=0.49\textwidth]{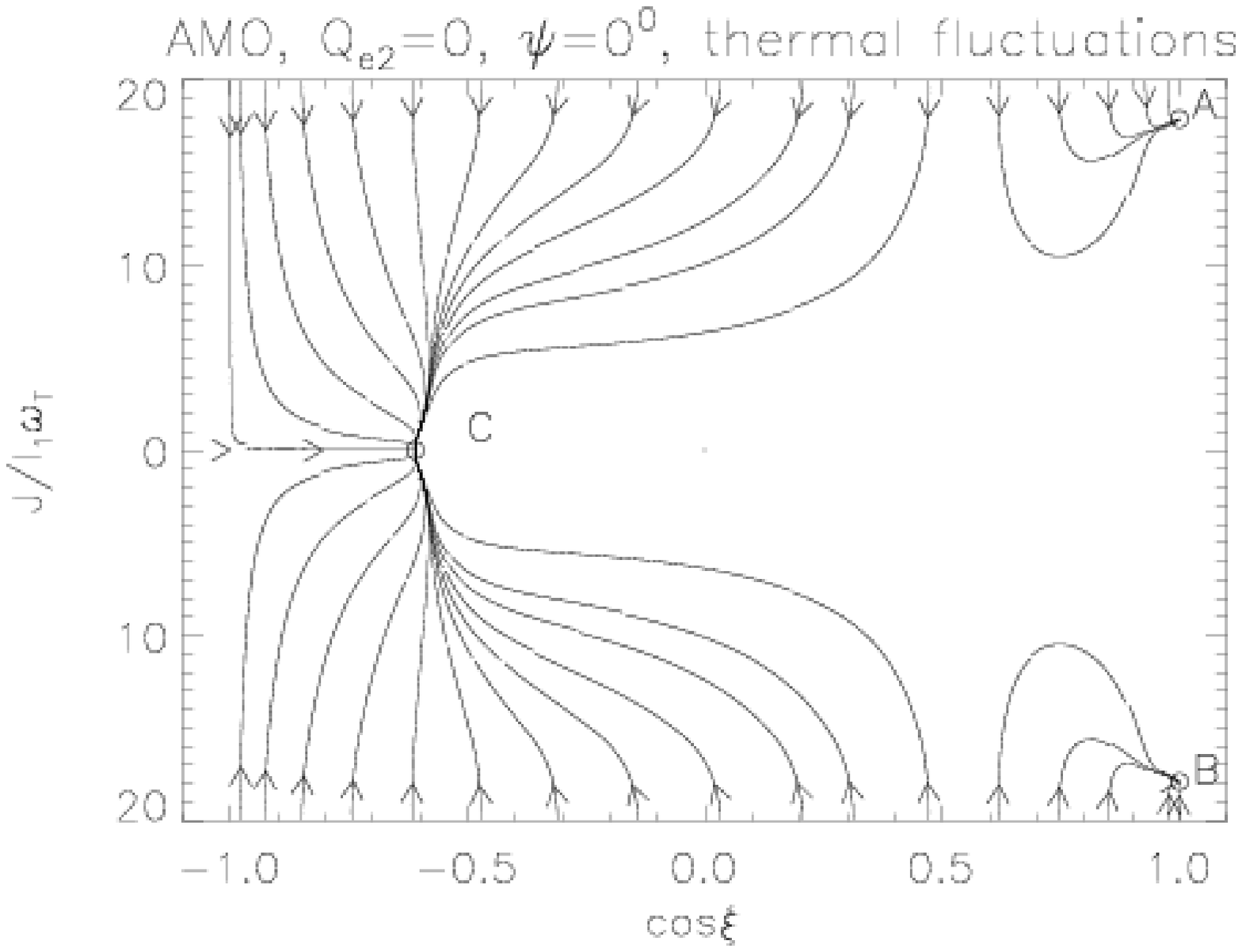} 
\caption{{\it Spheroidal} AMO: Phase trajectory maps for the alignment  by $Q_{e1}$ only. The upper panel shows the phase map in
  the absence of thermal fluctuations and the
  lower panel shows the phase map when thermal fluctuations are accounted for. The panels
  show that the position of the zero-$J$ attractor point at $\mc\xi=-0.6$ is not changed even in the presence of thermal fluctuations.}
\label{f5a}
\end{figure}

When the grain alignment is only driven by $Q_{e1}$ and $\psi=0^{\circ}$
(Fig. \ref{f5a}), we see that each phase map has two high-$J$ attractor points
A and B. In the absence of thermal fluctuations,
the torque $\langle H\rangle_{\phi} $ decelerates grains to the attractor
point C with $J=0$ (upper
panel). When thermal fluctuations are present, the attractor point C is not affected. This
result is consistent with our analytical predictions in \S 4.2.1 
\begin{figure}
\includegraphics[width=0.49\textwidth]{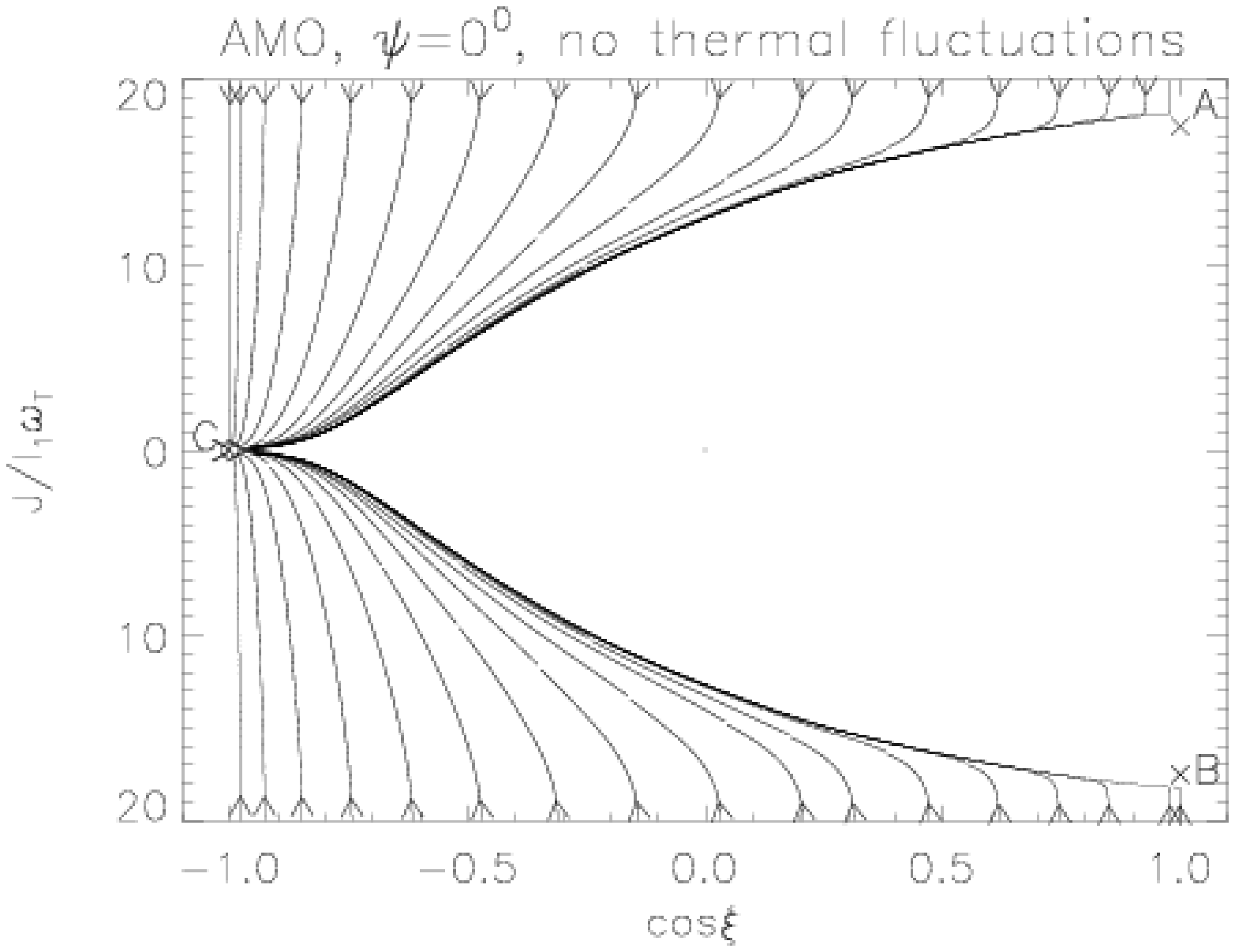} 
\includegraphics[width=0.49\textwidth]{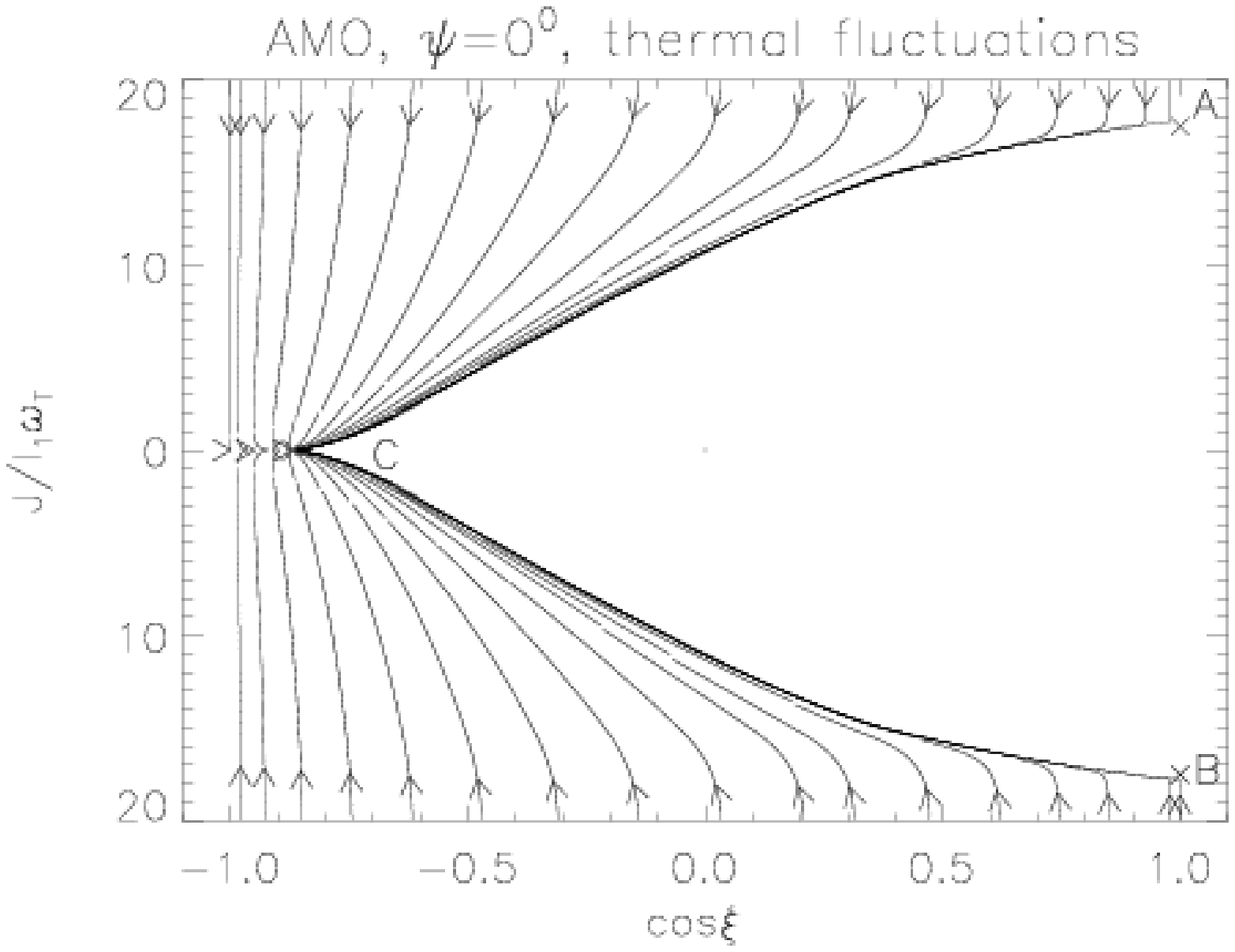} 
\caption{{\it Spheroidal} AMO: similar to Fig. \ref{f5a}, but grains are aligned by all torque components. Figs
  show the shift of the zero J attractor point from $\mc\xi=-1$ ({\it upper}) to
  $\mc\xi=-0.9$ ({\it lower}), but its angular momentum  remains equal zero even when thermal fluctuations are accounted for.}
\label{f5}
\end{figure}

When the components $Q_{e1}$ and $Q_{e2}$ act together, Fig. \ref{f5} shows that the angular momentum of low attractor point remain the same (i.e. $J$ remains equal zero). However, its position is slightly shifted to $\mc\xi=\pm0.9$.

\subsection{Alignment with respect to {\bf B}}\label{balign}
Below we consider grain dynamics when the magnetic field plays the role of alignment axis. As an example, the radiation direction $\psi=70^{\circ}$ is adopted.
\begin{figure}
\includegraphics[width=0.49\textwidth]{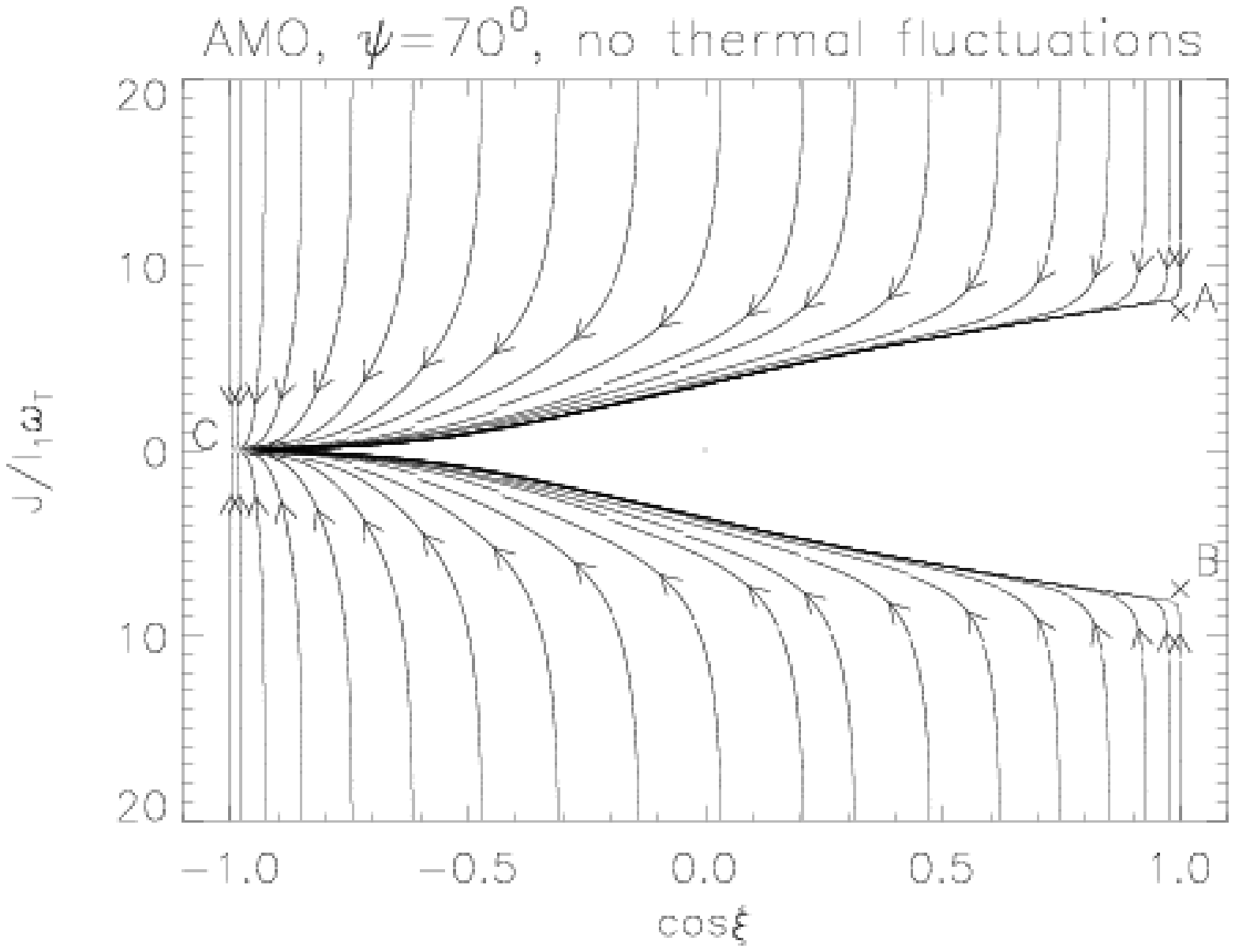} 
\includegraphics[width=0.49\textwidth]{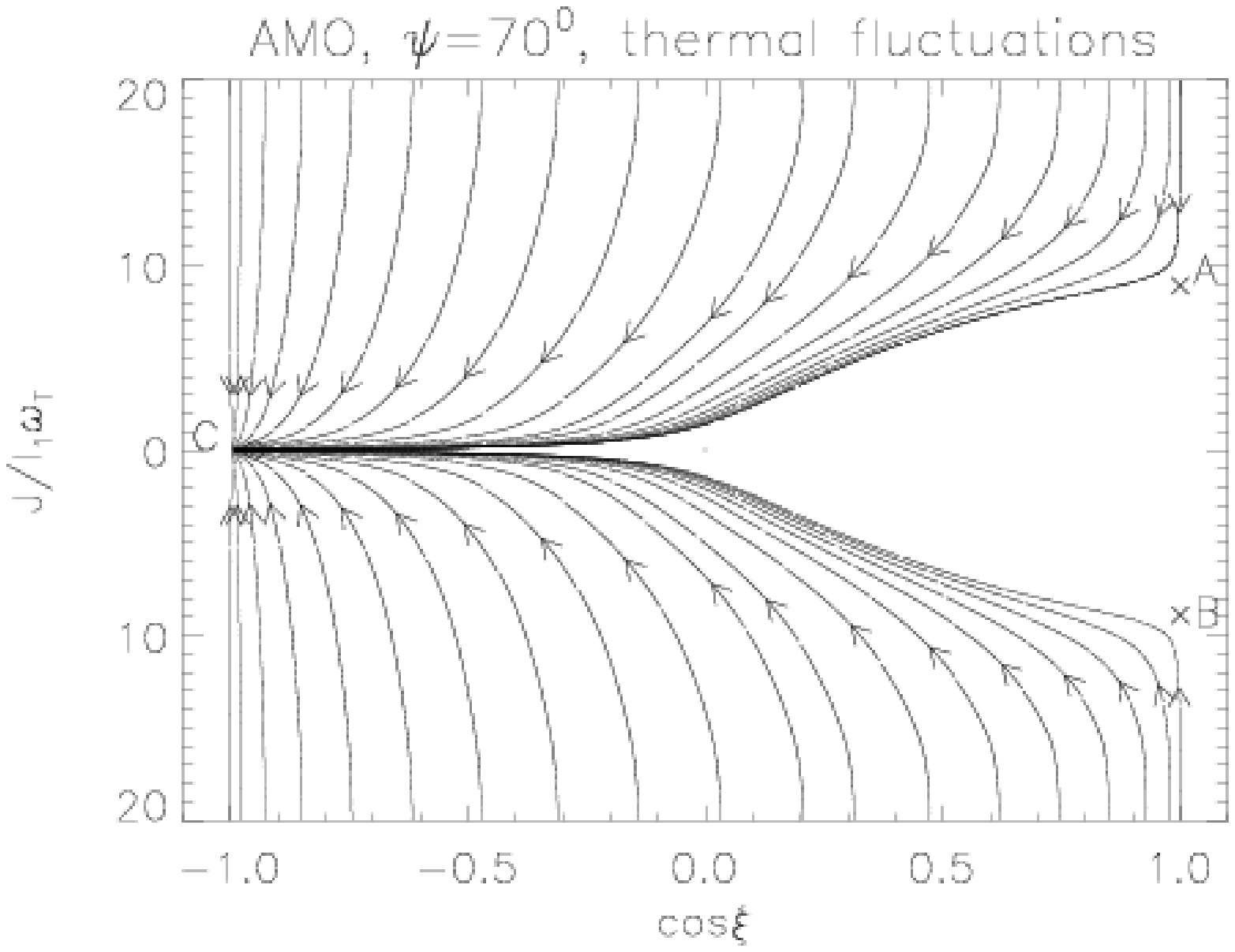} 
\caption{{\it Spheroidal} AMO: similar to Fig. \ref{f5}, but for $\psi=70^{\circ}$. Figs
  show that the zero J attractor point C at $\mc\xi=-0.9$ (upper panel)
  panel)  is unchanged in the presence of thermal wobbling (lower panel).}
\label{f6}
\end{figure}

Fig. \ref{f6} shows that thermal fluctuations do not increase the value of angular
momentum at the attractor point C for the case $\psi=70^{\circ}$. For other angles $\psi$, we also found that the zero-J
attractor point C is unchanged in the presence of thermal fluctuations. 

It is easy to see that in the assumption of ${\bf a}_{1}\|{\bf J}$, the irregular shape of the ellipsoid of inertia (i.e. $ I_{1}\ne I_{2}\ne I_{3}$), is not important for the grain dynamics. However, the irregularity in the grain shape becomes important in the case of a wobbling grain because the averaging of RATs over free-torque motion (see Fig. \ref{f2}) depends on its ellipsoid of inertia. To address such effects, the {\it spheroidal} AMO can be modified. For
instance, if the mirror is not weightless, the free motion of the {\it spheroidal} AMO will be of
a triaxial ellipsoid of inertia, rather than a spheroid. Further, in \S \ref{calign}, we replace the spheroid of the AMO by a triaxial ellipsoid.

\subsection{Alignment for an {\it ellipsoidal} AMO}\label{calign}
In this section, we study the alignment of the {\it ellipsoidal} AMO in which the spheroid body (see
Fig. \ref{f1}) is replaced by an ellipsoid with moments of inertia $I_{1}: I_{2}:I_{3}=1.745:1.62:0.876$, which is similar to shape 1 (see Fig. \ref{sh14}).

We recall that stationary points, which determine the alignment of grains
correspond to $\langle F \rangle_{\phi}=0$ (see equation~\ref{eq20}). Fig. \ref{f7} shows the torque components for $\psi=70^{\circ}$. It can be seen that for $J \gg
J_{th}$ (i.e., thermal fluctuations are negligible), $\langle F\rangle_{\phi}=0$ for  $\mc\xi=\mp
1$. This indicates that the {\it ellipsoidal} AMO produce two stationary points corresponding to
perfect alignment of ${\bf J}$ with ${\bf B}$. As $J \to 0.1 J_{th}$, there appears a
new stationary point at $\mc\xi=-0.8$. It means that thermal
fluctuations produce a new stationary point at $\mc\xi \sim -0.8$. However, this new stationary point is a repellor point as  $\left. \frac{d\langle F\rangle_{\phi}}{\langle H\rangle_{\phi}
  d\xi}\right|_{\mc\xi_{s}=-0.8} >0$ (see Fig. \ref{f7}). In addition, the lower panel shows the change in sign of the spinning torque from negative to positive as $J\sim 0.1 J_{th}$ for $\mc\xi \sim -1$. Thus, the  stationary point $\mc\xi=-1$ is an attractor because
$\left. \frac{d\langle F\rangle_{\phi}}{\langle H\rangle_{\phi}
  d\xi}\right|_{\mc\xi_{s}=-1} <0$ (see Fig. \ref{f7}). 
\begin{figure}
\includegraphics[width=0.49\textwidth]{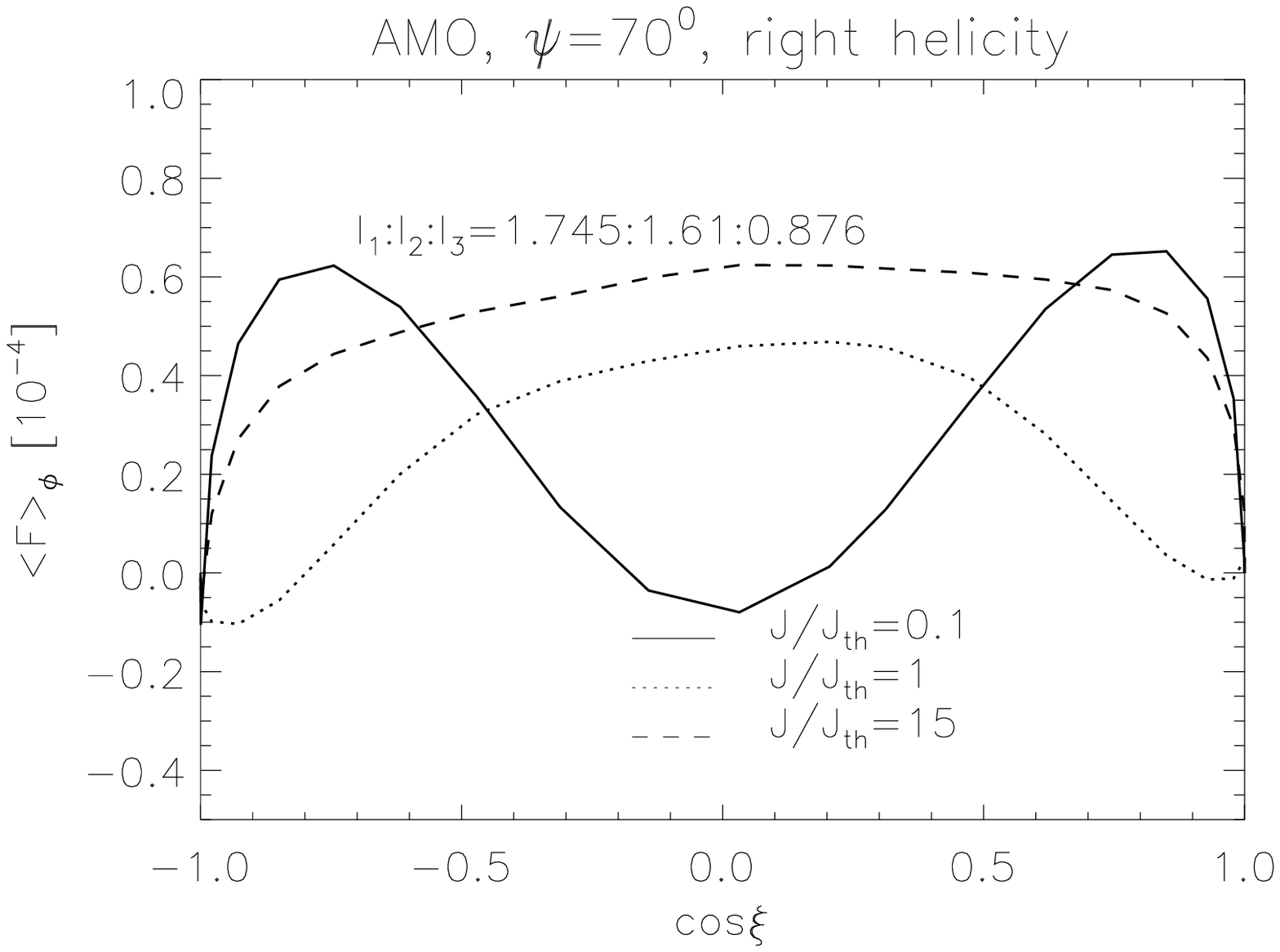} 
\includegraphics[width=0.49\textwidth]{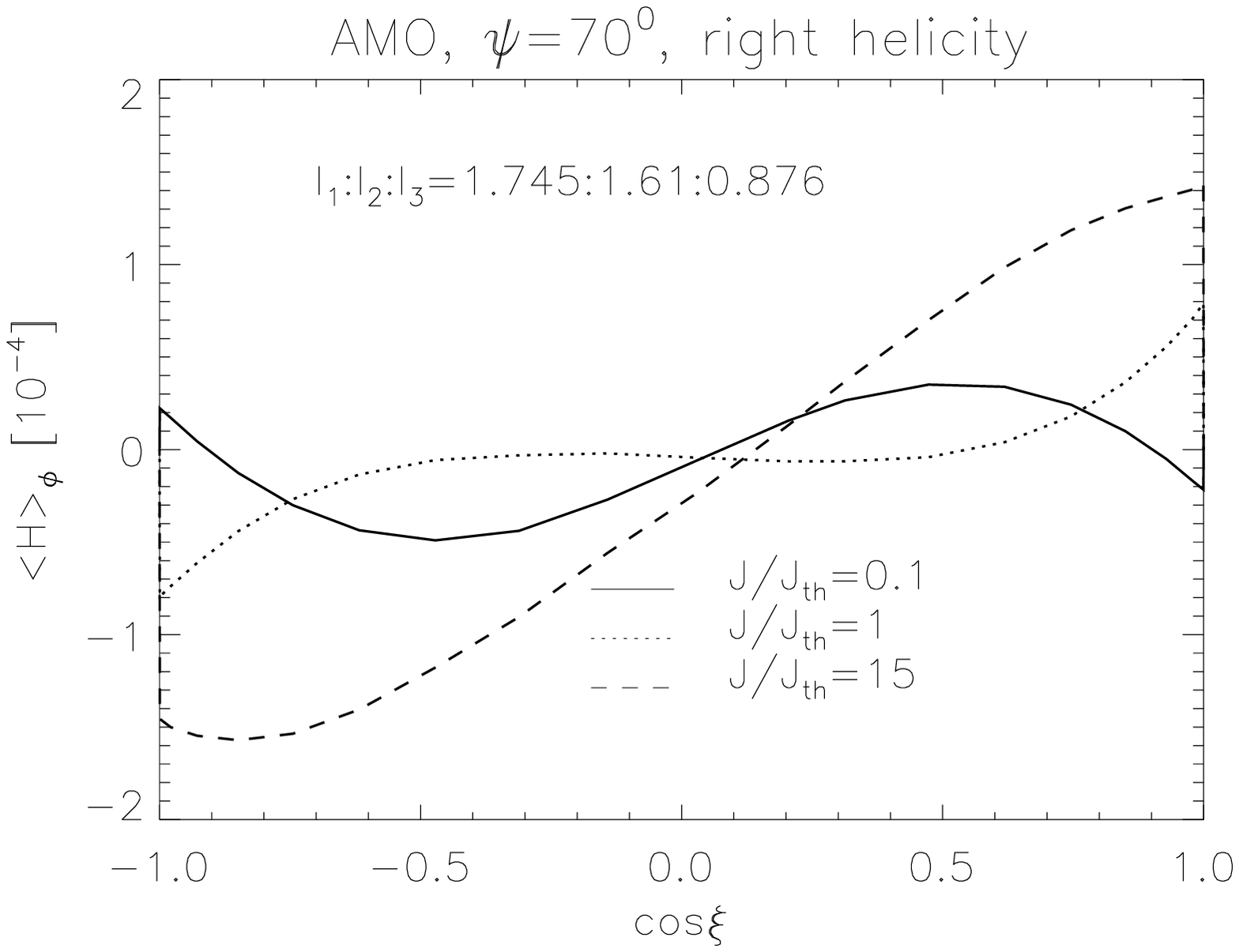} 
\caption{{\it Ellipsoidal} AMO: aligning and spinning torques averaged over thermal fluctuations as
  a function of $\xi$ for several angular momenta. {\it Upper panel} shows that as $J =1
  J_{th}$, one new zero point of $\langle F\rangle_{\phi}$ appears at $\mc\xi=-0.8$. While the change in sign of $\langle H\rangle_{\phi}$ in the vicinity of $\mc\xi=-1$ for $J=0.1 J_{th}$ is shown in lower panel.}
\label{f7}
\end{figure}
Therefore, thermal fluctuations within the {\it ellipsoidal} AMO modify the properties of RATs as $J
\rightarrow 0$, and split a zero-$J$ attractor point in absence of thermal wobbling into two low-$J$ attractor points. The corresponding trajectory maps for this case are shown in Fig. \ref{f8}
\begin{figure}
\includegraphics[width=0.49\textwidth]{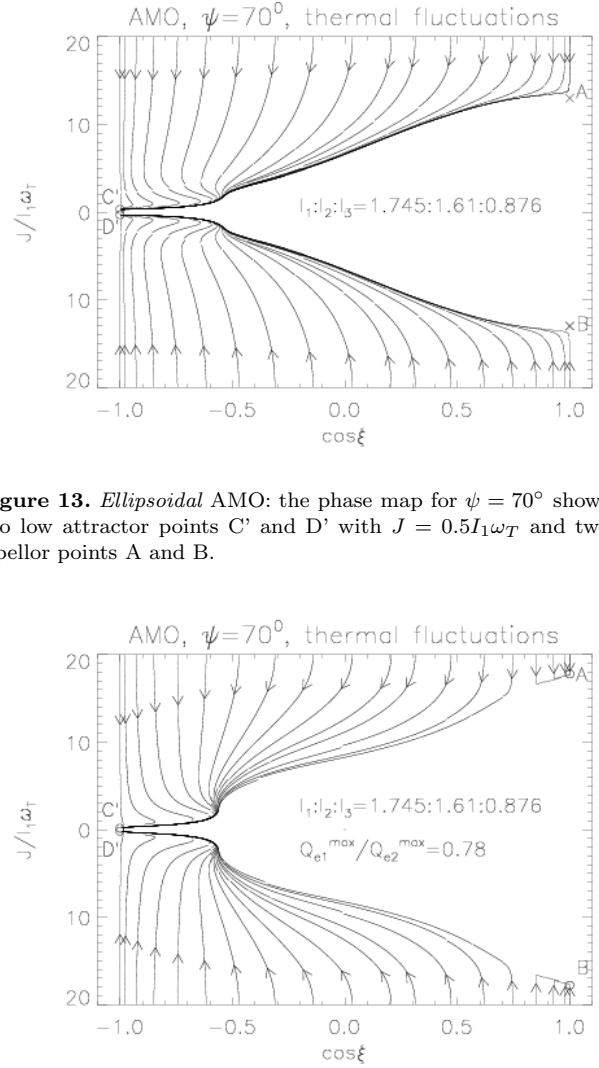} 
\caption{{\it Ellipsoidal} AMO: the phase map for $\psi=70^{\circ}$ shows two low attractor points C' and D' with $J=0.5 I_{1}\omega_{T}$ and two repellor points A and B.}
\label{f8}
\end{figure}

Now we rescale the amplitude of $Q_{e1}, Q_{e2}$ so that $Q_{e1}^{max}/Q_{e2}^{max}=0.78$. For this AMO, Paper I shows that the grains are aligned with two high-J attractor points and a zero-J attractor point. However, in presence of thermal fluctuations and for the {\it ellipsoidal} AMO with $I_{2}\ne I_{3}$, the zero-J attractor point becomes the attractor points with the value of angular momentum of the order of thermal value, as seen in Fig. \ref{f9a}.
 \begin{figure}
\includegraphics[width=0.49\textwidth]{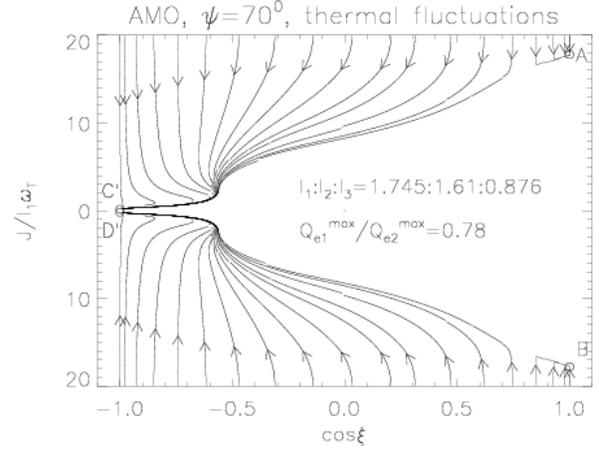} 
\caption{{\it Ellipsoidal} AMO: the phase map when the torque ratio $Q_{e1}^{max}/Q_{e2}^{max}$ is changed to the default value to $0.78$ for $\psi=70^{\circ}$ shows two low attractor points with $J=0.5J_{th}$ and two high J attractor points A and B with $J=18 J_{th}$.}
\label{f9a}
\end{figure}

\section{RAT alignment for irregular grains}
\subsection{Properties of averaged RATs}

We use the publicly available DDSCAT code (Draine \& Flatau 2004) to calculate
RATs for irregular grains (i.e. shape 1 and 3; see Fig. \ref{sh14}). The optical constant function for astronomical silicate is adopted (Draine \& Lee 1984). We 
computed RAT efficiency $ \bQ_{\lambda}(\Theta,\beta,\Phi)$ in the lab coordinate system (see Fig. \ref{f1}), over 32 directions of $\Theta$ in the range [$0, \pi$] and 33 values of $\beta$ between
$0$ and $2\pi$ for $\Phi=0$ (angle produced by ${\bf a}_{1}$ and the plane $\hat{e}_{1},\hat{e}_{3}$ where $\hat{e}_{i}$
are three unit vectors of the lab coordinate system; see Fig. \ref{f1}). The RAT efficiency $ \bQ_{\lambda}(\Theta,\beta,\Phi)$  for an arbitrary angle
$\Phi$ is easily calculated using equations (\ref{eq24})-(\ref{eq26}). In our
paper, we calculate RATs for irregular grains of size $a_{eff}=0.2\mu m$ induced by monochromatic radiation field of $\lambda=1.2 \mu m$ and the spectrum of the ISRF (see Mathis et al. 1983).

\begin{figure}
\includegraphics[width=0.49\textwidth]{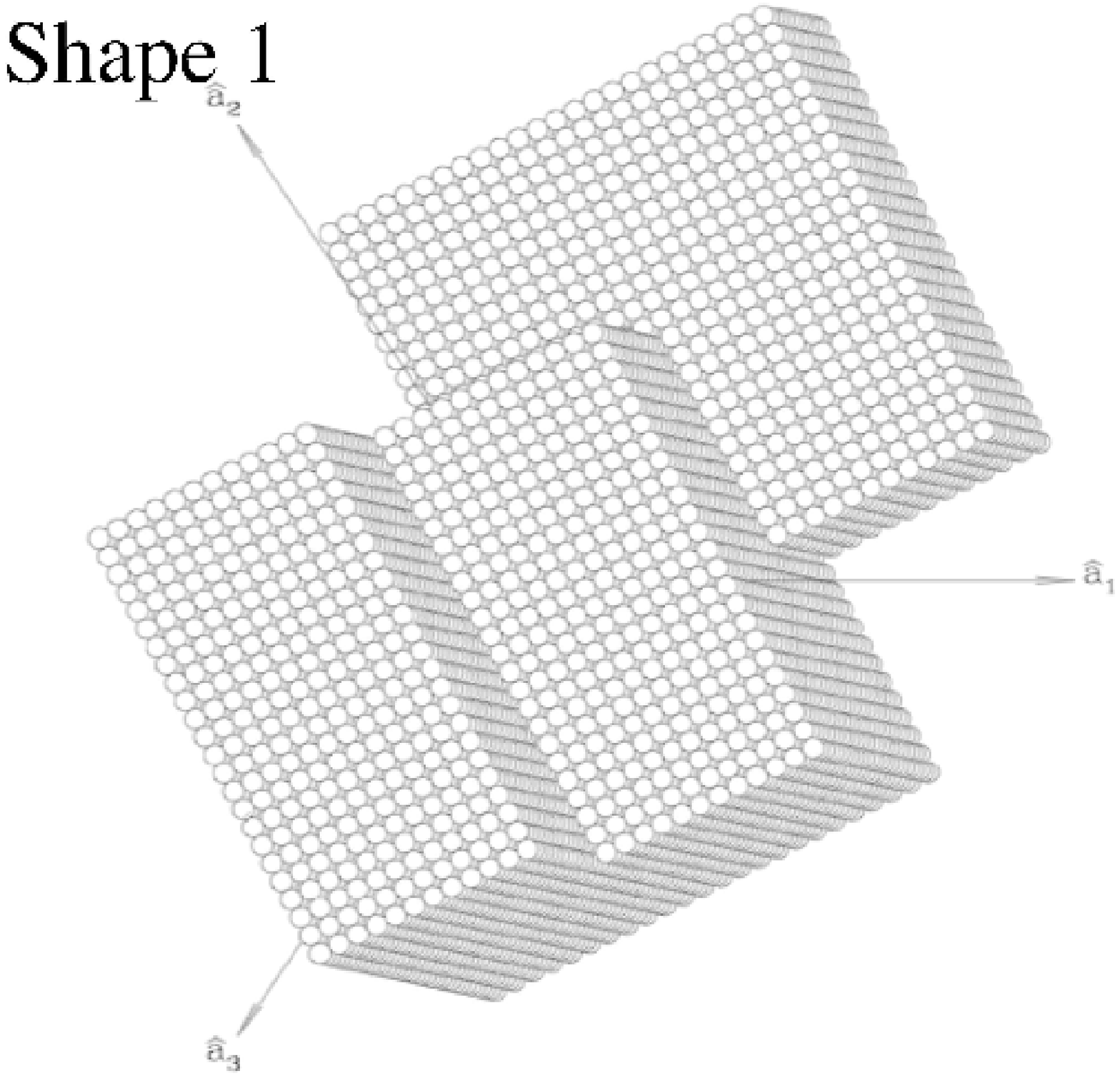}
\includegraphics[width=0.49\textwidth]{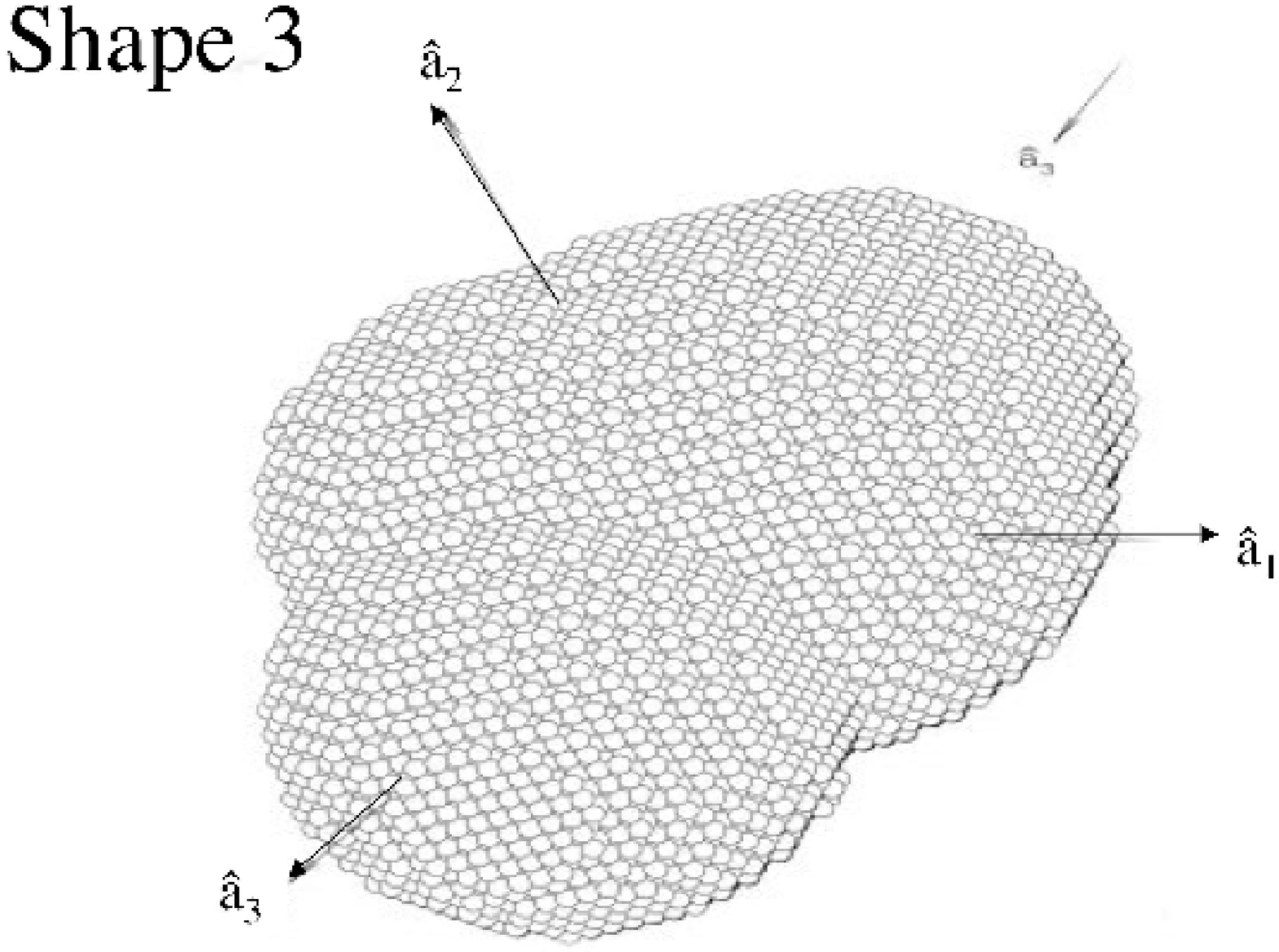} 
\caption{Irregular grains of shape 1 and shape 3 (see DW97) are taken to study RAT alignment by DDSCAT.}
\label{sh14} 
\end{figure}

\begin{figure}
\includegraphics[width=0.49\textwidth]{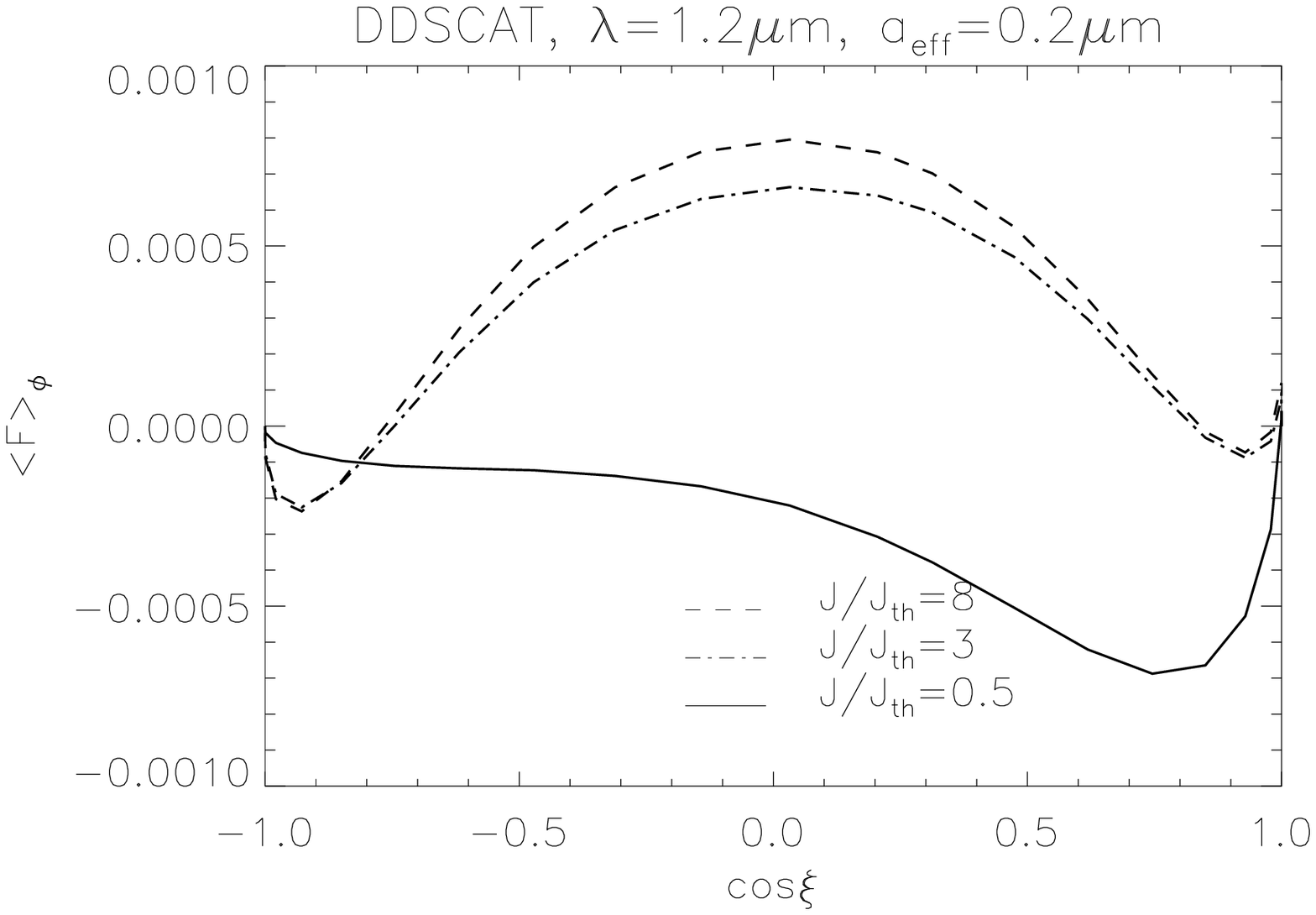}
\includegraphics[width=0.49\textwidth]{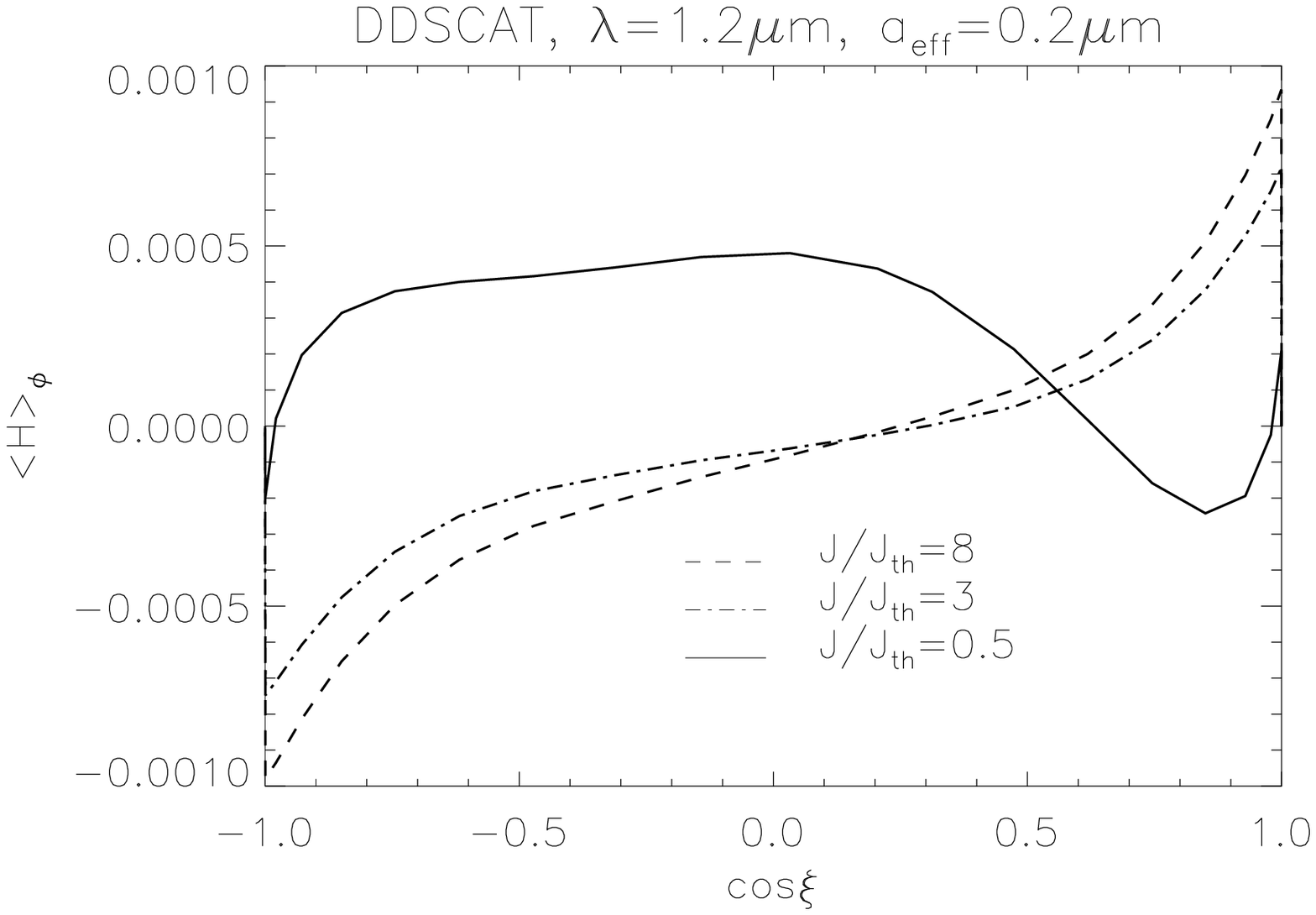} 
\caption{$\langle F\rangle_{\phi}$ and $\langle H\rangle_{\phi}$ for the monochromatic radiation field at $\psi=70^{\circ}$ as functions of the angle $\xi$ between ${\bf J}$ and ${\bf B}$ for three values of $J$.}
\label{f10} 
\end{figure}

Fig. \ref{f10} show $\langle F\rangle_{\phi}$ and $\langle H\rangle_{\phi}$ (i.e., aligning and spinning torques) obtained 
by averaging the corresponding expressions (see equations \ref{eq9} and \ref{eq11})
over thermal fluctuations for different angular momenta, and for the monochromatic
radiation at angle $\psi=70^{\circ}$ toward ${\bf B}$.

From Fig. \ref{f10} we see that for $J/J_{th}\gg 1$, $\langle F\rangle_{\phi}=0$ at $\mbox{cos}\xi=\pm 1, -0.65$,
corresponding to three stationary points. However,
when $J/J_{th}$ decreases (i.e., thermal fluctuations become stronger), the
intermediate stationary point shifts to the left, and disappears as $J/J_{th} \to 0$.
\begin{figure} 
\includegraphics[width=0.49\textwidth]{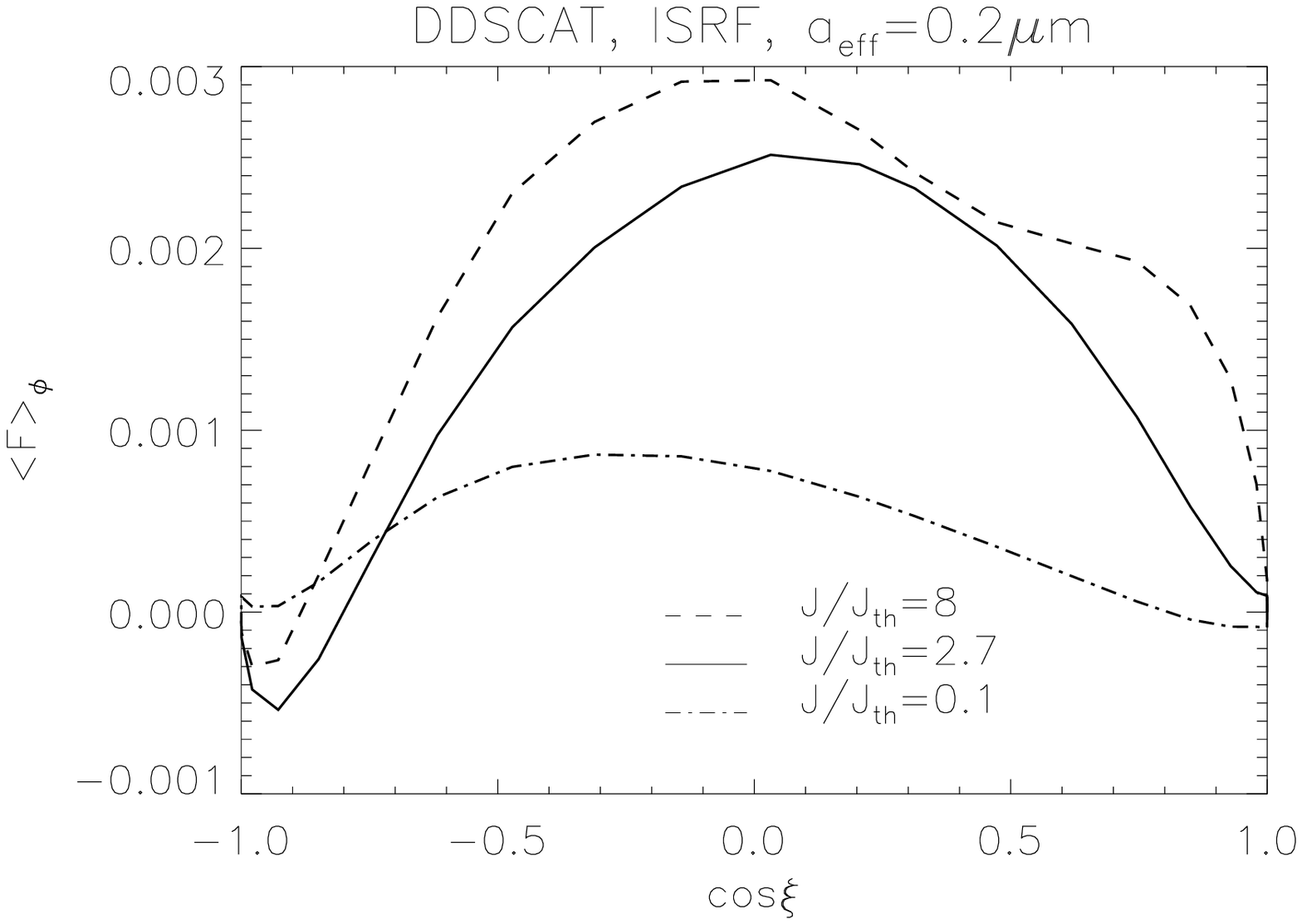} 
\includegraphics[width=0.49\textwidth]{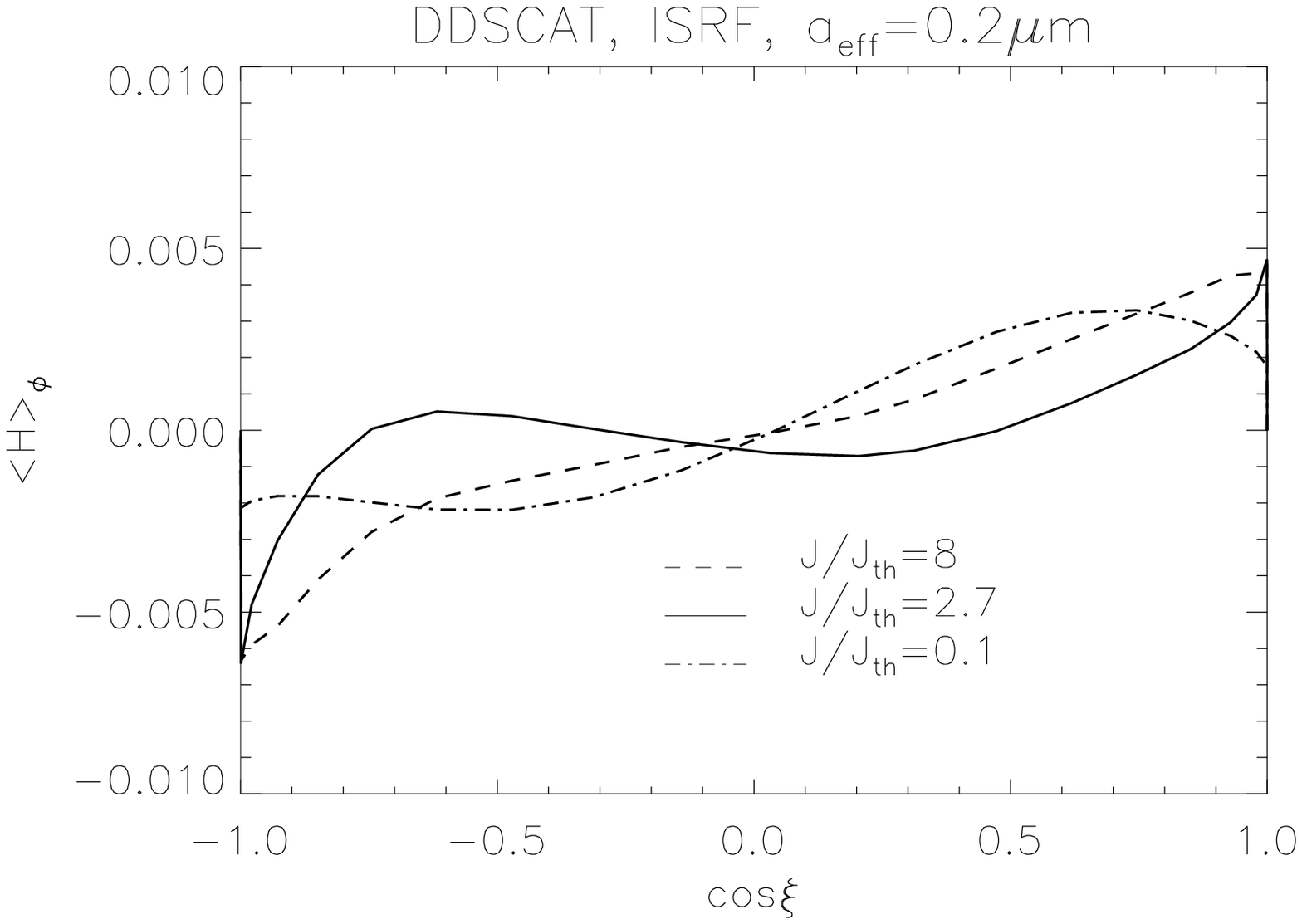} 
\caption{$\langle F\rangle_{\phi}$ and $\langle H\rangle_{\phi}$ for the ISRF at $\psi=0^{\circ}$ as functions of the angle $\xi$ between ${\bf J}$ and ${\bf B}$ for three values of $J$.}
\label{f11}
\end{figure}    

Fig. \ref{f11} is similar to Fig. \ref{f10}, but presents torques for the entire spectrum of the ISRF and
$\psi=0^{\circ}$. It
is also seen that the intermediate stationary point (i.e. point with $\langle F\rangle_{\phi}=0$) shifts to the left as $J/J_{th}$ decreases. It disappears when
$J/J_{th}=0$.
Furthermore, following both Fig. \ref{f10} and \ref{f11} (lower panels), we see that the
spinning torque  $\langle H\rangle_{\phi}$ decreases with $J$ decreasing. This is because 
for low 
angular momentum, thermal fluctuations become stronger, so the axis of grain 
${\bf a}_{1}$ fluctuates with a wider amplitude around ${\bf J}$. As a results, RATs
decrease (see analytical results for the {\it spheroidal} AMO in the upper panel of Fig. \ref{f41}). 

Now let us consider the property of the intermediate zero point of the
aligning torque $\langle F\rangle_{\phi}$ shown in the lower panel of Fig. \ref{f11}. We can 
check that this stationary point corresponding to $J/J_{th}\sim 2.5$ is an attractor point. According to equation (\ref{eq18_a}), the 
criteria in order for one stationary point to be an attractor point is 
$\left. \frac{d\langle F\rangle_{\phi}}{\langle H\rangle_{\phi}
    d\xi}\right|_{\xi_{s}}<1$. With the stationary point at
$\mbox{cos}\xi_{s}=-0.85$, we have $\langle H\rangle_{\phi} >0$ (see the upper panel), and
$\left. \frac{d\langle F\rangle_{\phi}}{d\xi}\right|_{\xi_{s}}
<0$ due to the decrease of $\langle F\rangle_{\phi}$ with $\xi$ in the vicinity of $\xi_{s}$ (see the upper panel of Fig. \ref{f11}). Thus, $\left. \frac{d\langle F\rangle_{\phi}}{\langle H\rangle_{\phi}
    d\xi}\right|_{\xi_{s}}<0$, which satisfies the criteria for an attractor point.
\subsection{Phase Trajectory Maps for shape 1}

Let us consider the trajectory map for grains of size $a_{eff}=0.2 \mu m$ driven by RATs produced by a
monochromatic radiation of $\lambda=1.2\mu m$, in the direction $\psi=70^{\circ}$,
which is similar to the setting in WD03. As we mentioned above, an important difference in our
treatment and that in WD03 is that we average RATs over $10^{3}$ rotation
periods rather than over only one period as WD03 did. As we discussed in \S 3.3, one of the consequences of this is that the contribution of $Q_{e3}$ component vanishes for
aligning and spinning torque components. 

\begin{figure} 
\includegraphics[width=0.49\textwidth]{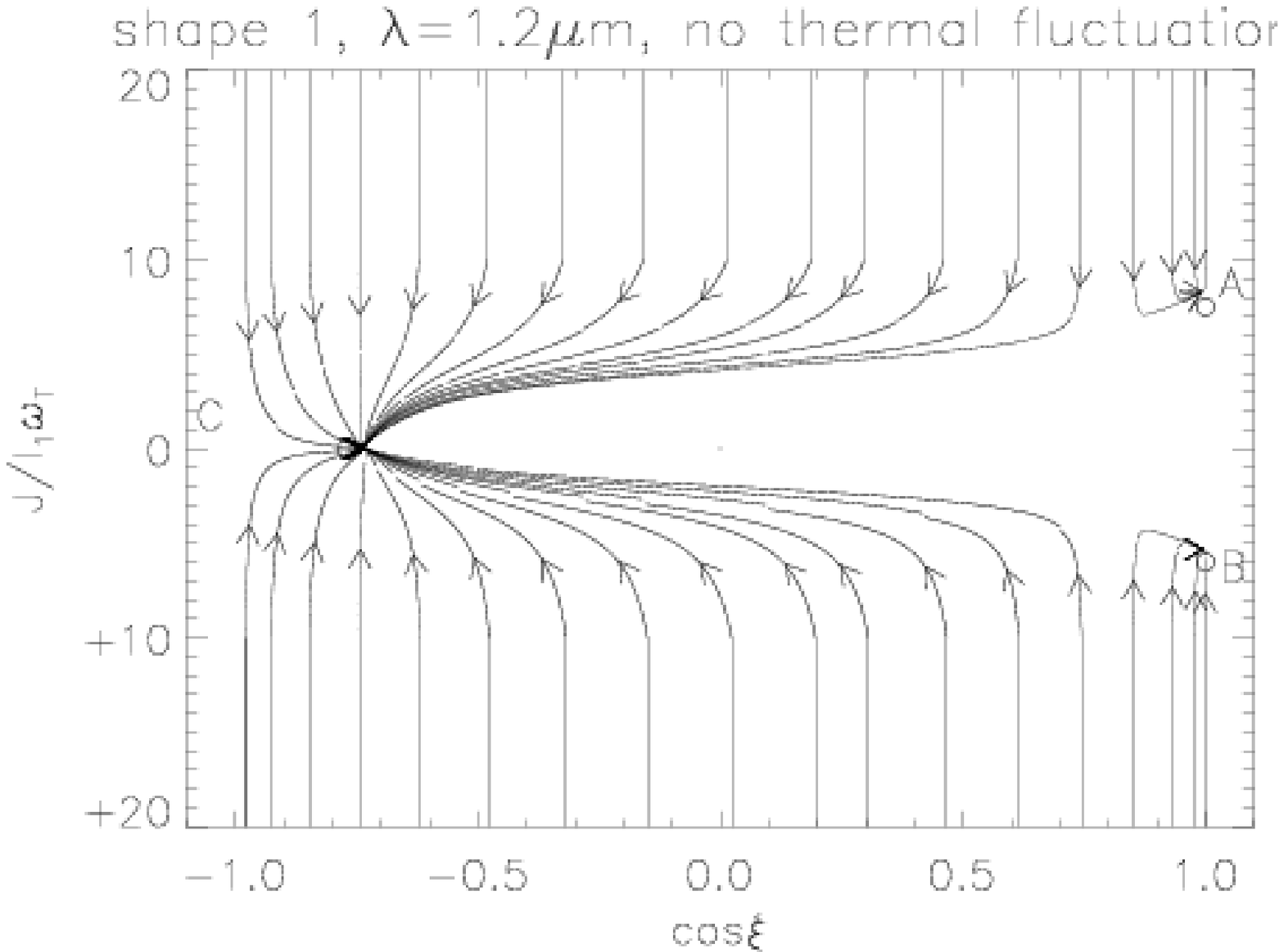} 
\includegraphics[width=0.49\textwidth]{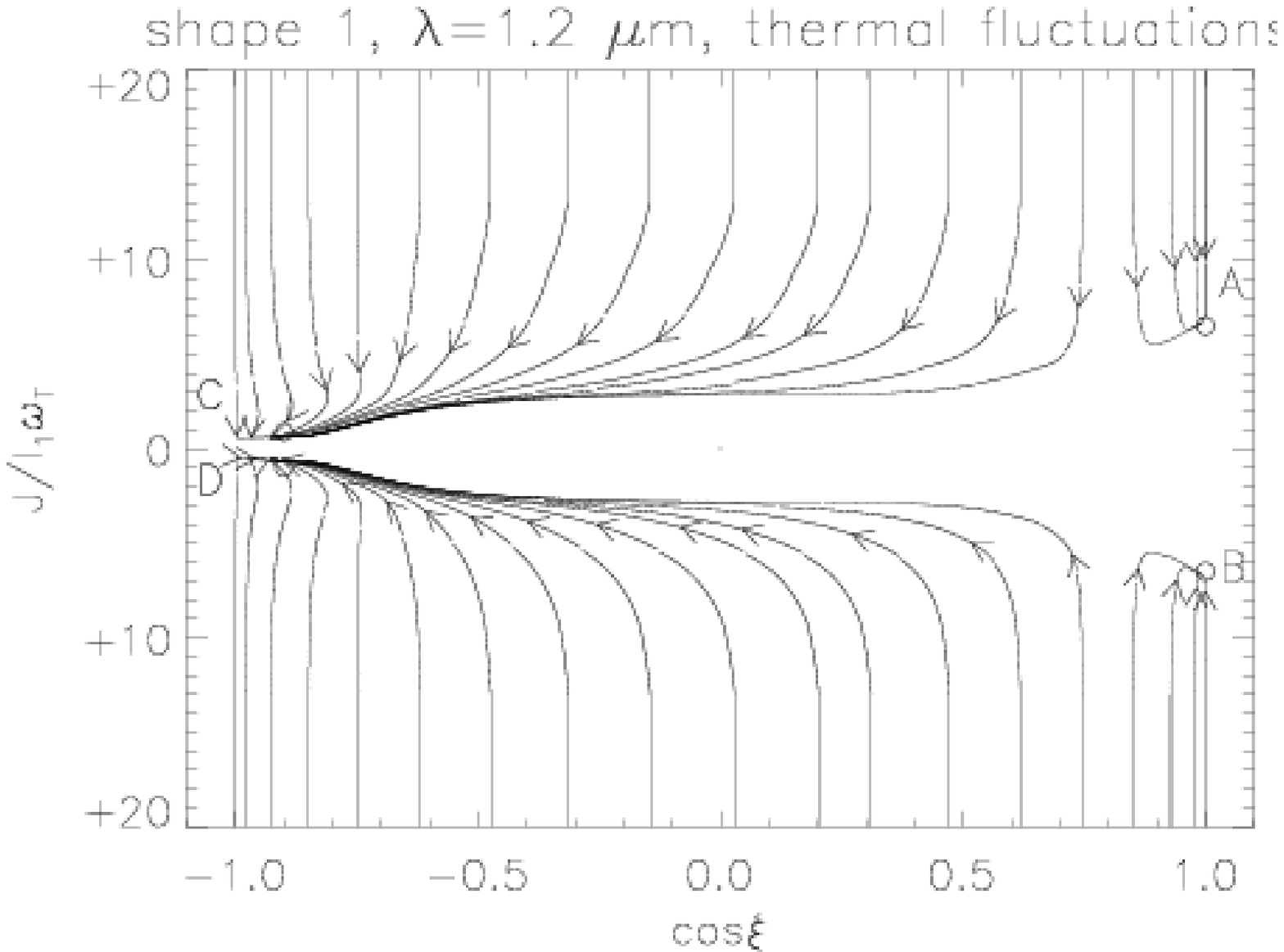} 
\caption{Phase trajectory maps for $\lambda=1.2 \mu m$ and $a=0.2\mu m$, and $\psi=70^{\circ}$ corresponding to no thermal fluctuations ({\it upper panel}) and 
  with thermal fluctuations ({\it lower panel}). For the case in which ${\bf J}$ is fixed, when grains
  are driven to 
  the crossover $J=0$, ${\bf a}_{1}$ flips over. So grains change frequently 
  their states corresponding to switching between the upper and lower frame in
  the map. }
\label{f12} 
\end{figure}
The upper panel in Fig. \ref{f12} shows the trajectory map for the case without thermal
fluctuations with two high-$J$ attractor
points (A, B) and one zero-$J$ attractor point C. When thermal fluctuations are accounted for, the lower panel reveals the split of the attractor point C to C' and D' as also seen in WD03. However, the value of angular momentum at C' and D' is $J=0.7J_{th}$, which is
lower that the value obtained by WD03. This difference stems from the fact
that the contribution of $Q_{e3}$ to the spinning and aligning torques is
completely canceled when we average RATs over a sufficiently long time. In fact, we
tested that, if $Q_{e3}$ is set to zero, the resulting map is the same with
the lower panel of Fig. \ref{f12}. Yet, if we adopt the same averaging as in WD03, we also get the low attractor point
with  similar value  $J$ as in WD03. We also see a correspondence between our results here with those obtained with the AMO (see \S 4.4).

The lower panel in Fig. \ref{f12} also shows that about $20\%$ of grains are aligned at two high-$J$ attractor points
A, B, and about $80\%$ grains are driven by RATs to the low-$J$ attractor points C and D. There, grains flip to the opposite flipping
state (i.e. from upper to lower frame) and back. However, immediately after
entering the opposite flipping state, grains are decelerated by RATs to that
point again and they flip back. So grains flip back and forth between C and D. In fact, C
and D are indistinguishable because grains flip very fast between them with the 
probability of finding grains on each point is $\sim 0.5$. In that sense, the points C and D are also crossover points.\footnote{These crossover points are different from those described in DW97, in which the grains were spinning up right after crossover. This stems from inaccurate treatment of crossovers in DW97, see more detail in Paper I.}

Figs \ref{f11a}  and \ref{f11b} show the phase maps of grain alignment for the
full spectrum of the ISRF instead of monochromatic radiation, and for two
directions of radiation $\psi=0^{\circ}$ and $90^{\circ}$. For the
direction $\psi=0^{\circ}$, the map in the lower panel consists of three attractor points A', B', C  in which the point C is the old 
attractor point at very high angular momentum, and $A', B'$ are low-$J$ attractor
points originated from A and B at $J=0$ for the case of no thermal
fluctuations (see the upper panel). In other words, averaging over thermal fluctuations
of grains for the ISRF and $\psi=0^{\circ}$ also changes the zero-$J$ attractor point
to the new thermal $J$ attractor point. 

\begin{figure} 
\includegraphics[width=0.49\textwidth]{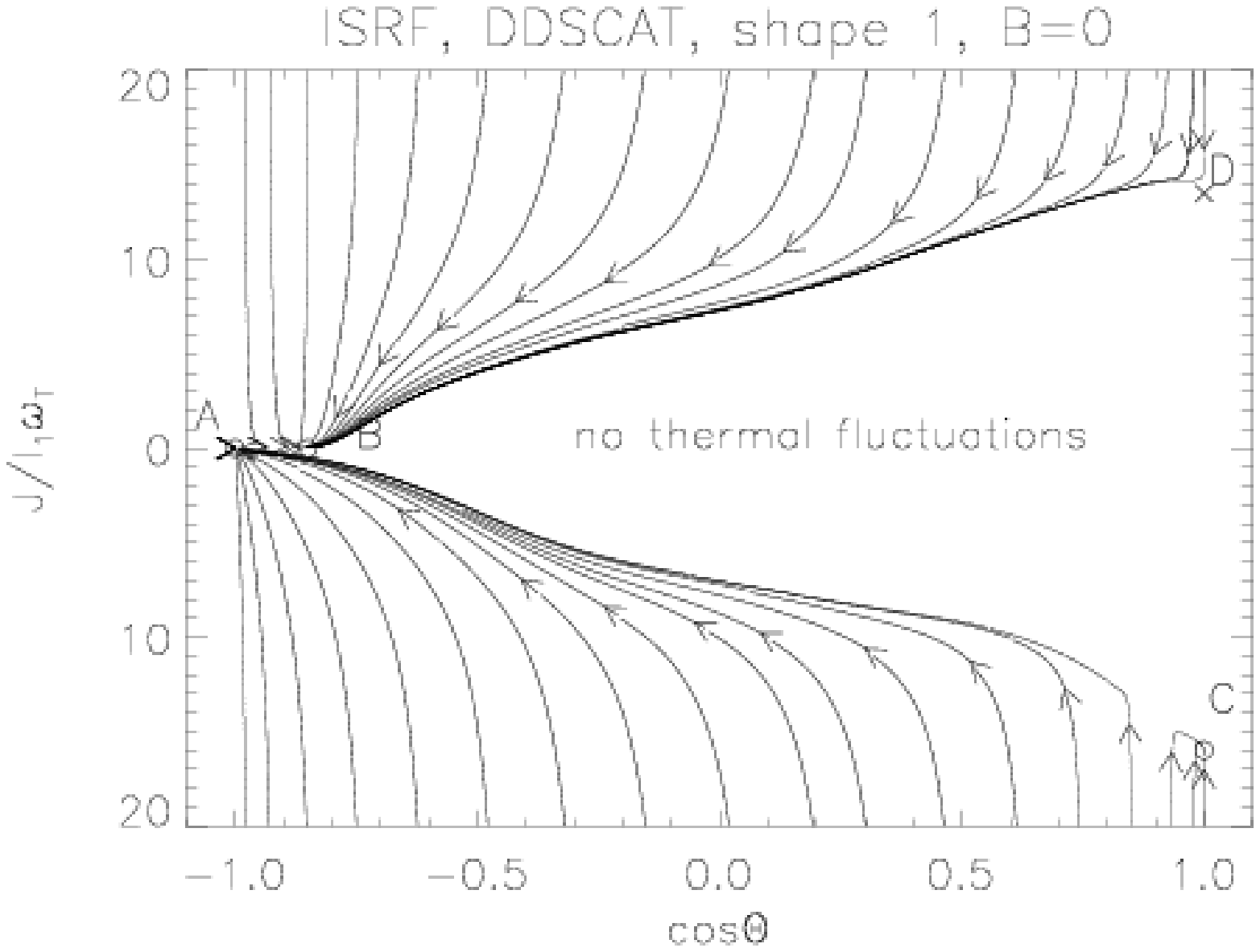} 
\includegraphics[width=0.49\textwidth]{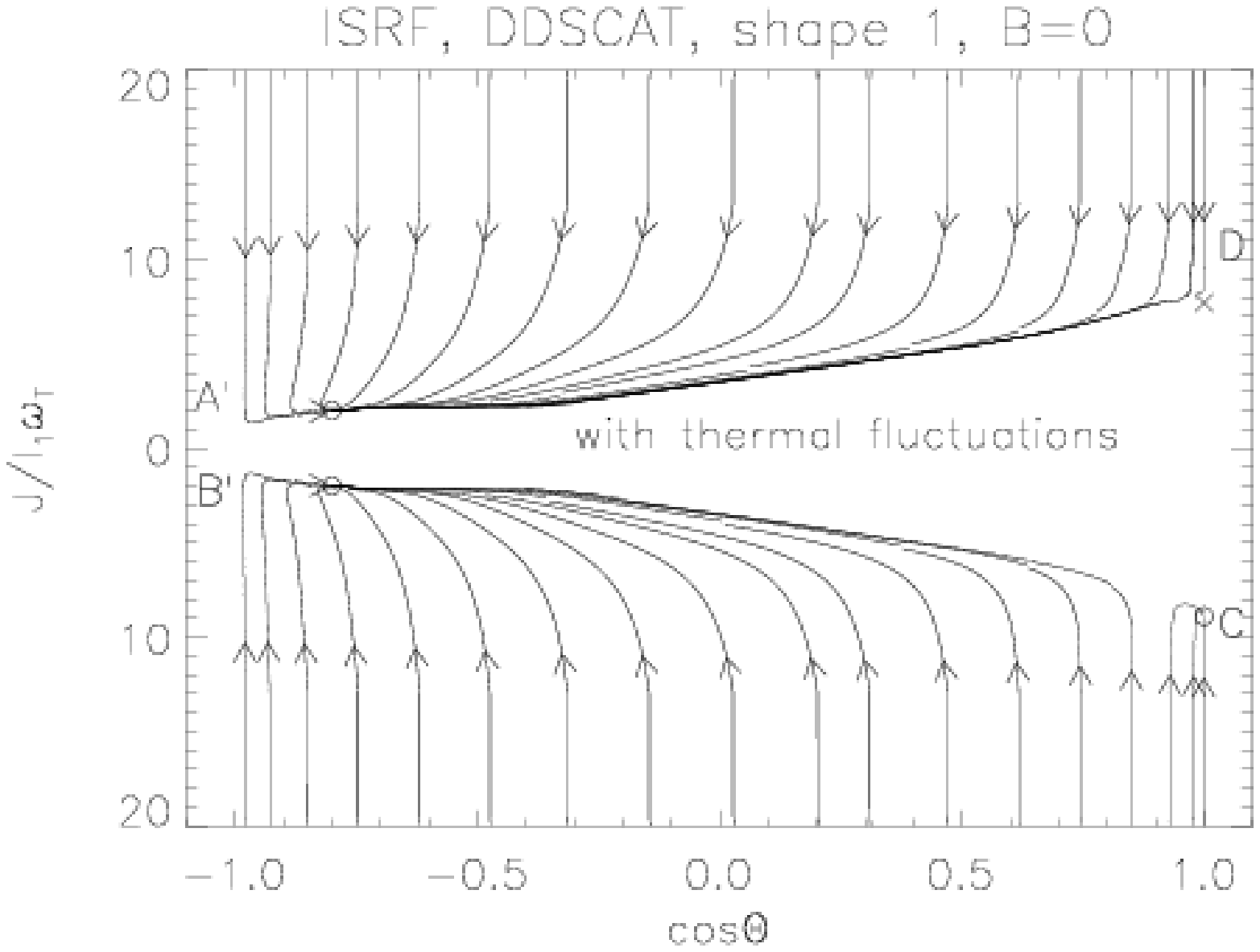} 
\caption{Trajectory maps for the entire spectrum of the ISRF and irregular grain of size $a_{eff}=0.2 \mu
  m$ when thermal fluctuations are absent (upper panel) and when thermal fluctuations and thermal flipping are included (lower panel),
  corresponding to the alignment with respect to ${\bf k}$ or $\psi=0^{\circ}$ (equivalent to the grain alignment in the absence of magnetic field).}
\label{f11a}
\end{figure}
\begin{figure} 
\includegraphics[width=0.49\textwidth]{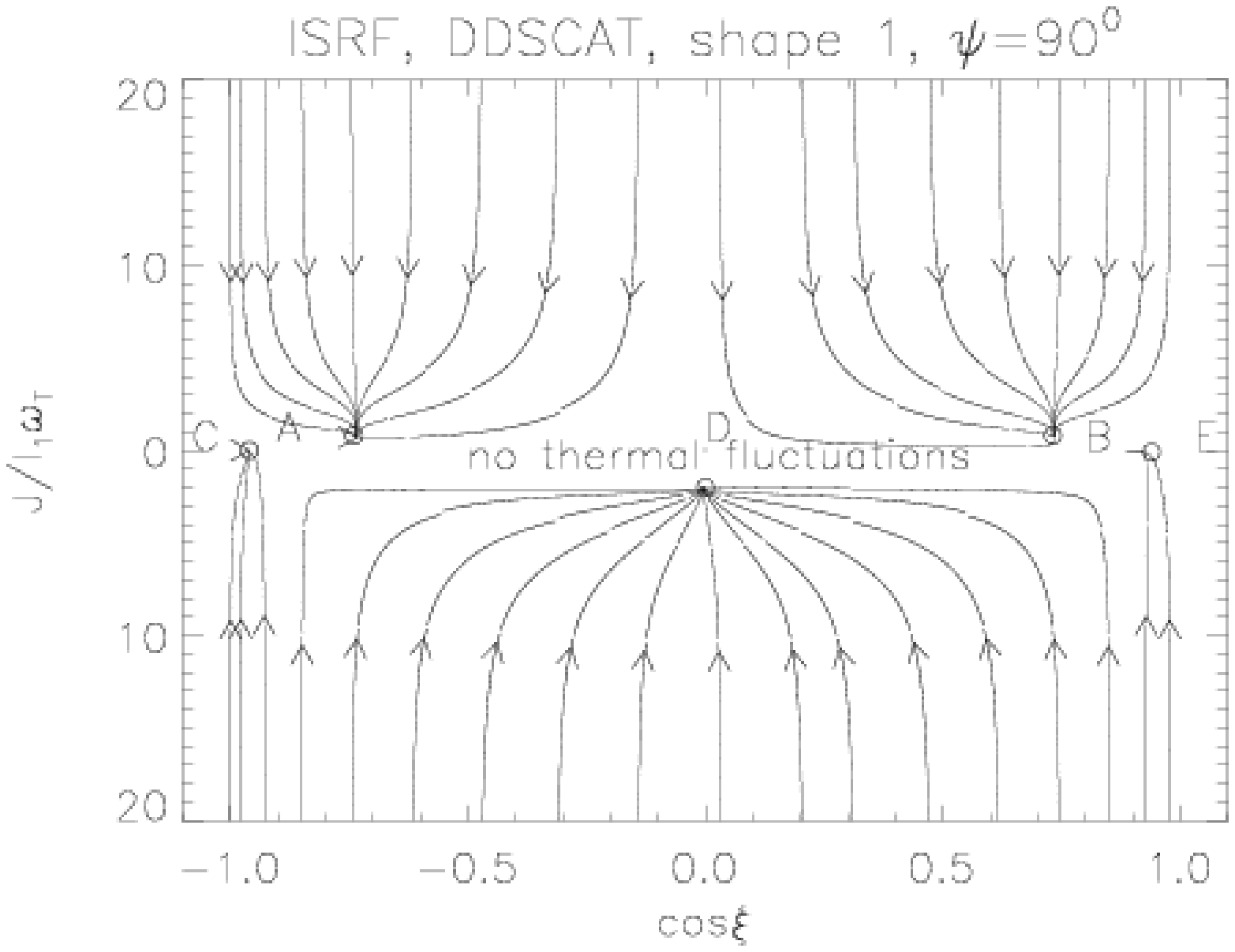}
\includegraphics[width=0.49\textwidth]{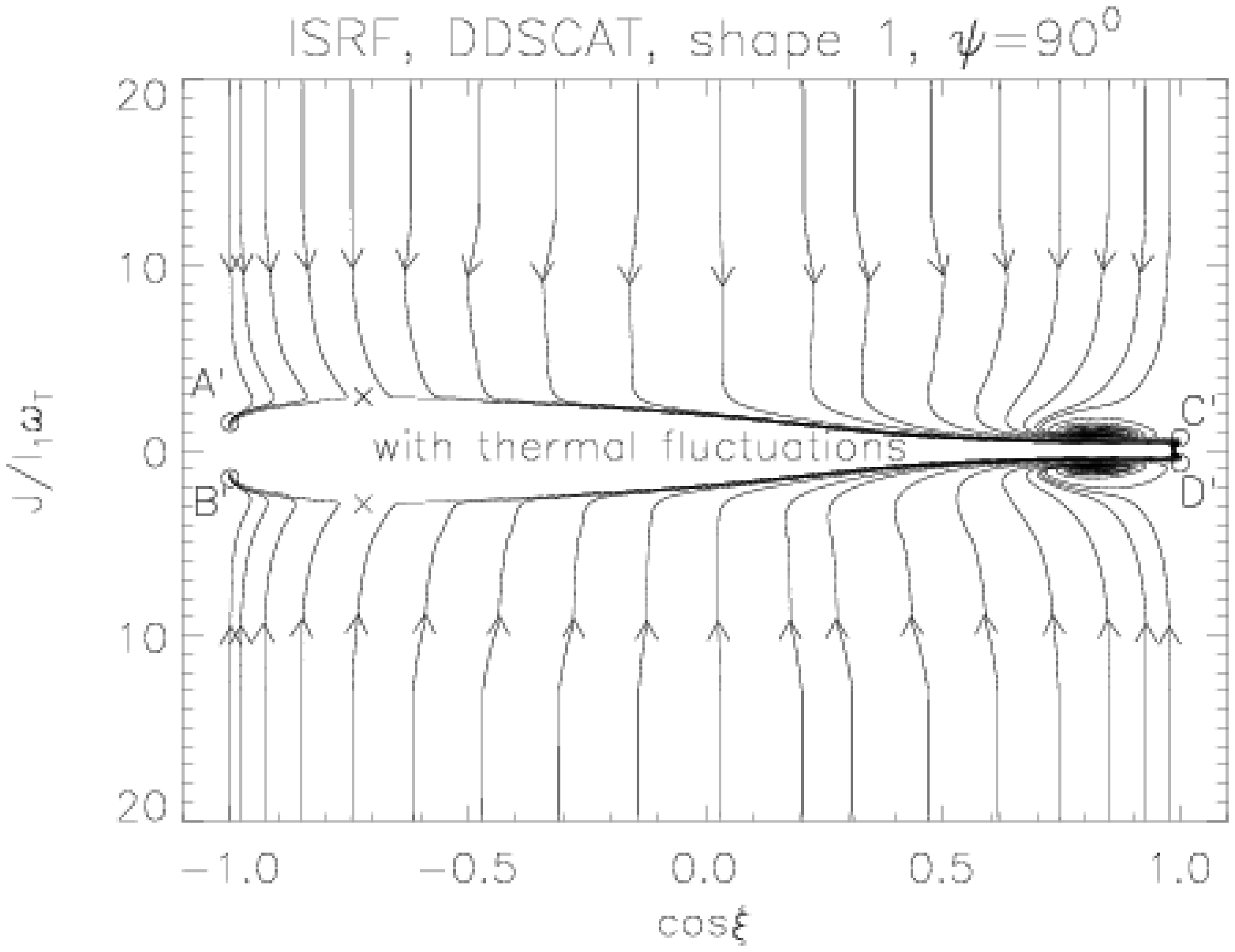} 
\caption{Similar to Fig. \ref{f11a}, but for $\psi=90^{\circ}$. {\it Upper 
  panel}: the map with two high J attractor point A, B, two zero-J attractor points C, E and a ``wrong''
  attractor point D at $\mc\xi=0$. {\it Lower panel}: with thermal fluctuations
  and the ``wrong'' attractor point D disappears.} 
\label{f11b}
\end{figure}

In Paper I we pointed out that at $\psi$ close to $90^{\circ}$ for both the AMO and
irregular grains, the grain alignment tends to take place with ${\bf a}_{1}\perp {\bf
  B}$, which is in contrast to the classical Davis-Greenstein (1951) expectations. We
termed this  ``wrong'' alignment.
An attractor point D at "wrong" alignment angle (i.e. $\xi=90^{\circ}$) shown in Fig. \ref{f11b}{\it upper} for the direction $\psi=90^{0}$ together with four other attractor points A, B, C and E, in the absence of thermal wobbling (see also Paper I). However, in the presence of thermal wobbling, the "wrong" attractor point D is no longer existing. Instead, we obtain four new attractor points A', B', C' and D', corresponding to  
$J\sim J_{th}$ and $\mc\xi=\pm 1$ (see the lower panel in Fig. \ref{f11b}). It means that the ``wrong''
alignment is indeed eliminated by averaging induced by thermal wobbling (see also \S 2). 

\subsection{Dynamics of shape 3}
For the sake of completeness, let us study the effect of thermal fluctuations on RAT alignment for another irregular grain (shape 3 in DW97). As an example, we consider a particular grain size $a_{eff}=0.2 \mu m$ and one light direction $\psi=70^{\circ}$.

Fig. \ref{f22} shows the trajectory maps obtained for this grain corresponding to the case without (upper panel)
and with thermal fluctuations (lower panel). It is shown that
for $\psi=70^{\circ}$, this grain produces a phase map with two high-$J$
attractor points A, B and two low-$J$ attractor points C' and D' in the presence of thermal fluctuations (see lower panel of Fig. \ref{f22}). The lifting of
the low-$J$
attractor point C (see the upper panel of Fig. \ref{f22}) from $J=0$ to $J=2 J_{th}$ (C', D') when thermal fluctuations are taken into account, is also seen for this shape. However, grains can rapidly flip back and forth between C' and D'.
\begin{figure}
\includegraphics[width=0.49\textwidth]{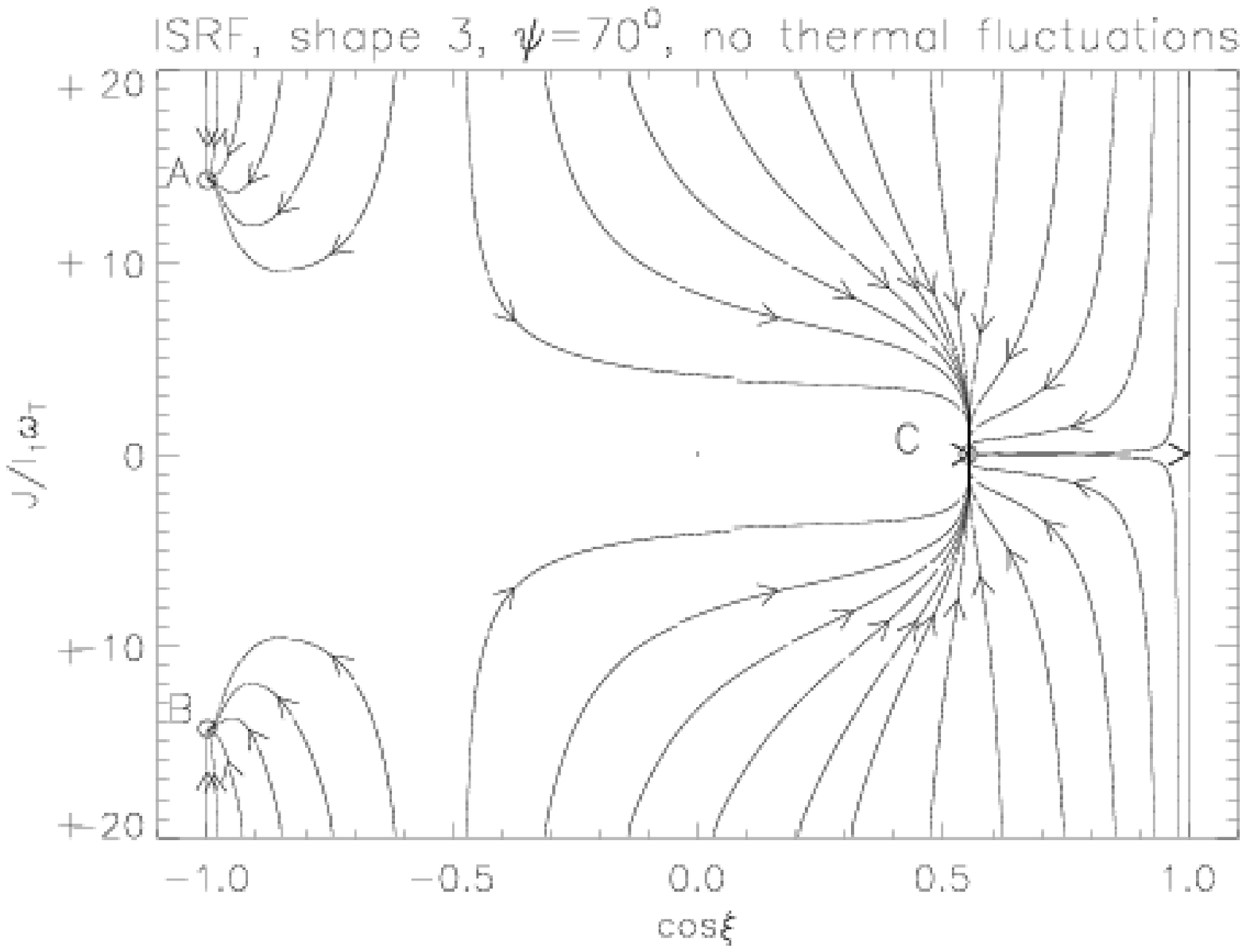}
\includegraphics[width=0.49\textwidth]{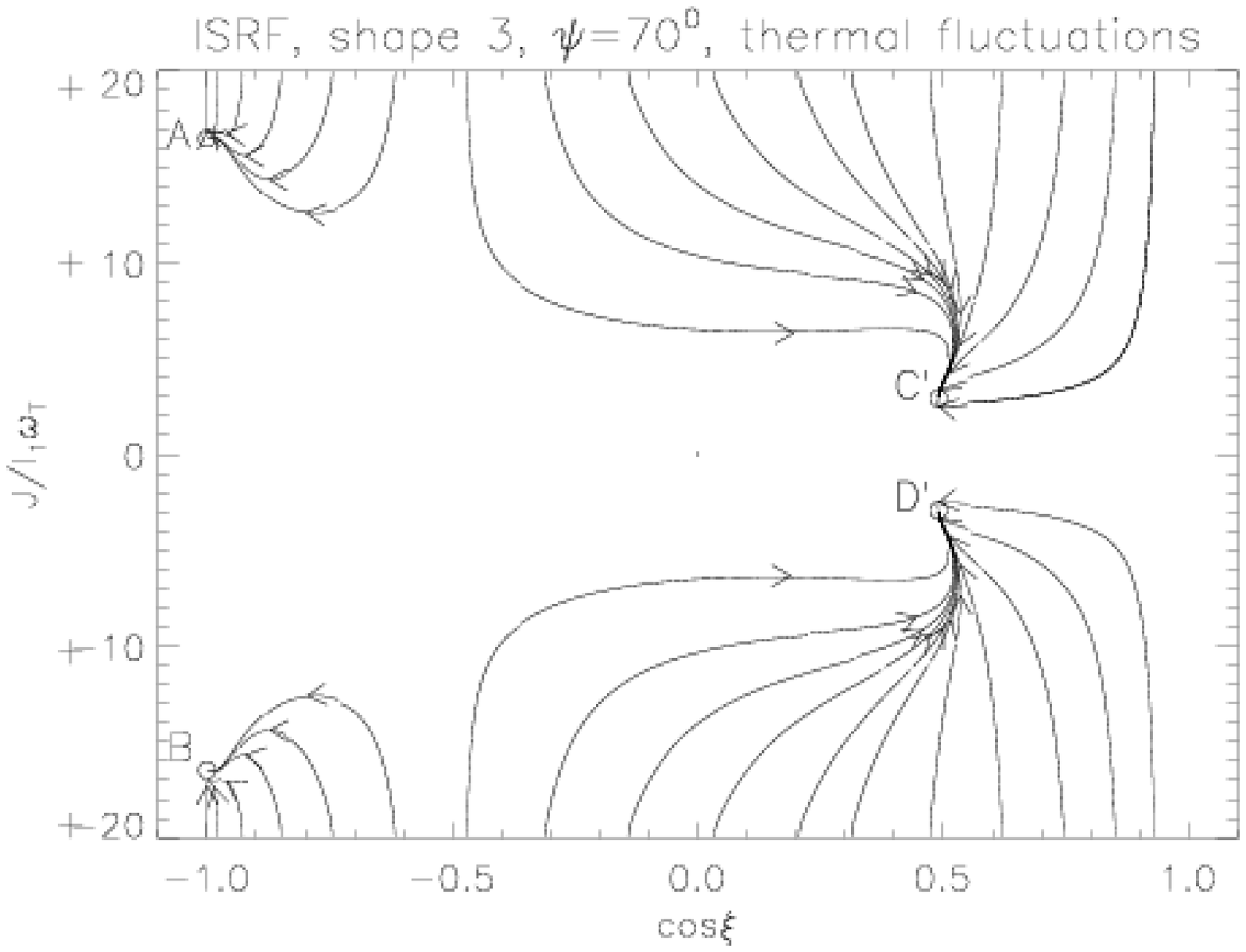}
\caption{Phase maps for the grain shape 3 and the ISRF with two high-$J$ attractor
  points corresponding to the cases without thermal fluctuations ({\it upper panel}) and with thermal fluctuations ({\it lower panel}). Thermal fluctuations shift the zero-$J$ attractor point C to C', D' with $J=2 J_{th}=2 I_{1}\omega_{T}$. }
\label{f22}
\end{figure}


\section{Low $J$ attractor points} 
Our study indicates that a high fraction of grains are aligned at low-$J$ attractor points. Therefore, the origin and stability of this low-$J$ attractor point are important for grain alignment. We address these questions below.

\subsection{Origin}\label{lowJ}
The results above show that thermal fluctuations produce a new low-$J$ attractor
point from the zero- $J$ attractor point, for both the ellipsoidal AMO and irregular grains, but this effect has not been seen for the {\it spheroidal} AMO. Now let us explain why this occurs by using the AMOs, but the explanation can be applicable to irregular grains.

First of all, let us study RATs averaged over the free-torque motion for the {\it spheroidal} and {\it ellipsoidal} AMOs. Following equation (\ref{ap2}) (see Appendix D), the average over thermal fluctuations can be rewritten as
\begin{align}
\langle A\rangle &\sim\int_{0}^{1} ds A e^{(-q \frac{J^{2}}{J_{th}^{2}})},\label{eq23a}
\end{align}
where $A$ is the torque component that arises from the average over free-torque motion and  {\bf $q=2I_{1}E/J^{2}$ }with $E$ is the total kinetic energy.
It is convenient to define 
\begin{align}
A(s,J)&=A e^{(-q \frac{J^{2}}{J_{th}^{2}})},\label{eq23b}
\end{align}
which represents the torque resulting only from the average over free-torque motion as a function of $J$ and the density of states in phase space $s$ (see Appendix D). 

In Fig. \ref{f21d} we show torque components defined by equation (\ref{eq23b}) for {\it spheroidal} and {\it ellipsoidal} AMOs with $J=10 J_{th}$ and $J=0.5 J_{th}$, corresponding to cases in which the role of thermal fluctuations is marginal and important, respectively.

It can clearly be seen that for $J =10 J_{th}$ (i.e. $J\gg J_{th}$), the obtained torques are nearly similar for irregular and axisymmetric grains. Yet the torques drop very rapidly to zero as $s$ (note that $s=\ms\gamma$ for the spheroid) increases (see dashed lines in Fig. \ref{f21d}). The former stems from the fact that for the suprathermal rotation, a good coupling between the maximal inertia axis and the angular velocity is achieved. Therefore, there is no difference between the free-motion of an irregular and axisymmetric grains.

However, the averaged torques become very different as $J$ decreases. In fact, Fig. \ref{f21d} (see the curves with $J=0.5 J_{th}$) reveal that for axisymmetric grains, RATs decrease regularly with respect to $s$. Whereas, RATs for irregular grains drop rapidly, change the sign and rise again with $s$ increasing. 

The effect of such a variation on the average of RATs over thermal fluctuations (see equation \ref{eq23a}) is evident.  In Fig. \ref{f21b} we show the corresponding torque components  $\langle F\rangle_{\phi}$ and $\langle H\rangle_{\phi}$  obtained by averaging $F$ and $H$ using equation (\ref{eq23a}), as functions of $J$ for several angles $\xi$ between  ${\bf J}$ and ${\bf k} $ for $\psi=0^{\circ}$. It can be seen that for $J\gg J_{th}$, the averaged torques of irregular grains are similar with those of axisymmetric grains, and they become saturated in both cases. However, their averaged torques become very different when $J$ decreases. For instance, the averaged torques for the former case change their sign at $J \sim J_{th}$, while the torques for the later case do not. We note that for the angle $\mc\xi<0$, the torque $\langle H\rangle_{\phi}$ is negative (i.e. torques tend to drive grains to zero-$J$ attractor point) for $J\gg J_{th}$, equivalent with the absence of thermal fluctuations. As a result, the change of sign of $\langle H\rangle_{\phi}$ for $J\sim J_{th}$ in the presence of thermal fluctuations reveals that grains can be spun up again for $J \sim J_{th}$. To some value of $J$, $\langle H\rangle_{\phi}$ continues to reverse its sign, and grains are decelerated.  
In other words, the irregular grains can be aligned at low attractor points with angular momentum $J\sim J_{th}$.

\begin{figure} 
\includegraphics[width=0.49\textwidth]{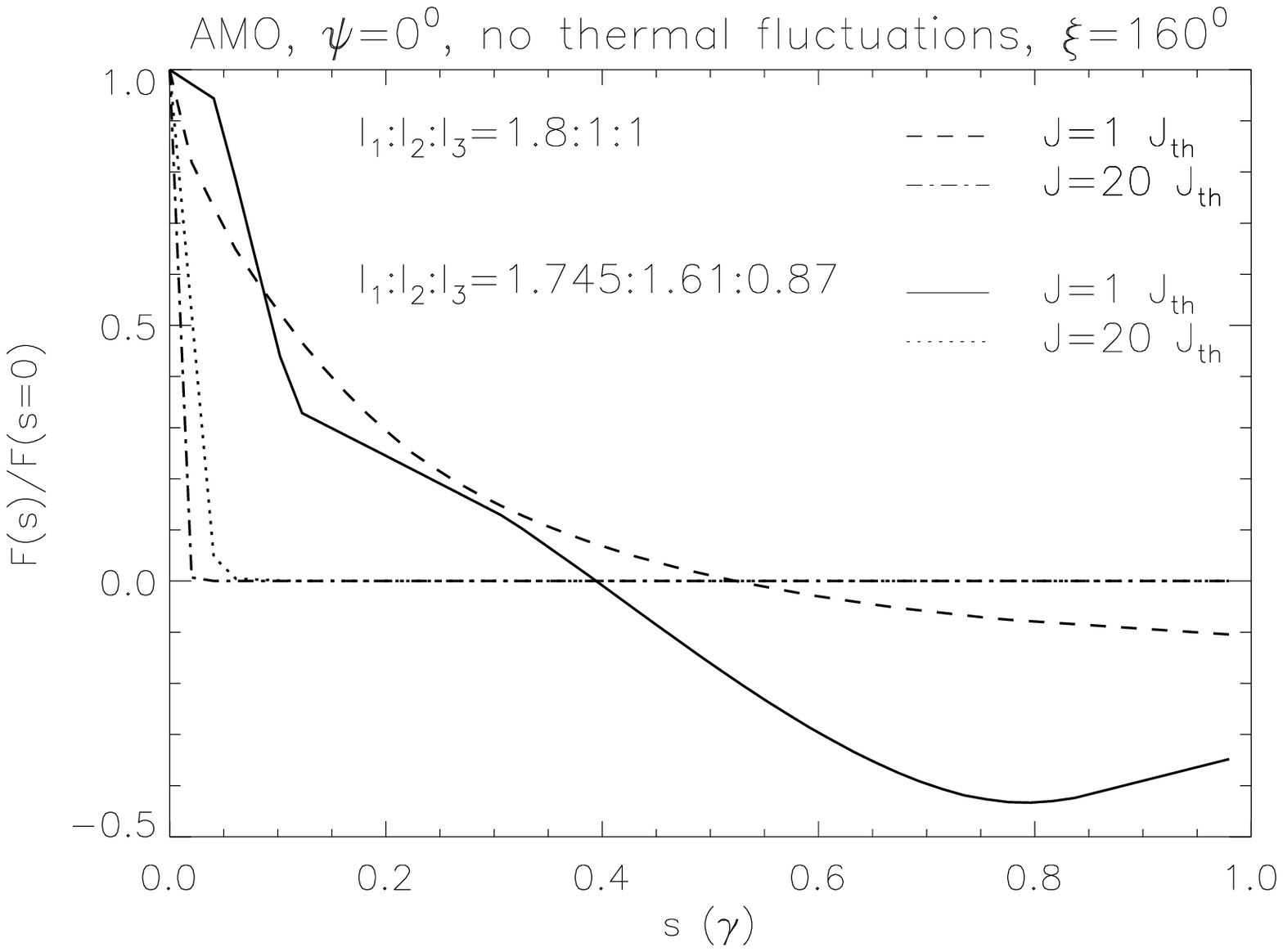}
\includegraphics[width=0.49\textwidth]{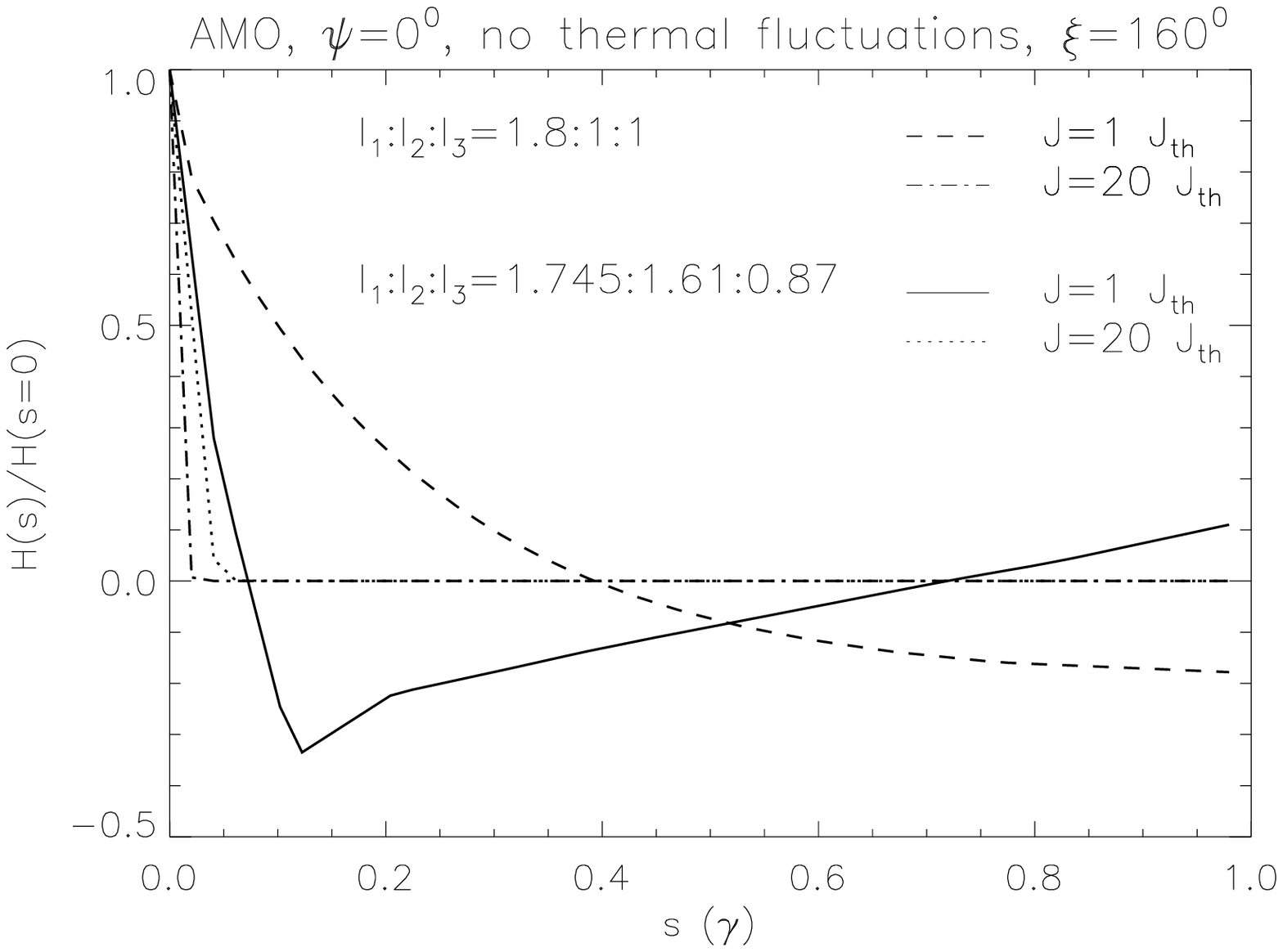}
\caption{Spin-up and alignment torques as functions of $s(\gamma)$ for two values of angular momentum $J=20 J_{th}$ and $J= J_{th}$ corresponding to an angle $\xi=160^{\circ}$ between angular momentum and the magnetic field, and the precession angle $\phi=180^{\circ}$. The fast drop of $ F$ and $H$ for small $J$ is observed that gives rise to the fact that the sign of their averaged value over $s$ is opposite to its sign at $s=0$.} 
\label{f21d}
\end{figure}

\begin{figure} 
\includegraphics[width=0.49\textwidth]{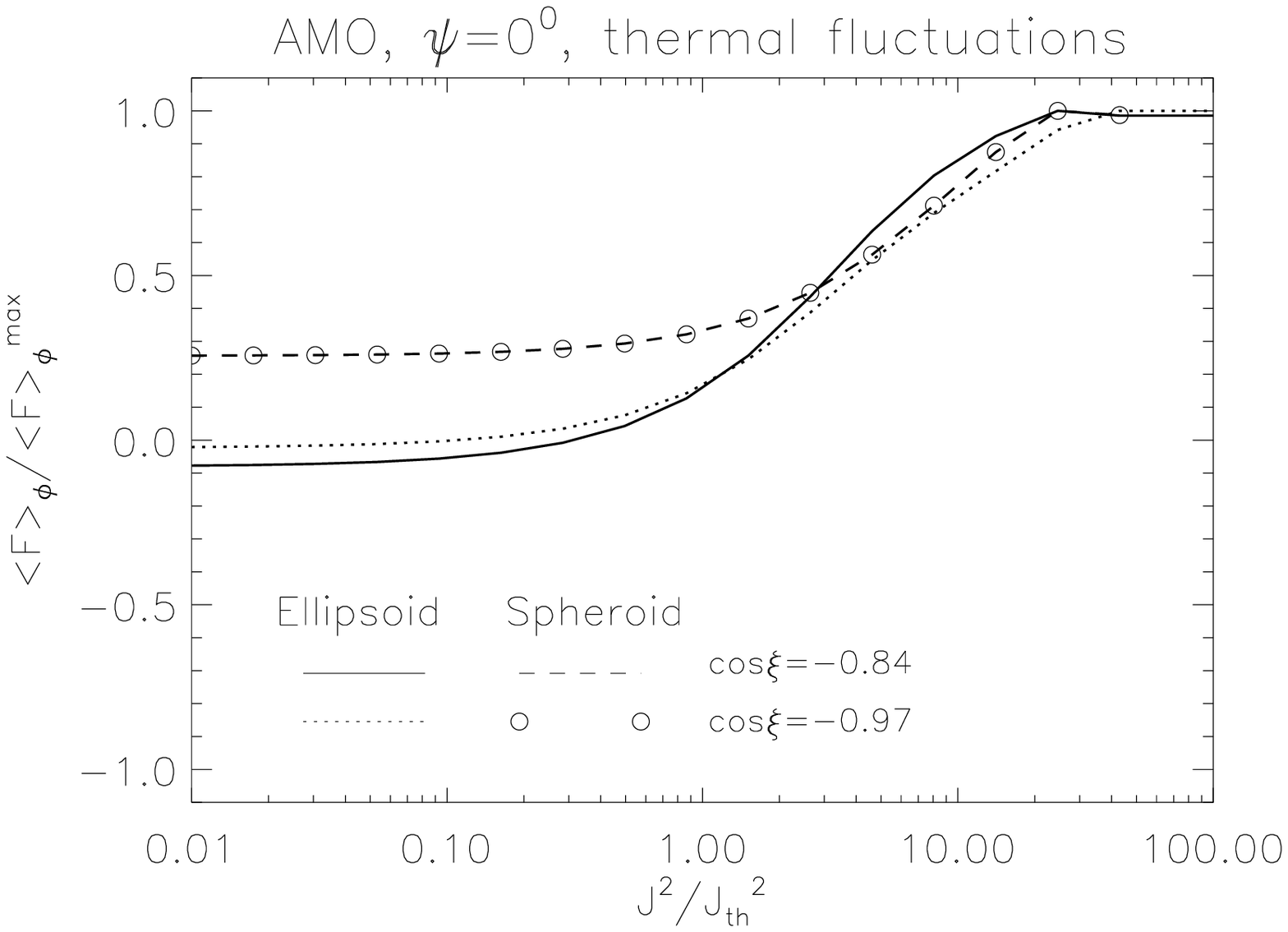}
\includegraphics[width=0.49\textwidth]{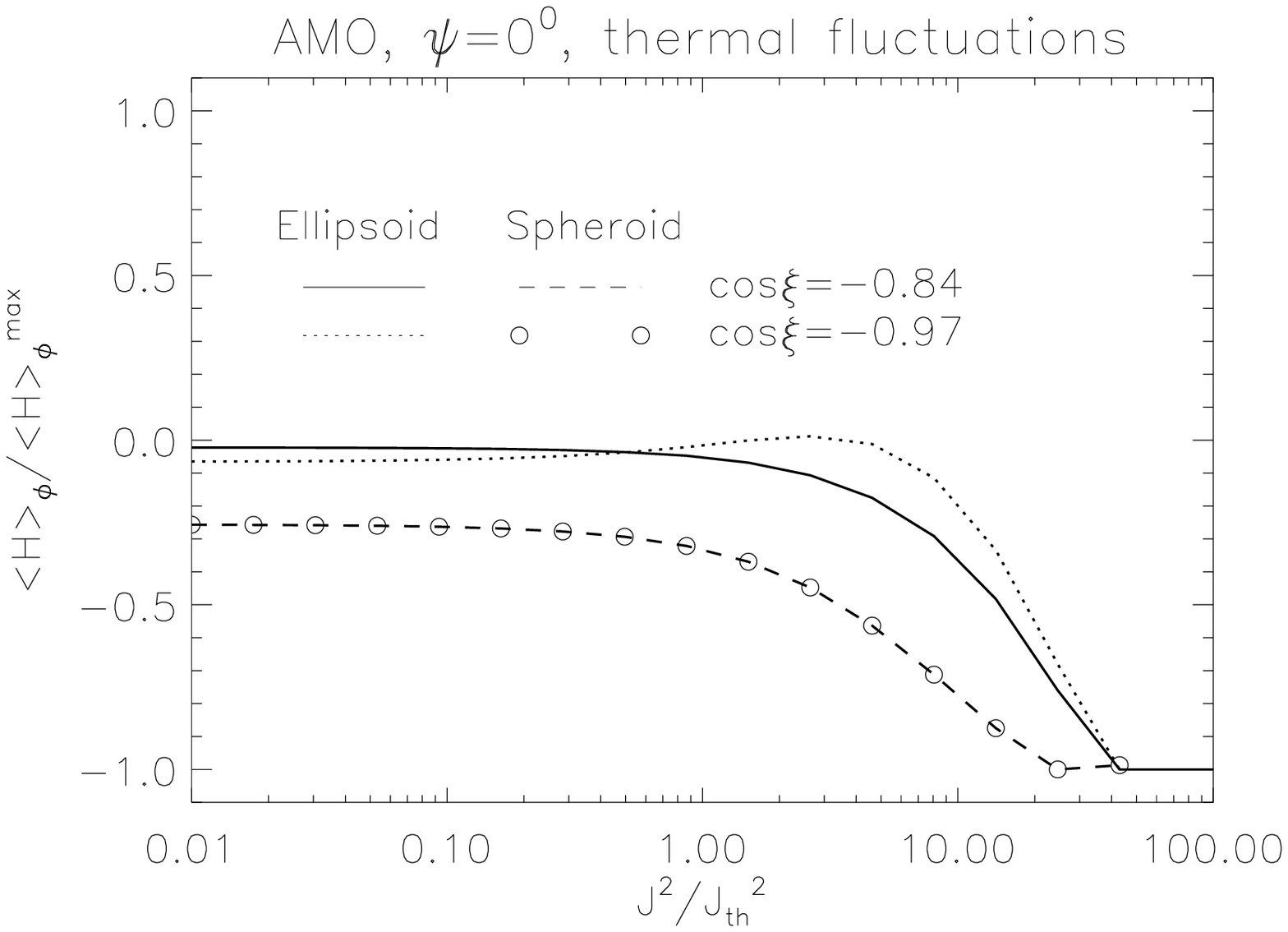} 
\caption{Aligning $\langle F\rangle_{\phi}$ and spinning $\langle H\rangle_{\phi}$ torque components averaged over the Larmor precession angle $\phi$ and thermal fluctuations (i.e. over $s$ using equation \ref{eq23a}), normalized over their amplitude values for three angles $\xi$, as functions of angular momentum for the spheroid and ellipsoid AMOs. Plots show the change of sign of both torques when $J^{2}/J_{th}^{2}$ decreases (i.e., thermal fluctuations increase) for the ellipsoid AMO, but not seen for spheroid. For $J\gg J_{th}$, the torques are constant.} 
\label{f21b}
\end{figure} 

Consider now a stationary point $\xi_{s}$. As discussed previously, the stationary point $\xi_{s}$ is either an attractor or a repellor point depending on the sign of the spin-up torque and the derivative of the alignment torque at that point; it is an attractor point if  $\left.\frac{1}{\langle H\rangle_{\phi} }\frac{d\langle F\rangle_{\phi}}{d\xi}\right|_{\xi_{s}}<0$. Therefore, the change in sign found above for $\langle F\rangle_{\phi} $ and $\langle H\rangle_{\phi}$ can give rise to the thermal attractor points observed in maps for the {\it ellipsoidal} AMO and irregular grains .

In summary, the analytical analysis for the AMO results indicate that as thermal fluctuations become more important (i.e. $J$ small), RATs as a function of $\xi$ can have different forms (sign and magnitude) that increase the angular momentum of the low attractor points from $J=0$ to $J \approx J_{th}$. We also see a radical difference between torques averaged over free precession of spheroidal and the free wobbling of ellipsoidal grains.

\subsection{Stability of low and high-$J$ attractor points} 

From Section 6.1, we see that the effect of thermal fluctuations on 
RATs is to produce attractor points at low $J$. In addition, there are also attractor points at high $J$ to which thermal
fluctuations have 
marginal influence. It can be shown that most of grains in the ensemble tend to get in low-$J$ attractor points. However, having low $J$, they may be significantly influenced by 
randomization processes (e.g. collisions by gaseous atoms). 

For high-$J$ attractor points present in the trajectory map (see Figs \ref{f5a}, \ref{f9a} and \ref{f12} ), the angular momentum is determined by
the radiation energy density $u_{rad}$ (see equation \ref{eq20}), that is $J/I_{1}\omega_{T} \sim u_{rad} \langle H\rangle_{\phi}$. As a result, depending on the
distance to a given radiation source, the angular momentum of the high-$J$
attractor points changes.

However, for low-$J$ attractor points, because the angular momentum
depends on thermal fluctuations (see Hoang \& Lazarian, in preparation), it does not change when radiation intensity varies. So 
grains of the same size, near and far from the pumping source, have low-$J$ attractor 
points of similar angular momentum. Therefore, due to thermal wobbling (see Lazarian 1994), even near the radiation source, the alignment degree is not high for the case of trajectory maps without high-$J$ attractor point. Consequently, even
in regions close to the strong radiation source (e.g., star forming regions)
most of grains may rotate rather slowly.

Furthermore, the angular momentum of low attractor points depends on the grain size $a_{eff}$ because internal
relaxations are a function of $a_{eff}$ (scales as $a_{eff}^{7}$; see also Table 1). So we expect that for large grains that
have weak internal fluctuations, the value of angular momentum at low-$J$ attractor points is very close to zero as a result of the
deceleration action of RATs. In contrast, small grains having strong internal 
relaxations, can have low-$J$ attractor points of
higher angular momentum.
 
\section{Crossover dynamics}
\subsection{Description}
The randomization of grains during crossovers has been studied by
Spitzer \&McGlynn (1979) and Lazarian \& Draine (1997). In Paper I we studied
the crossover for a {\it spheroidal} AMO. It is shown that grains move to the attractor point
$\mc\xi=-1, J=0$ while undergoing multiple crossovers. In this section, we
first consider large grains, so that $t_{c} < t_{Bar}$ and thermal flipping
is not important for both axisymmetric and irregular grains.  Therefore, we can solve the equations of motion for both
$J, \xi$ and the component $J_{\|}$ of {\bf J} along ${\bf a}_{1}$. Then we incorporate the thermal fluctuations and thermal flipping into our treatment.

In Paper I, to simplify our treatment, we disregarded all internal
relaxation processes during the crossover. This is justifiable as for $a_{eff}\gg a_c$ the crossover occurs on the time
scale shorter than the internal (e.g.Barnett and  nuclear) relaxation time. However, for typical interstellar grains with $a_{eff} < a_{c}$, thermal flipping occurs very fast, and thus it must be taken into account.
Because the mirror is assumed to be weightless, 
for the {\it spheroidal} AMO, the dynamics of free rotation coincides with that of a spheroid. As a result, for a given $J$, ${\bf a}_{1}$ precesses around  ${\bf J}$ with a constant angle $\gamma$. The state of the grain is completely determined
by describing ${\bf J}$ in the lab and body systems. Equations of motion for
 this case are (see Paper I)
\begin{align}
\frac{d{\bf J}}{dt}&={\bf \Gamma}-\frac{{\bf J}}{t_{gas}},\label{cr1}\\
\frac{J d\mc\gamma}{dt}&=-\frac{dJ}{dt}\mc\gamma+\frac{dJ_{\|}}{dt},\label{cr2}
\end{align}
where ${\bf \Gamma}$ is the RAT, and $J_\|=J \mc\gamma$ is the
component of angular momentum along the maximal inertia axis.

Assuming that both precession time-scales of ${\bf a}_{1}$ around ${\bf J}$ and ${\bf J}$ around ${\bf B}$ are much smaller than the crossover time, we obtain 
\begin{align}
\frac{dJ}{dt}&=M(\langle H\rangle _{+}f_{+}+\langle H\rangle_{-}f_{-})-\frac{J}{t_{gas}},\label{cr3}\\
\frac{d\xi}{dt}&=\frac{M}{J}(\langle F\rangle_{+} f_{+}+\langle F\rangle_{+}f_{-}),\label{cr4}\\
\frac{J~d\mc\gamma}{dt}&=(\langle Q_{a1}\rangle_{+}f_{+}-\langle
Q_{a1}\rangle_{-}f_{-}) -\frac{dJ}{dt}\mc\gamma-(1-\mc\gamma)J ,\label{cr5}
\end{align}
where $M=\frac{\gamma u_{rad} a_{eff}^{2}\overline{\lambda}}{2}$, $\langle Q_{a1}\rangle$ is the torque component along the axis ${\bf
a}_{1}$, $f_{+}, f_{-}$ are probability of finding grains in the positive and
negative flipping states, respectively (see WD03).

In equation (\ref{cr5}), the component $Q_{a1}$ which is along ${\bf a}_{1}$ also involves the flipping probability because after each flipping, $\gamma \rightarrow \pi-\gamma$ for which $Q_{a1}$ changes the sign, i.e., $Q_{a1}(\pi-\gamma)\sim -Q_{a1}(\gamma)$ while equation (\ref{cr5}) describes the variation of $\gamma$ only due to RATs.

\subsection{Crossover: no averaging  over thermal fluctuations}
Let us consider first a grain with $a_{eff}\gg a_{c}$ for which the
crossover time $t_{c} \ll t_{tf}$.
the grain studied is 0.2 micron, and this phrase is not correct.
In this case, as the thermal flipping is
negligible, the motion of grains comprises the alignment of ${\bf J}$ with
respect to ${\bf B}$ and the motion of ${\bf a}_{1}$ about ${\bf J}$. We solve
the equations of motion with the initial condition $J_{0}=2 J_{th}$ and
$\gamma_{0}=\pi/4$. Clearly,
when the alignment time corresponding to the former case is shorter than the
crossover time, then the grains are quickly driven to the attractor
points. In particular, grains that become aligned at high J attractor points do not undergo
crossovers, but others that are driven to low-$J$ attractor points do. The trivial
consequence of this is that, as $\gamma$ changes, it can alter
$\langle F\rangle_{\phi}$ and $\langle H\rangle_{\phi}$ because they depend on $\Theta$, which is a
function of $\gamma$ and $\xi$ (see Appendix A).

\subsubsection{Only low-J attractor points}
If we consider the crossover of a grains (e.g. the {\it spheroidal} AMO) for the radiation direction $\psi=30^{\circ}$ for which only low-J attractor points are present in the phase trajectory map. The grain dynamics of this grain is similar to the case
studied in Paper I.

\subsubsection{High-J and low-J attractor points coexist}
Now let us consider a situation when both high-J and low-J attractor points are present in the phase trajectory map (e.g. $\psi=0^{\circ}$ for shape 1). As discussed earlier, irregular grains undergo {\it irregular} wobbling (cf. thermal wobbling) which
induces averaging RATs over three Euler angles. Therefore, the obtained RATs
are functions of $s$ (see Fig \ref{f21d}). For simplicity, we assume that the dependence on $\gamma$ of averaged RATs
for irregular grains is similar to that of axisymmetric grains, and thus we can
solve equations of motion (\ref{cr3})-(\ref{cr5}) for phase trajectories. In
fact, the above assumption is feasible when the slightly irregular grains with $I_{2}
\sim I_{3}$.

The results for irregular grains are shown in trajectory maps (see Fig. \ref{f9}). 
\begin{figure} 
\includegraphics[width=0.49\textwidth]{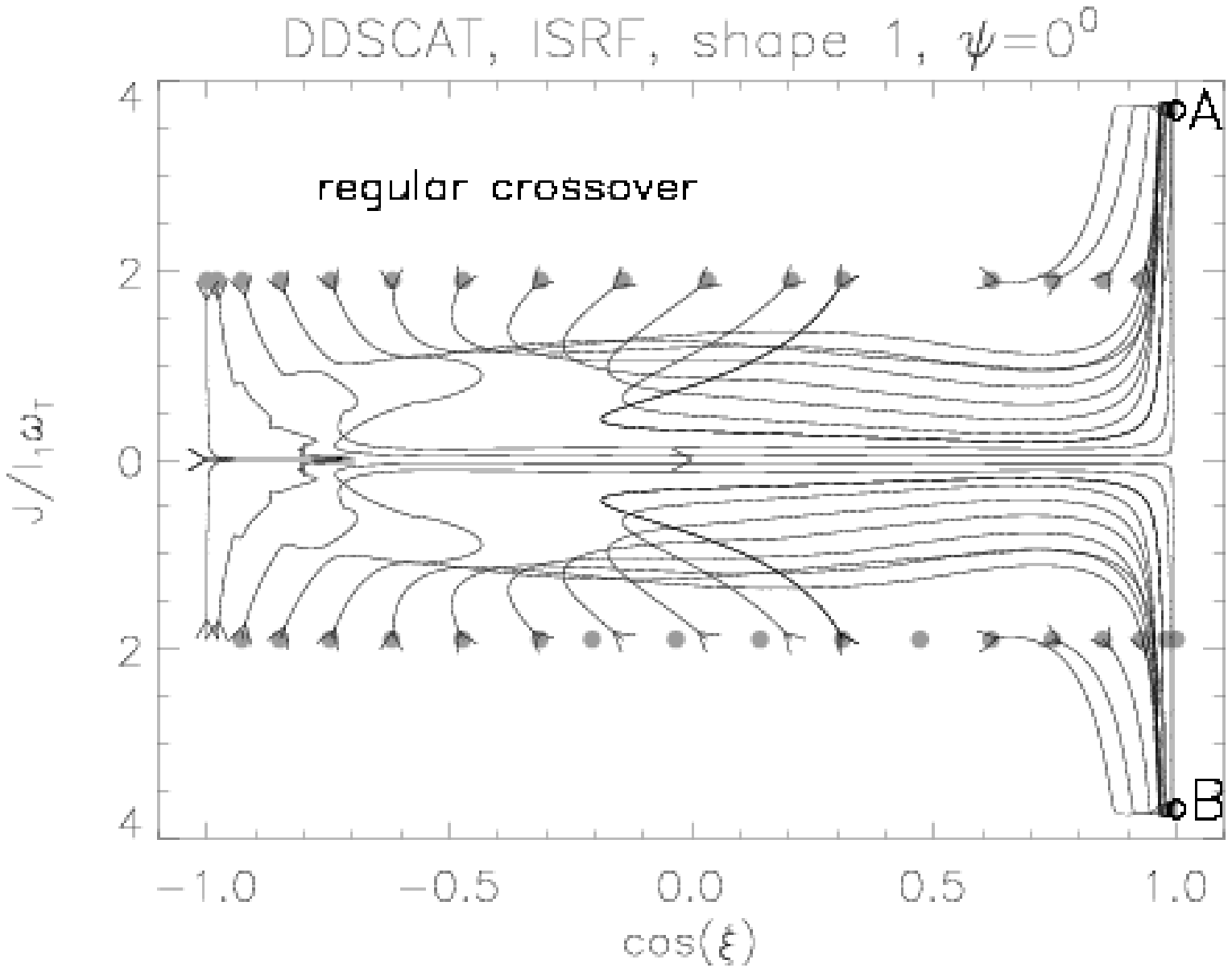} 
\includegraphics[width=0.49\textwidth]{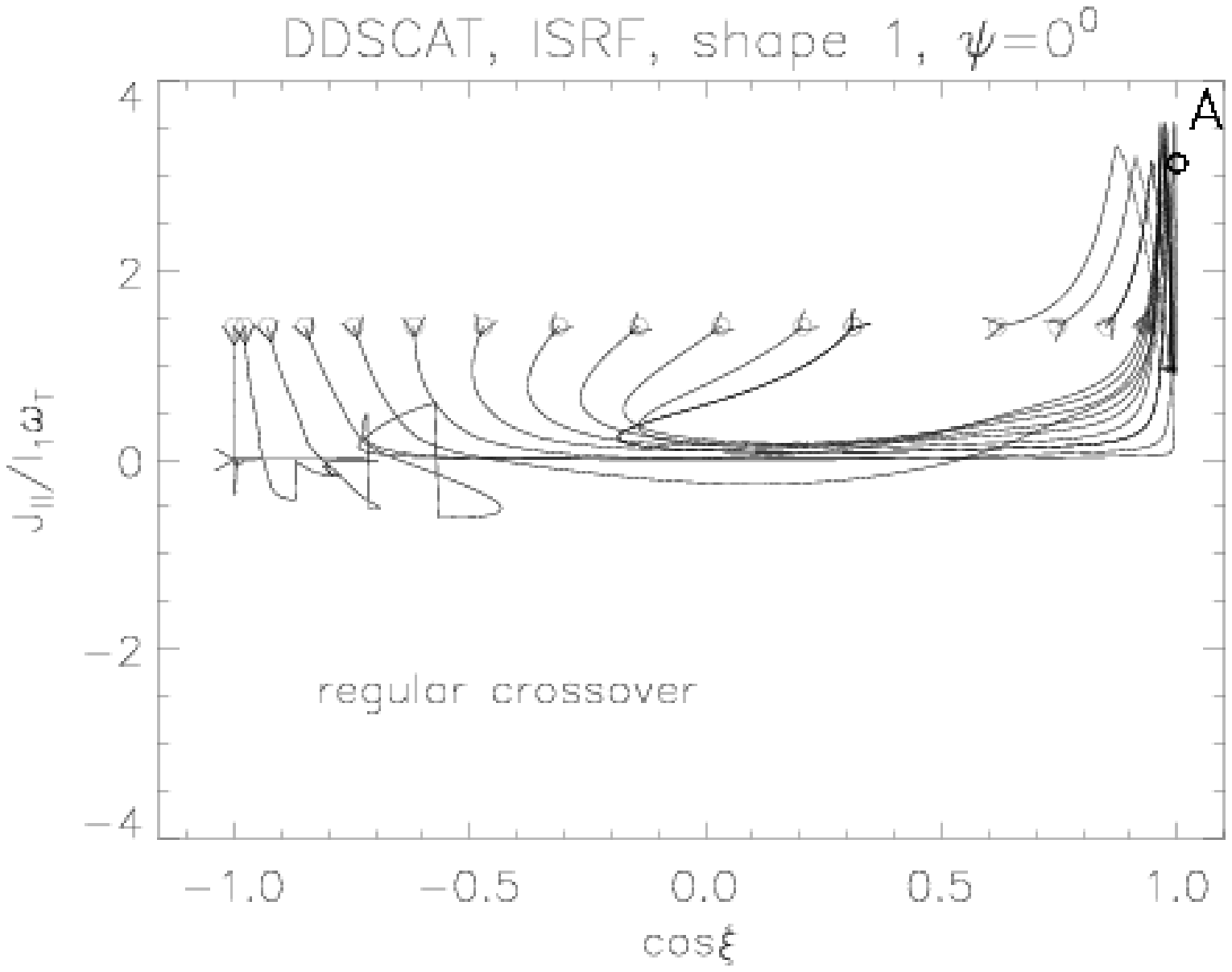}
\caption{The phase maps for the angular momentum $J$ ({\it Upper panel}) and the component of {\bf J} along the maximal inertia axis ${\bf a}_{1}$ ({\it Lower panel}). The lower panel shows that
  grains experience multiple crossovers.}   
\label{f9} 
\end{figure}
One result that can be seen directly from Fig. \ref{f9} is that grains
do not reach the low attractor point with $J=0$ as seen in the phase maps in the absence of thermal fluctuations. Instead, they first tend to reach $J<J_{th}$, and then reverse their direction to head for the high-$J$ attractor point. This stems from the property of RATs. Indeed, the torques are functions of $\Theta$, angle between ${\bf a}_{1}$ and ${\bf k}$. Therefore, we can qualitatively explain the effect based on the relationship of $\xi, \Theta$ and $\gamma$.
Firstly, for grains with initial angles $\xi<\pi/2, \gamma=0$, the spinning up
torques $H<0$ and $F>0$ (see Fig \ref{f21b}). Therefore, the angular momentum of grains decreases
while their alignment angle $\xi$ increases. In a short time, $\gamma$ increases, which
gives rise to the change in sign of $F$ and $H$ (see also \S
\ref{lowJ}). Thus, grains start to reverse their orientation ($\xi$ decreases)
and they are spun up to the high-$J$ attractor points. Finally, a perfect alignment at the high attractor point is established. 

We see here the difference in the crossover for irregular grains and the {\it spheroidal} AMO. For example, the difference is evident from the different variation of RATs averaged over
free-torque motion for the spheroidal and irregular grains (see Fig. \ref{f21d}). There it is shown that RATs for irregular grains ({\it ellipsoidal} AMO) change the sign
at small deviation angle $\gamma$ compared to $\gamma=\pi/3$ for the {\it spheroidal} AMO. It turns
out that when grains are driven to low-$J$ attractor points, they can be spun up
again as seen in Fig. \ref{f9} (cf. Fig. \ref{f9a}).

Fig. \ref{f9} (lower panel) represents the phase map for the parallel component
$J_{\|}$. It reveals that $J_{\|}$ evolves tightly with $J$. However, grains
undergo multiple crossovers (i.e. $J_{\|}=0$) before being spun up to high J attractor
points. 
Fig. \ref{f9} also indicates that during the crossovers, the angular momentum of grains is about the thermal angular momentum associated with the
grain temperature.

One appealing feature seen in the phase map is that without thermal
fluctuations, grains are not trapped at low-$J$ attractor points. They can be
spun up after crossovers. However, as thermal fluctuations are considered, the
results in \S 4 and 5 show that grains are still trapped at low-$J$ attractor points.

\subsection{Crossover: Averaging over thermal fluctuations}

For grains of effective size comparable to the critical size, the thermal
fluctuations are important because of the dependence of the Barnett relaxation time on
the grain size (see Table \ref{tab1}). In this case, we can consider
the grain dynamics in two different regimes. First, assuming that for $J$
greater than $J_{th}$, the crossover time is shorter than the Barnett
relaxation time, then the crossover can be treated as in \S 7.2. As a result,
grains undergo multiple crossovers (see Fig. \ref{f9b} for the phase
trajectories with $J > J_{th} $). As $J \le J_{th}$, thermal fluctuations
occur faster than the crossover time,
therefore, RATs must be averaged over thermal fluctuations. The dynamics of
grains in this stage is the same as studied in \S 4 and
5 (i.e. grains are trapped at low-$J$ attractor points).

\begin{figure} 
\includegraphics[width=0.49\textwidth]{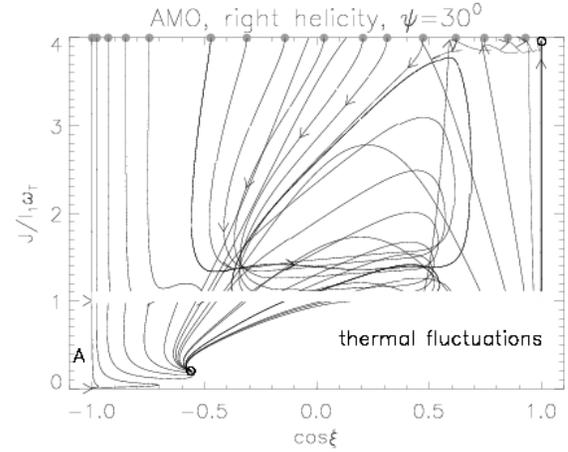} 
\caption{The phase map of grains into different regimes
  corresponding to the cases with and without thermal fluctuations. For $J>
  J_{th}$, this figure represents phase trajectories without averaging over thermal
  fluctuations; grains undergo multiple crossovers. For $J\le J_{th}$, this
  figure shows the regime in which RATs are averaged over thermal
  fluctuations; it is shown that grains get aligned at the attractor point A,
  which is similar to what obtained in \S 6.}   
\label{f9b} 
\end{figure}

The critical size $a_{c}$ depends on the radiation energy in the anisotropic
flux as $u_{rad}^{-1/2}$. Thus the alignment near stars and in the diffuse ISM is perhaps different.

\section{Fast alignment}
In Paper I we showed that the grains can be aligned over a time scale much smaller than the gas damping time. Now let us consider this problem when thermal wobbling and flipping are important. 

 Fig. \ref{f22b} shows the trajectory map constructed from an {\it ellipsoidal} AMO with $Q_{e1}^{max}/Q_{e2}^{max}=0.78$ and the light direction $\psi=70^{\circ}$. The arrow represents a time interval $\Delta t=10 t_{phot}$ where $t_{phot}$ defines the time-scale over which RATs decelerate grains from $J=J_{th}$ to $J=0$ in the absence of gas damping. It can be seen that the angular momentum of the low attractor point is the same as the case in which the gas damping is included (cf. Fig. \ref{f9a}).
\begin{figure}
\includegraphics[width=0.49\textwidth]{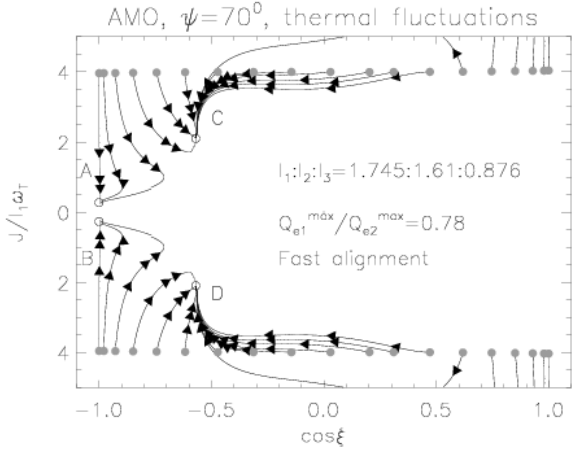}
\caption{Similar to Fig. \ref{f9a} but the gas damping is neglected. The arrow represents a time interval $10t_{phot}$.}
\label{f22b}
\end{figure}
We observe that a significant number of grains are aligned on low-$J$
attractor points A, B, C and D over $t_{ali}\sim 40 t_{phot}$ (see Fig. \ref{f22b}). 

As discussed in Paper I, this type of fast alignment can occur in a diffuse
medium with high radiation intensity, such as in the supernovae vicinity or comet wakes. This
stems from the fact that strong radiation can drive grains faster to the low-
J attractor points where the thermal fluctuations act to maintain the stability of grain
alignment.

\section{Influence of random bombardment by atomic gas}

In Paper I we showed that in most cases  RATs align grains with respect to magnetic fields
while decreasing the grains's angular momentum to $J=0$. The results above indicate that thermal fluctuations can change
a zero-$J$ attractor point to a $J\sim J_{th}$ attractor point. In addition, atomic collisions can affect the alignment established by
RATs in the presence of thermal fluctuations. In this section, we first briefly discuss the method
of studying gas bombardment based on the Langevin equation. Then, we show the
influence of this process on RAT alignment.

\subsection{Method} 
 
The effects of collisions in the framework of paramagnetic alignment have been studied 
by many authors (Jones \& Spitzer 1967; Purcell 1971; Purcell \& Spitzer 1971; 
Lazarian 1997) using the Fokker-Planck equations. 
However, the Langevin approach was used to study this
problem numerically in Roberge, Degraff \& Flatherty (1993) and Roberge \& Lazarian (1999). 
The afore-cited papers used the equivalence of the 
Fokker-Planck and the Langevin equations to simulate the evolution of
angular momentum of grains in a gas coordinate system. They derived explicit 
diffusion coefficients for random torques acting on spheroidal grains. According to the above works, an increment of angular 
momentum resulting from gas-grain collisions within an infinitesimal time interval
$dt$ is (see Roberge et al. 1993) 
\bea 
d J_{i}=A_{i}(t)dt+B_{ij}(J,t)dw_{j},~ i=x,y,z,\label{eq24b}
\ena
where $d\omega_{j}$ are Wiener coefficients,  $A_{i}, B_{ij}$ are diffusion coefficients, given by 
\begin{align}
A_{i}&=\langle \Delta J_{i} \rangle,~ i=x,y,z, \label{eq25b}\\ 
(BB^{T})_{ij}&=\langle \Delta J_{i} \Delta J_{j}\rangle,i,j=x,y,z,\label{eq26b}
\end{align} 
where $B^{T}$ is the transposal matrix of ${\bf B}$. Similar to Roberge et al. (1993), 
diffusion coefficients are first calculated in the grain body frame,
 and then transformed to the lab system. The diffusion coefficients are averaged over the precession of $\ma_{1}$ around ${\bf J}$ and over the Larmor precession angle of ${\bf J}$ about ${\bf B}$, given by 
\begin{align}
\langle{\Delta J\rangle}_{i}&=-\frac{J_{i}}{t_{gas}},\label{eq27}
\end{align} 
where $t_{gas}$ is the gaseous damping time. For a spheroidal grain, the gaseous damping time is 
\begin{align} 
t_{gas}&=\frac{3}{4\sqrt{\pi}}\frac{I_{zz}^{b}}{nmv_{th}b_{m}^{4}\Gamma_{\parallel}(e\ 
_{m})},\label{eq28b} 
\end{align}
where $I_{zz}^{b}$ is the moment of inertia of the grain 
along $zz$ axis (i.e., along the maximal inertia axis $\ma_{1}$), $v_{th}=\sqrt{2mkT_{gas}}$ is the thermal velocity of atom, and
$b_{m} $ is the semi-axis of the grain. $\Gamma_{\parallel}, \Gamma_{\perp}$ 
are factors characterizing the geometry of grain given by 
\begin{align} 
\Gamma_{\parallel}&=\frac{3}{16}{3+4(1-e_{m}^{2})g(e_{m})-e_{m}^{-2}[1-(1-e_{m}^{2}\
)^{2}g(e_{m})]},\label{eq29b}\\ 
\Gamma_{\perp}&=\frac{3}{32}[[7-e_{m}^{2}+(1-e_{m}^{2})g(e_{m})+(1-2e_{m}^{-2})\nonumber\\
&{(1+e_{m}^{-2})^{2}[1-(1-e_{m}^{2})g(e_{m})]}]].\label{eq30b} 
\end{align}
$g(e_{m})$ is related to the eccentricity of the grain through the expression:
\begin{eqnarray}
g(e_{m})=\frac{1}{2e_{m}}\mbox{ln}(\frac{1+e_{m}}{1-e_{m}}),\label{eq31} 
\end{eqnarray}
where $e_{m}=\sqrt{1-(a/b)^{2}}$. 
Diffusion coefficients are diagonal and given by 
the following expressions in the lab coordinate system in which the $z$ axis is
along the magnetic field (Roberge et al. 1993) 
\begin{align}
\langle(\Delta J_{x})^{2}\rangle&=\frac{\sqrt{\pi}}{3}nm^{2}b_{m}^{4}v_{th}^{3}(1+\frac{T_{d}}{T_{g}})\nonumber\\ 
&\times[(1+\mbox{cos}^{2}\xi)\Gamma_{\perp}+\mbox{sin}^{2}\xi\Gamma_{\parallel}],\label{eq32}\\ 
\langle(\Delta J_{y})^{2}\rangle&=\langle(\Delta
J_{x})^{2}\rangle,\label{eq33}\\ 
\langle(\Delta 
J_{z})^{2}\rangle&=\frac{2\sqrt{\pi}}{3}nm^{2}b_{m}^{4}v_{th}^{3}(1+\
\frac{T_{d}}{T_{g}})\nonumber \\ 
&\times[\mbox{sin}^{2}\xi\Gamma_{\perp}+\mbox{cos}^{2}\xi\Gamma_{\parallel}].\label{eq34}
\end{align} 
Note that the above diffusion coefficients are
derived by assuming perfect internal alignment of ${\bf a}_{1}$ with ${\bf
  J}$ and for spheroidal grains. However, for the sake of simplicity, we can adopt these diffusion coefficients for studying the influence of gas bombardment on the alignment of irregular grains.

\subsection{Results}
First, we solve the Langevin equation (i.e. equation \ref{eq24b}) for grains subjected to random 
torques as a result of gas collisions, assuming that the diffusion coefficients remain the
same for irregular grains. We use the initial condition $J=J_{th, gas}=\sqrt{2I_{1}k T_{gas}}$, $\xi$ is generated from a uniform distribution in the range $0$ to $\pi$ and $\phi$ is a free parameter. Wiener coefficients $d\omega$ in equation (\ref{eq24b}) are
generated from a Gaussian distribution function at each time-step. Then the resulting solution $J_{x}, J_{y}$ and $J_{z}$ are taken as input parameters for solving the equations of motion of grains driven by RATs in the spherical coordinate system described by $J, \xi, \phi$ (see equations (15)-(17)) to obtain new values of $J$ and $\xi$.\footnote{Here we average over the Larmor precession angle $\phi$.} This process is performed over $N=10^{6}$
time-step $\Delta t=10^{-4} t_{gas}$. Other physical parameters for the ISM are
taken from Table \ref{tab1}. The alignment angle $\xi$ is used in averaging
over the total time to obtain the degree of external alignment, and $J$ is used to
calculate the internal alignment, assuming that thermal fluctuations follow a Gaussian distribution. 

For the AMO, we consider our default model of $\alpha=45^{\circ}$ (see Paper I). For DDSCAT, we study the entire
spectrum of the ISM for grain shapes 1 and 3 with size $a_{eff}=0.2\mu m$.

\subsubsection{Influence of randomization to the  phase trajectory map in the presence of high$-J$ attractor point} 
We can see that collisions have an interesting effect on grain dynamics when the high attractor point is present. Random collisions are very
efficient for low-$J$. So, the motion of grains is disturbed by random collisions
when they approach the low-$J$ attractor point. After a time interval, grains
enter the region of $\mc\xi>0$ (i.e., positive spinning radiative torque) for which RATs can spin up to the high J
attractor point. Therefore, for this case, random collisions {\it increase
the degree of alignment}. The percentage of grains present in the vicinity of
the high-$J$ attractor points as a function of time is shown in the upper panel in
Fig. \ref{f14}. The respective degree of alignment is shown in the lower
panel.

From the upper panel (Fig. \ref{f14}), we find that, during the time interval
$t=t_{gas}$, only about $10\%$ of grains are present at the high $-J$ point. Then it
increases with time and attain the saturated value of $75\%$ after $t=40
t_{gas}$. 

Following the lower panel in Fig.~\ref{f14}, we see the rise of the degree of
alignment with time. It get to a significant value after $t=10 t_{gas}$,
and the alignment is nearly perfect with $R=0.8$ after $t=80 t_{gas}$.
\begin{figure}
\includegraphics[width=0.49\textwidth]{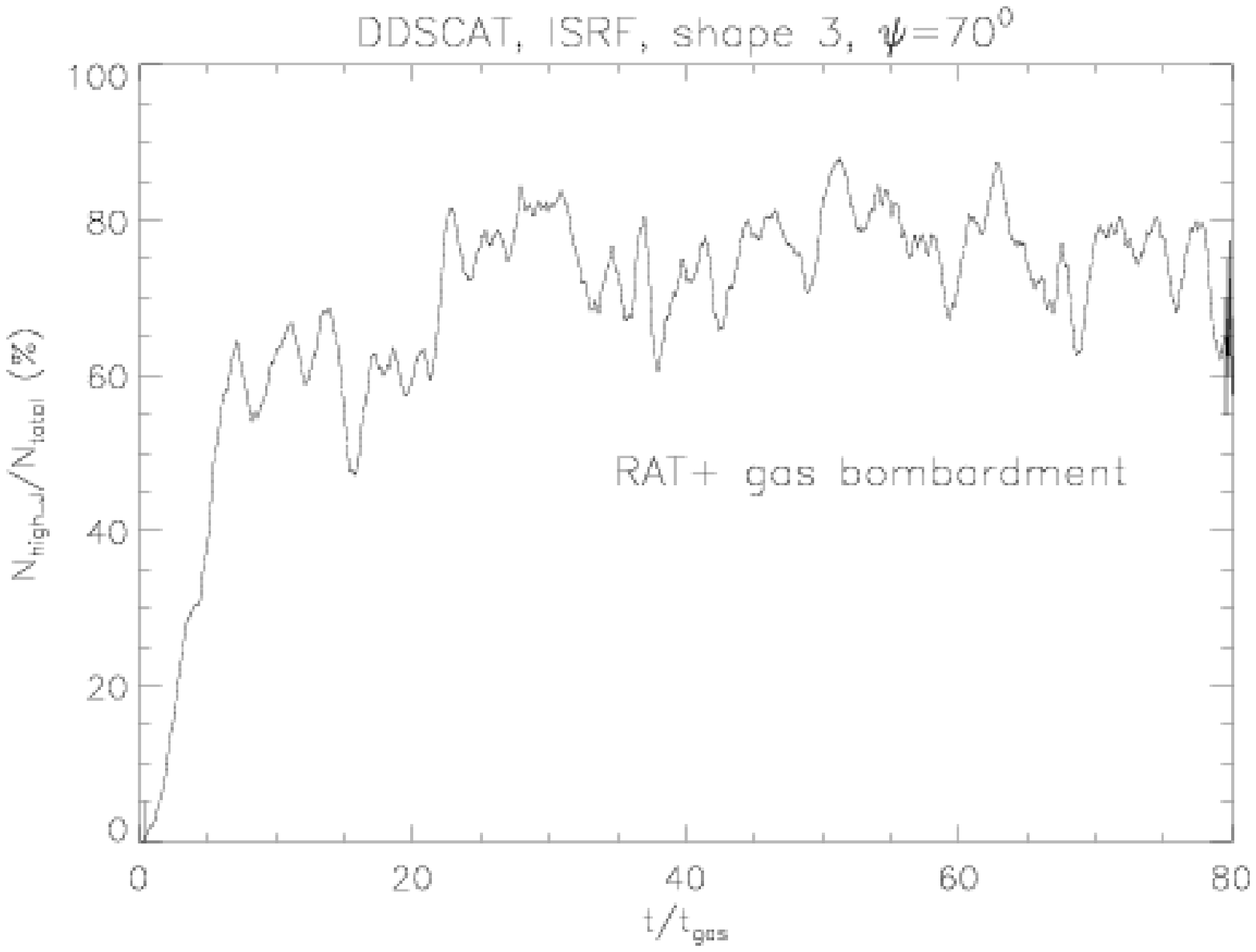}
\hfill
\includegraphics[width=0.49\textwidth]{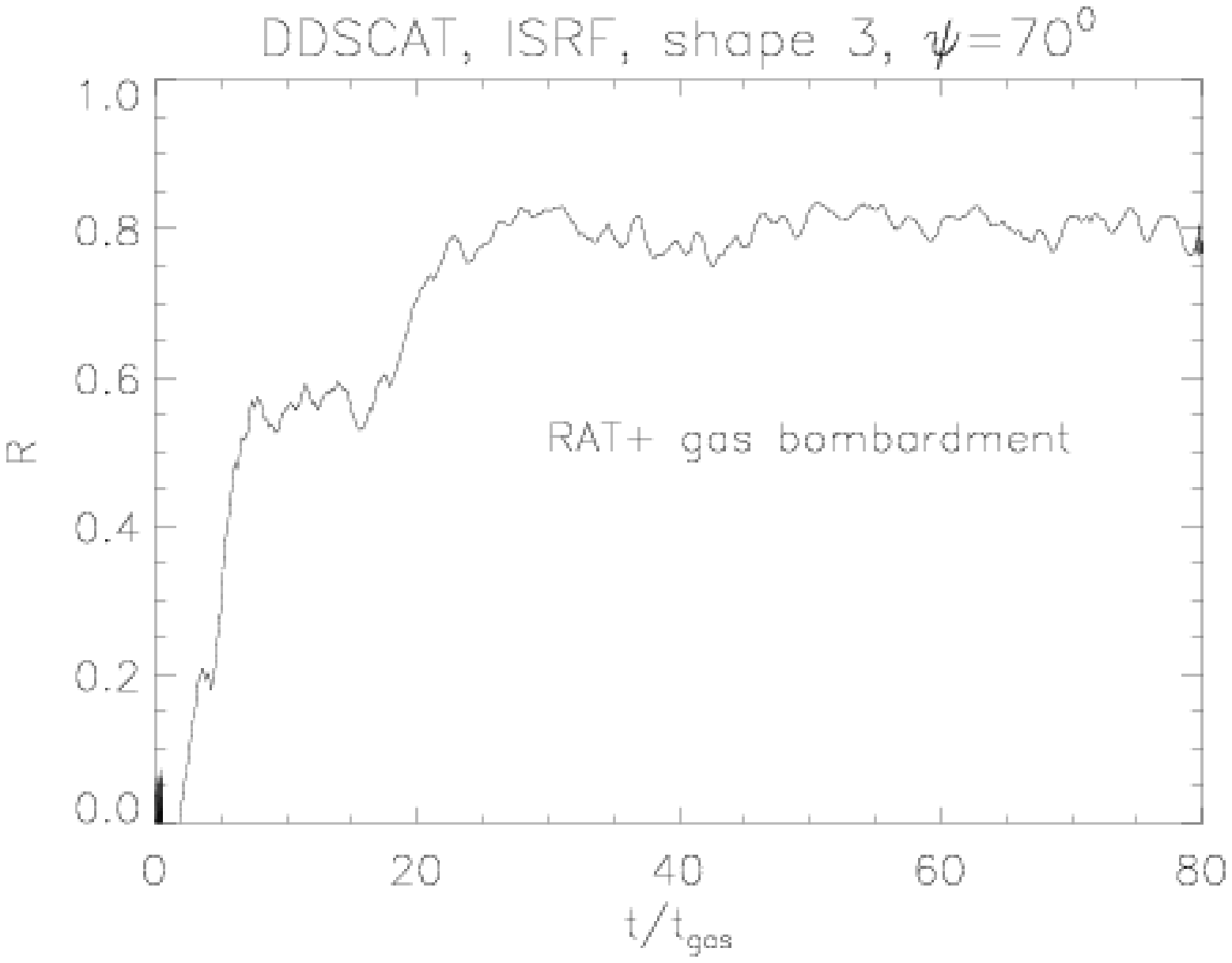}
\caption{For the ISRF, shape 3, and $\psi=70^{\circ}$:{\it Upper panel} shows the variation of the percent of grains present
  in the vicinity of the high J attractor point A, B in time, and {\it lower
  panel} shows the Rayleigh reduction factor for this case.}
\label{f14} 
\end{figure}

Due to the stochastic properties of gas bombardment, the degree of alignment depends on the angular momentum of the high attractor point $J_{high}(\psi)$. A detailed study of the degree of alignment with $J_{high}(\psi)$ in Hoang \& Lazarian ( ; in preparation) shows that the alignment at high attractor points is nearly perfect if $J_{high}(\psi)\ge 3 J_{th, gas}$ where $J_{th,gas}$ is the angular momentum corresponding to the temperature of the ambient gas.

\subsubsection{Influence of randomization to the phase trajectory map in absence of high-$J$ attractor point} 
The degree of alignment for the {\it spheroidal} AMO and irregular grains in the absence of high-$J$ attractor points is shown in Fig. \ref{f14b}.
For these cases, random collisions act to decrease the alignment. This arises from the fact that random
collisions remove grains from the low-$J$ attractor points.
We have found that if the angular momentum of the high-$J$ stationary point, $J_{high}(\psi)$, larger than $J_{th, gas}$, RATs move grains to the vicinity of the high-$J$ stationary points, and decelerate them
again. It is seen that although R is
decreased by gas bombardment, it is still not negligible (e.g., about $0.1$ and $0.2$ for
AMO and irregular grains; see Fig. \ref{f14b}). 

For $J_{high}(\psi) <  J_{th, gas}$, grains are in a fully thermal regime. Therefore,
the phase map of grains is mostly random. From the lower panel in Fig.~\ref{f14b}, it
follows that the degree of alignment is marginal. However, a more elaborate treatment in Hoang \& Lazarian (in preparatiom) does not use equation (2) and demonstrate higher degree of alignment. 
\begin{figure}
\includegraphics[width=0.49\textwidth]{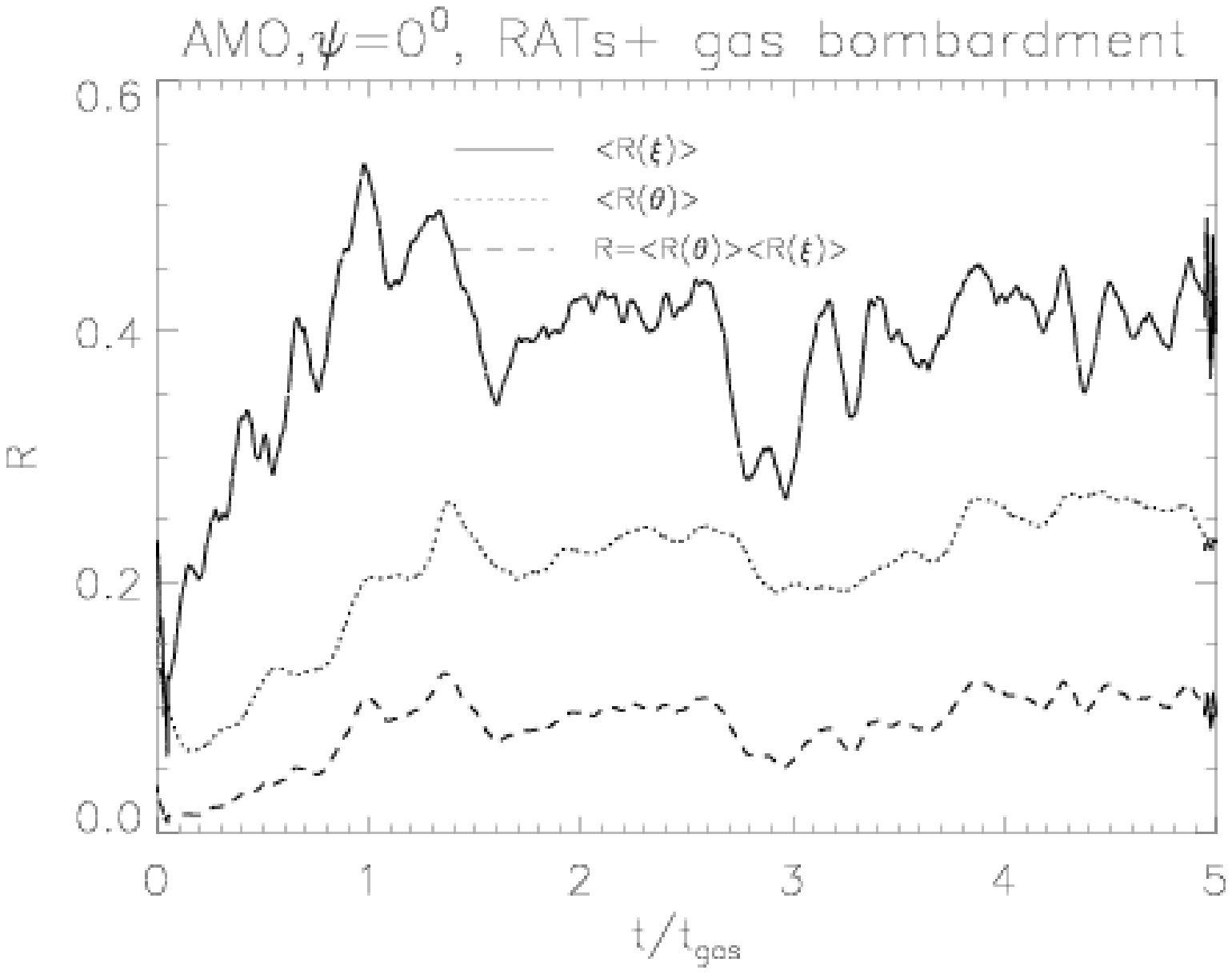}
\includegraphics[width=0.49\textwidth]{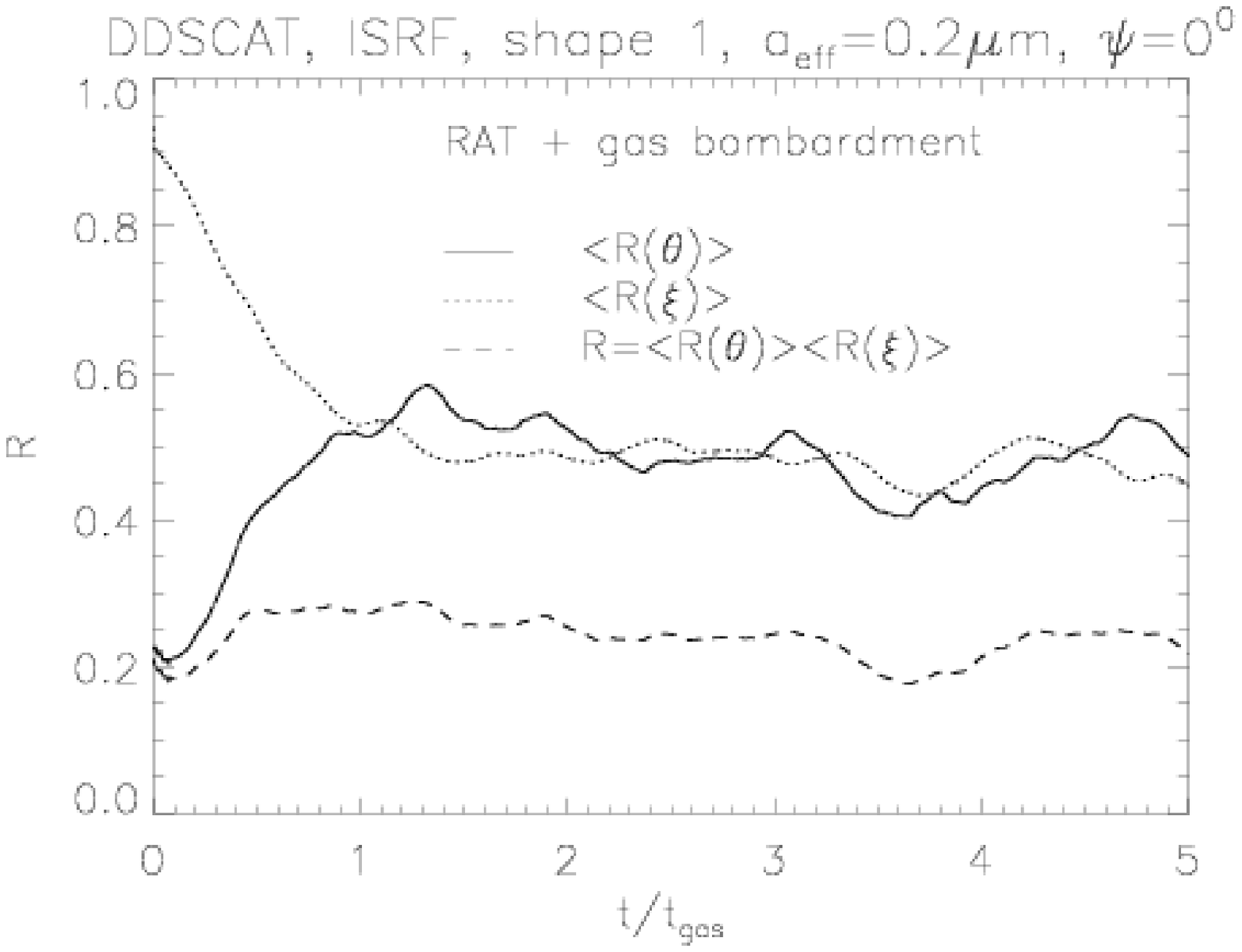} 
\caption{Figs show dynamics of external, internal degree of alignment and
  Rayleigh reduction factor in time corresponding to RATs from the AMO ({\it upper 
    panel}) and DDSCAT ({\it lower panel}) when the phase trajectory maps do not have high-J attractor points.}
\label{f14b} 
\end{figure}

\section{Influence of $h_{2}$ formation torques on grain alignment} 
When a hydrogen atom sticks to a 
grain, it starts its diffusion over the grain surface. In general, the grain surface is never uniform, and there are
always certain special catalytic sites where hydrogen atoms can be trapped (see Purcell 1979; Lazarian 1995). A wandering H 
atom on the  
grain surface may encounter another atom dwelling at the catalytic site and a reaction takes 
place, producing a $H_{2}$ molecule. The ejected $H_{2}$ molecule acts as miniature rocket thrusters. Averaging over 
the grain surface (e.g. a brick surface), $H_{2}$  rockets produce a net angular torque
that is parallel to the maximal inertia axis (Purcell 1979). However, resurfacing or poisoning by
accretion  of heavy elements can destroy 
the catalytic sites and create new ones. As a result, $H_{2}$ torques change
both magnitude and direction over a definite time
scale $t_{L}$, so-called resurfacing time.

\subsection{Method}

Purcell (1979) used the Monte Carlo method to simulate the variation of the torque 
as a result of grain resurfacing. Roberge \& Ford (2000) used the equivalence of the Langevin and the Fokker-Plank
equations to simulate the $H_{2}$ torque. They model $L_{z}$ as a Gaussian 
process that is averaged to zero in time, and the correlation function that is
exponential: 
\begin{align}
\overline{L_{z}^{b}(t)}&=0,\label{eq35}\\ 
\overline{L_{z}^{b}(t)L_{z}^{b}(t-\tau)}&=\langle(\Delta L_{z})^{2}\rangle 
\mbox{exp}(-\tau/t_{L}).\label{eq36} 
\end{align}
In equation (\ref{eq36}), $\langle(\Delta L_{z})^{2}\rangle$ is the magnitude of $H_{2}$ torque given by
\begin{align}
\langle(\Delta L_{z})^{2}\rangle^{1/2}&=\frac{1}{3}(\frac{\pi}{3})^{1/6}f n(H) (2EkT_{gas})^{1/2}a_{eff}^{2}l,\label{eq36a}
\end{align}
where $f$ is the fraction of H atoms absorbed by the grain and converted to $H_{2}$, $n(H)$ is the H density, $E$ is the kinetic energy of each departing $H_{2}$ and $l$ is the side of the individual catalytic site (Purcell 1979; Lazarian \& Draine 1997).

In an interval of time $dt$, the torque along the axis ${\bf a}_{1}, L_{z}^{b}$ , can be simulated  by a Gaussian process with the 
correlation time scale $t_{L}$, given by the Langevin equation (Roberge \& Lazarian 1999)
\begin{eqnarray} 
dL_{z}^{b}= -L_{z}^{b}\frac{dt}{t_{L}}+[2\frac{\langle(\Delta
  L_{z})^{2}\rangle}{t_{L}}]^{1/2}dw,\label{eq38} 
\end{eqnarray}
where $dw$ is a random variable, independent of time, sampled from a 
Gaussian distribution function, and
$L_{z}^{b}(t)$ is the instant torque at time $t$. This equation allows us to 
follow the evolution of $H_{2}$ torque in time.

 When the internal relaxations (Purcell 1979; Lazarian \& Draine 1999b) are taken into account, the angle $\gamma$ between
the maximal inertia axis ${\bf a}_{1}$ and the angular momentum fluctuates in
time. However, as the fluctuation time-scale is much shorter than the alignment time-scale, we do
not follow the evolution of this angle. Instead, we average over this. Following Spitzer \& McGlynn (1979), the mean $H_{2}$ torque for a spheroid is given by 
\begin{eqnarray}
{\bf \Gamma_{H_{2}}}=L_{z}^{b}\overline{\mc\gamma} \frac{{\bf J}}{J},\label{eq39} 
\end{eqnarray} 
where  $\overline{\mc\gamma}$ denotes the average of $\mc\gamma$ over thermal fluctuations as defined by equation (\ref{eq12}), which defines the value of the projection of ${\bf L}_{z}^{b}$ onto the angular momentum axis
${\bf J}\|\hat{Z}$. For irregular grains, $\overline{\mc\gamma}$ in equation (\ref{eq39}) is replaced by the thermal average, $\overline{C}$, of $C$ which is given by 
\bea 
C=(\frac{I_{1}-I_{3}q}{I_{1}-I_{3}})^{1/2} \frac{\pi}{2F(\pi/2,k^{2})}
\mbox{ for } q < \frac{I_{1}}{I_{2}},\label{eq40}    
\ena   
and   
\begin{align}
C&=(\frac{I_{3}(q-1)}{I_{1}-I_{3}})^{1/2}\frac{\pi}{2F(\pi/2,k^{-2})},
\mbox{ for } q> \frac{I_{1}}{I_{2}},\label{eq41}
\end{align}
where $k^{2}$ and $F$ are given in Appendix C (see WD03).

Because of the grain flipping, equation (\ref{eq39}) becomes
\begin{align}
{\Gamma}_{H2}^{J}&=L_{z}^{b}\overline{\mc\gamma}(f_{+}-f_{-}).\label{eq41a}
\end{align}
where $f_{+}$ and $f_{-}$ are probability of finding the grain in positive and negative flipping states, respectively.

The equations of motion (i.e. equations \ref{eq16}-\ref{eq17}) of grains by RATs and $H_{2}$ torque read
\begin{align}
\frac{d\xi}{dt}&=\frac{\gamma u_{\mbox{rad}}a_{\mbox{eff}}^{2}\overline{\lambda}}{2J}\langle F(\xi, \phi, \psi,
J)\rangle_{\phi},\label{eq42a}\\ 
\frac{dJ}{dt}&=\frac{1}{2}\gamma u_{\mbox{rad}}a_{\mbox{eff}}^{2}\overline{\lambda} \langle H(\xi, \phi, \psi,
J)\rangle_{\phi}-\frac{J}{t_{gas}}+{\Gamma_{H_{2}}^{J}},\label{eq42b}
\end{align}
where we have averaged RAT components  over the Larmor precession angle $\phi$.

\subsection{Results} 
We solve the equations of motion for grains subjected to RATs and $H_{2}$ torques 
using the Runge-Kutta method with a finite time-step, $\Delta t=10^{-4} t_{gas}$ and $N=10^{5}$ time-steps . We use the initial condition $J= 20 J_{th}$, and $\xi$ is generated from a uniform angle distribution in the range from $0$ to $pi$. At each time-step, we first solve
the Langevin equation (equation \ref{eq38}) for components of $\mbox{H}_{2}$ torques. Then, by subsituting the resulting $H_{2}$ torques into equations {\ref{eq42a})-(\ref{eq42b}), we solve the obtained equations for $J$ and $\xi$. The resulting solutions $J$ and $\xi$ are used to construct trajectory maps and calculate the degree of alignment.
\subsubsection{Trajectory maps}

Fig. \ref{f15} shows the map when the variation of $H_{2}$
torques is accounted for (lower panel), compared with the map driven by RATs
(upper panel)
produced by a radiation field of $\lambda=1.2 \mu m$ for a grain with size $a_{eff}=0.2 \mu m$. We
assume that the resurfacing process on the grain surface occurs rapidly, so
we model the variation of $H_{2}$ torques with the
correlation time-scale $t_{L}=0.1 t_{gas}$.
\begin{figure}
\includegraphics[width=0.49\textwidth]{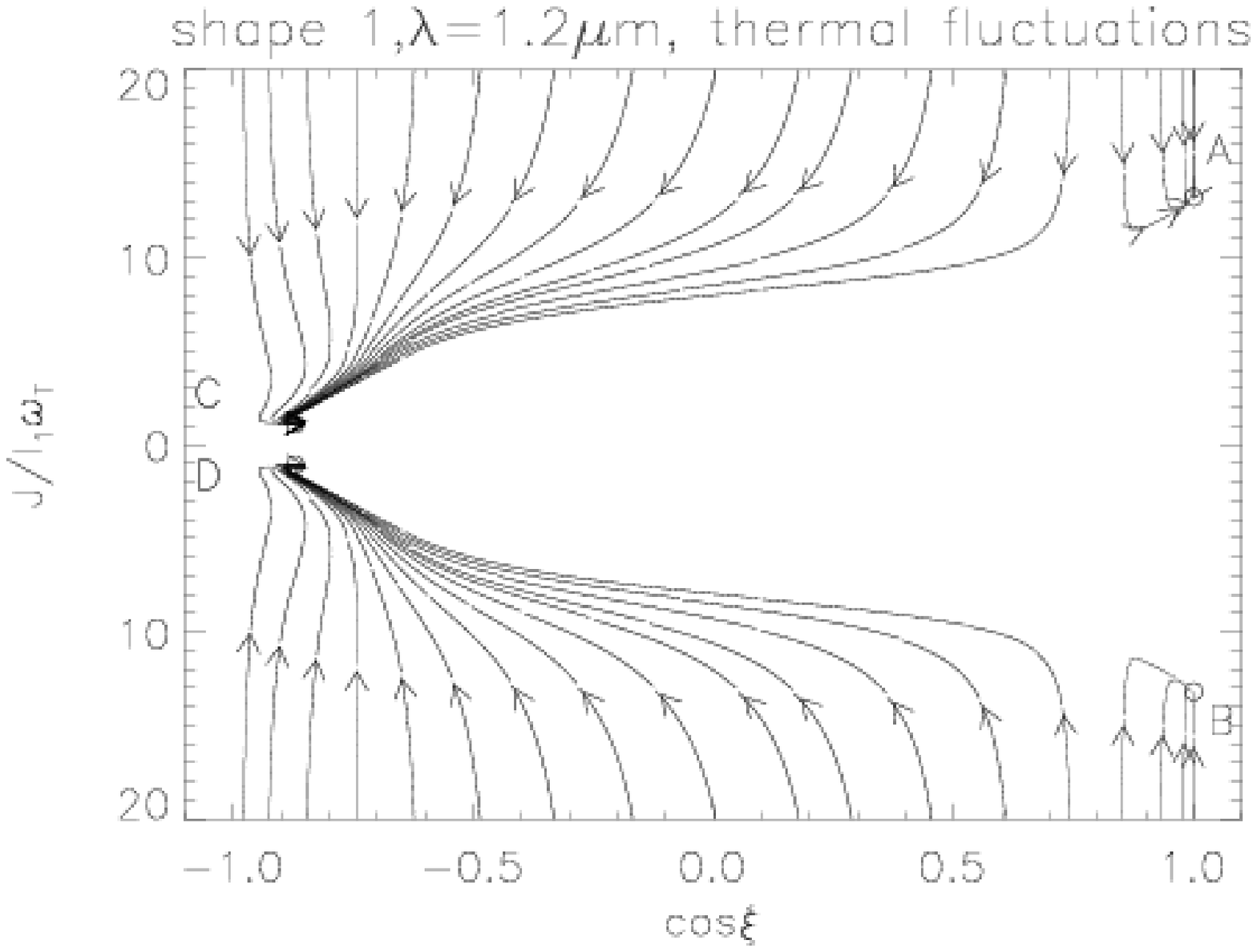}
\includegraphics[width=0.49\textwidth]{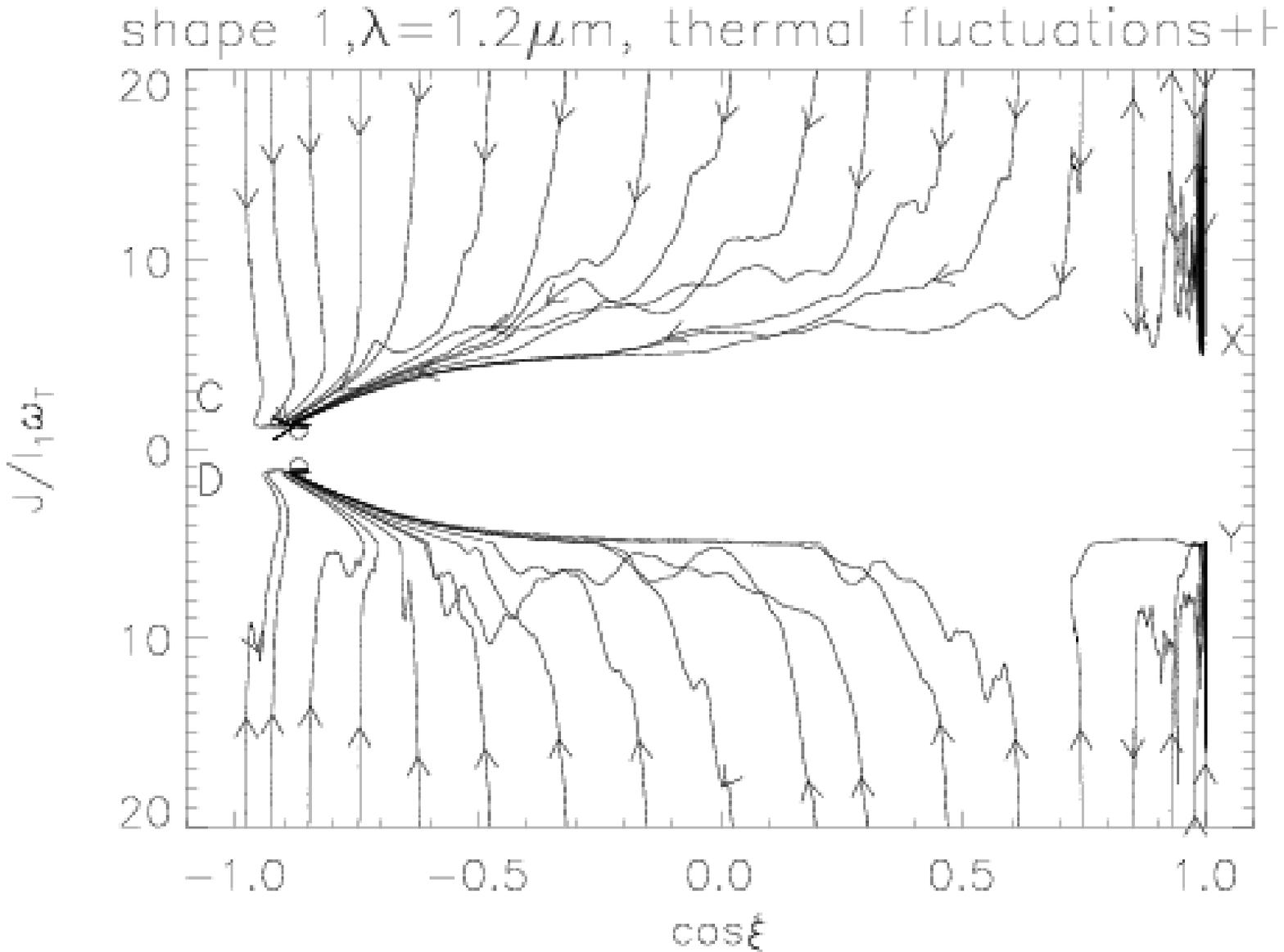}
\caption{{\it Upper panel:} map trajectory of grain $a_{eff}=0.2 \mu m$ and
  $\lambda=1.2\mu m$ by RATs; {\it lower panel:} when $H_{2}$
  torque varies due to resurfacing of active sites on grains surface. The results show
  that, during the spin-down, grains are driven by RATs to
  thermal angular momentum. However, because grains flip very fast, $H_{2}$ torques become inefficient
  to spin up grains at low-J attractor points.}
\label{f15}
\end{figure}
In the absence of $H_{2}$ formation torques, RATs drive grains to
attractor points. A significant fraction of grains are driven to low attractor
points marked by C, D, at $J/I_{1}\omega_{T} \sim 1$, and some grains are
aligned at attractor points A, B corresponding to high angular momentum (see the
lower panel in Fig. \ref{f15}).

Consider now a case of when $H_{2}$ torque amplitude is larger than that of RATs (e.g. consider the radiation direction $\psi=70^{\circ}$). It is obvious that RATs
still dominate the Purcell torque for this case, and drive a significant number of grains to the low attractor points C and D. In fact, at the initial phase of $J \gg J_{th}$, grains
are spun up by the Purcell torque;  however, during the spin-down, RATs lead grains to low-$J$ attractor points. Because of grains having low angular
momentum at the low-$J$ attractor points, grains flip fast and the Purcell torques are averaged to
zero (see the lower panel in Fig. \ref{f15}) similar to what is described in Lazarian \& Draine (1999a). Therefore, the trajectory map with  two attractor points  C, D of low angular momentum is determined by
RATs as in the case without $H_{2}$ torques (see the upper
panel in Fig. \ref{f15}). However, for the attractor points A, B with high angular momenta, the thermal flipping does not occurs because of $J \gg J_{th}$. In other words, during spin-down
period, grains are driven by $H_{2}$ torques to reach the low angular
velocity regime. As $J$ decreases, RATs become dominant and drive grains to attractor points C, D, as shown in the lower panel of Fig. \ref{f15}. However, as $J $ increases,
$H_{2} $ formation torques become dominant and do not allow grains to have
attractor points. Therefore, attractor points A, B become unstable, and grains are spin
up and down by $H_{2}$ torques and RATs in the range between X and Y (see the lower panel in Fig. \ref{f15}), corresponding to $J/J_{th}=5$ to $40$. The range of such a variation depends on the relative amplitude of $H_{2}$.

When the correlation time-scale $t_{L}=10t_{gas}$, our results show that
trajectories of grains ending at high-$J$ attractor points C, D are marginally affected by $H_{2}$ torques,
because RATs eventually drive grains to these attractor points at which the fast thermal flipping
destroys completely the influence of $H_{2}$ torques. In contrast, grains that
are driven to X and Y experience the same effect as in the case $t_{L}=0.1 t_{gas}$. 

For $t_{L} \to \infty$, $H_{2}$ torques act to spin grains up. Therefore, the low-$J$ attractor points become high -$J$ attractor points, provided that $H_{2}$ torques dominate RATs. 

For larger grains, thermal flipping is less important (see Lazarian \& Draine 1999a) , so $H_{2}$ torques
are important even during the slow rotation period, influencing significantly
the radiative alignment. We provide the treatment of this case elsewhere.

\subsubsection{Dynamics of degree of grain alignment}
Fig. \ref{f16} shows
the variation of the degree of alignment $\langle R\rangle$ in time for the cases in which only
RATs are considered (upper panel) and RATs plus $H_{2}$ torques with the correlation time-scale $t_{L}=0.1 t_{gas}$ are taken into account (lower panel).

It is seen from Fig. \ref{f16} that the degree of alignment $\langle R\rangle$ is the same for the case $\psi=70^{0}$ because of the
    dominance of RATs in driving grains to attractor points. At
    $\psi=30^{0}$, the alignment degree exhibits small variations. In addition, for
    both directions, $\langle R\rangle$ does not increases smoothly to a stable value as in
    the case of alignment by only RATs as grain dynamics is affected by the stochastic
    variation of $H_{2}$ torques. When grains are driven to the attractor points
    with thermal angular momentum A and B, the fast thermal flipping of grain
    wipes out the effect of $H_{2}$ torques, which are fixed in the grain body
    coordinate system. Therefore, the grain alignment is mostly determined by RATs.
\begin{figure}
\includegraphics[width=0.49\textwidth]{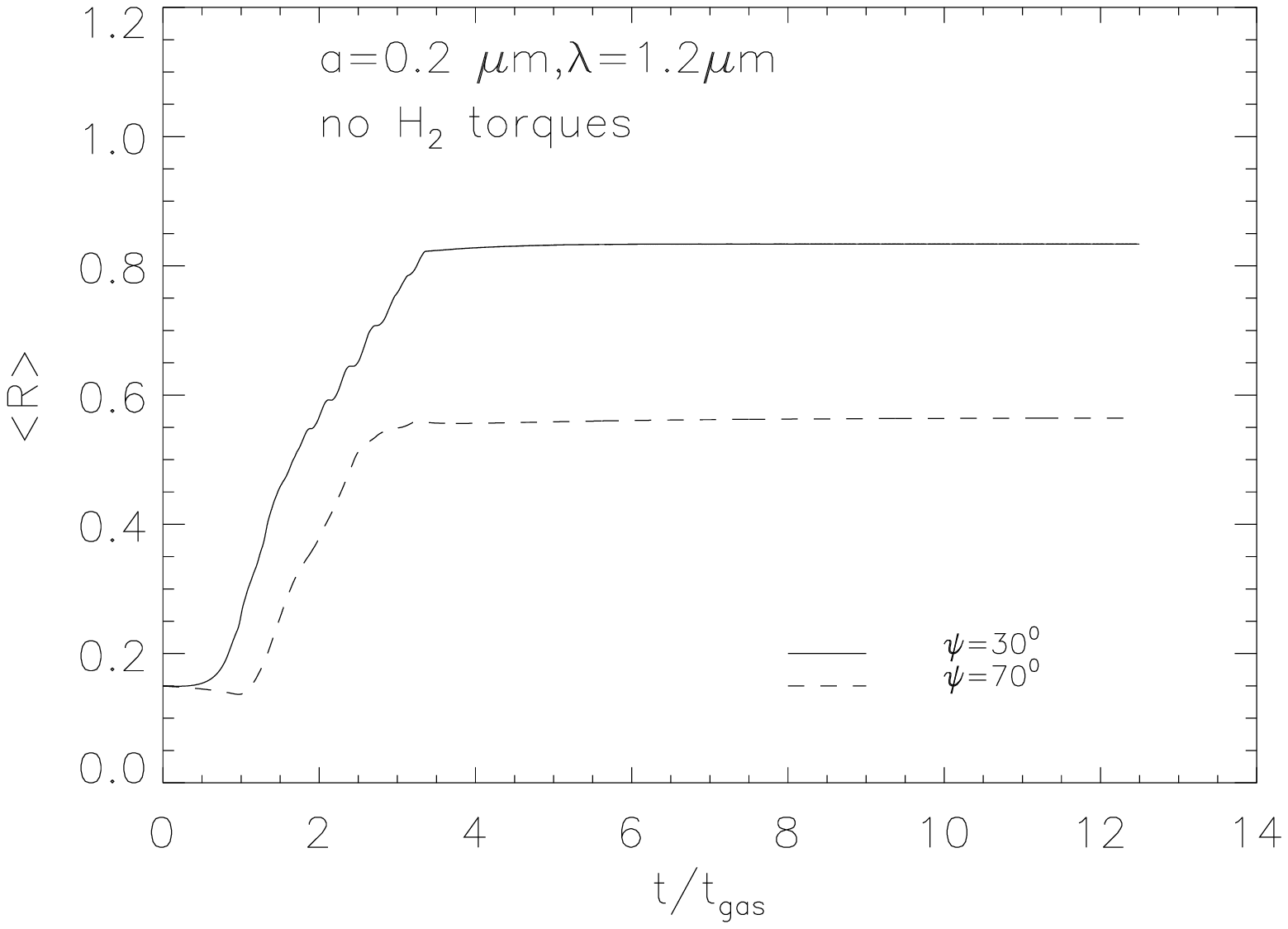}
\includegraphics[width=0.49\textwidth]{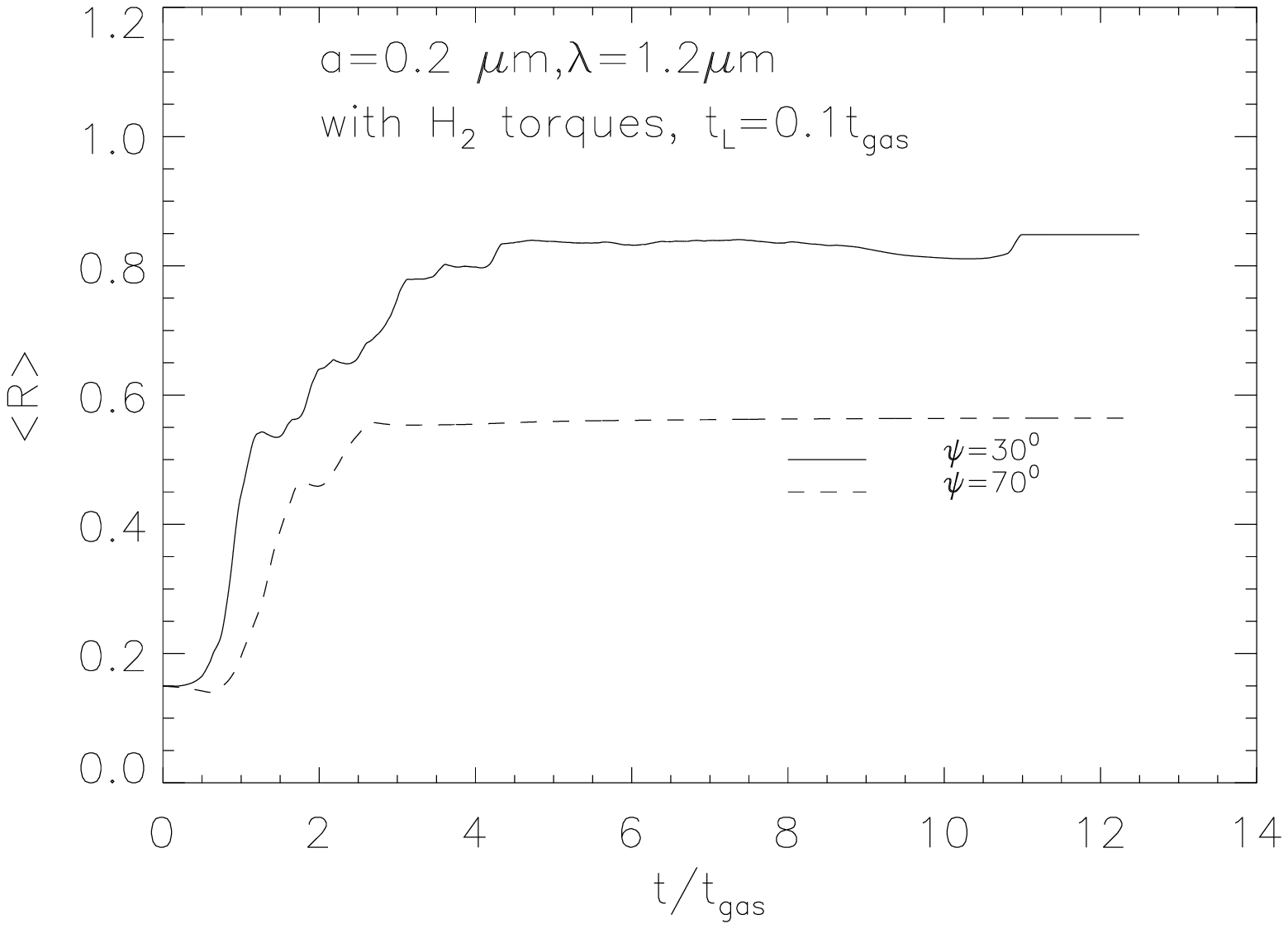}
\caption{Panels  show the variation of degree of alignment $\langle R\rangle$,  of ${\bf J}$ with ${\bf B}$ in time for the
  monochromatic light illuminating grain in two directions $\psi=30^{\circ}$ and $70^{\circ}$ . {\it upper
    panel:} only RATs align grains and they have stable
    orientation; {\it lower panel:} the same as the lower panel, but $H_{2}$
    torques  with a correlation time-scale $t_{L}=0.1 t_{gas}$ are included.}
\label{f16}
\end{figure}

\subsection{Alignment by RATs, $H_{2}$ formation torques and gas bombardment}
 In most cases, RATs play the key role in aligning
grains with a magnetic field, while $\mbox{H}_{2}$ is only important to spin up
grains. Moreover, for grains smaller than $\sim 10^{-4}\mbox{cm}$, thermal flipping
is very fast, and thus the influence of {\bf $H_{2}$} torques is rather marginal. In contrast, for
large grains $ \geq 10^{-4}$ cm, $H_{2}$ torques could be significant for
the grain alignment . On the other hand,
random collisions affect the alignment of grains mostly when grains rotate
at thermal velocity.
\begin{figure}
\includegraphics[width=0.49\textwidth]{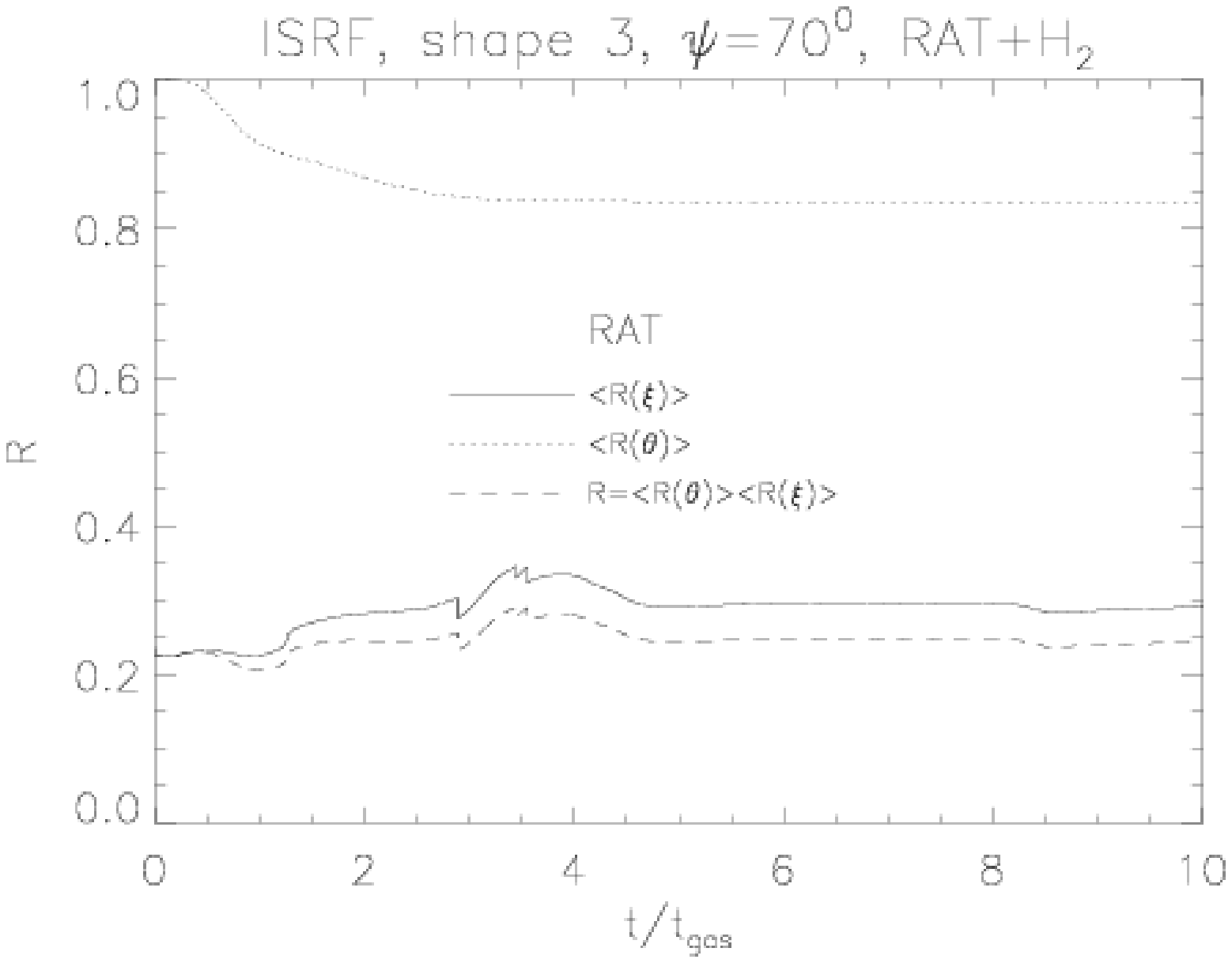}
\includegraphics[width=0.49\textwidth]{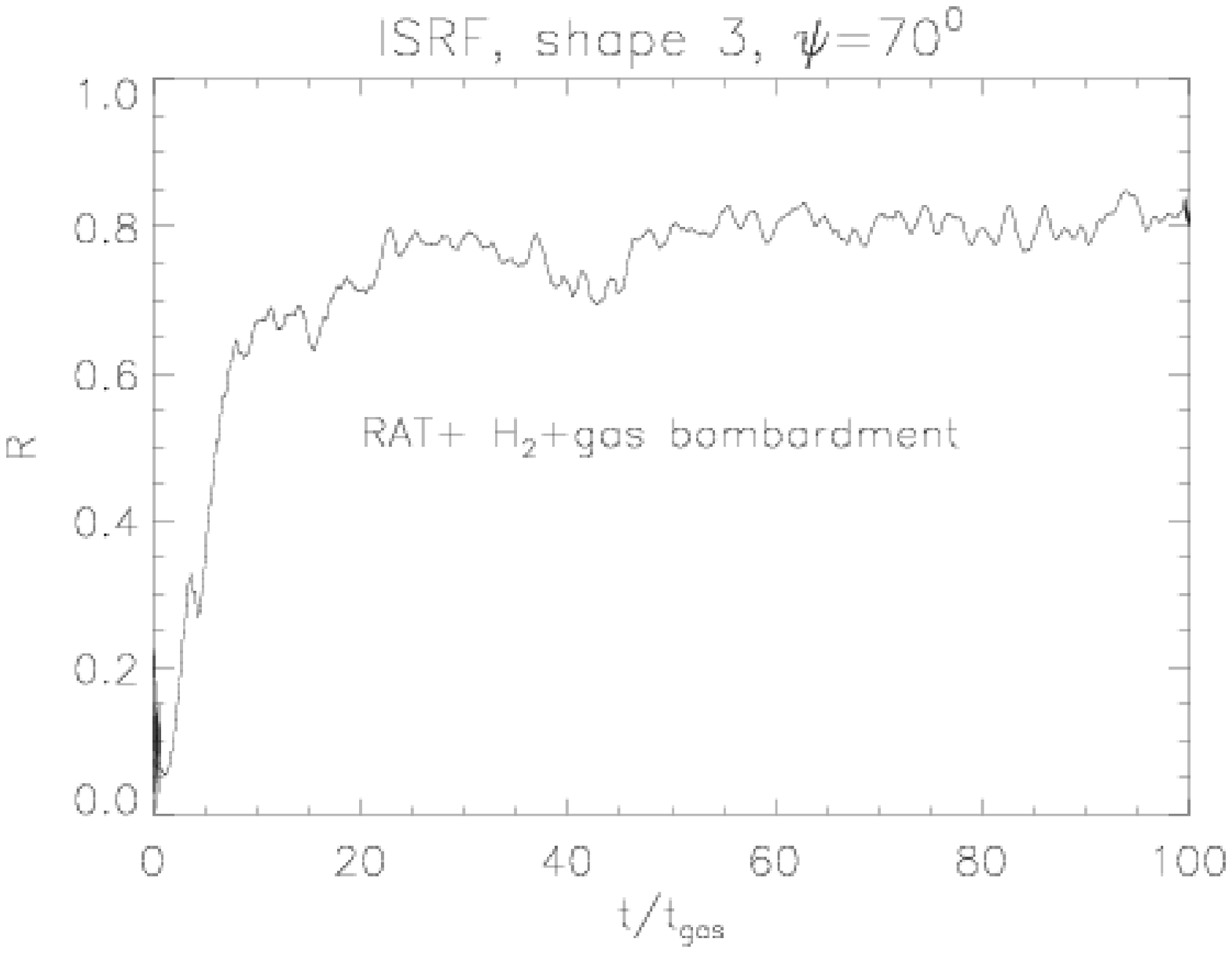}
\caption{The Rayleigh reduction factor for grain $a_{eff}=0.2\mu m$ aligned by RATs,
  $H_{2}$ formation torques ({\it upper panel}) and  by RATs, $H_{2}$ and gas
  bombardment ({\it lower panel}).}
\label{f17}
\end{figure}
The upper panel in Fig. \ref{f17} shows both internal and external degree of
alignment as well as the Rayleigh reduction factor for the alignment induced by RATs
and $H_{2}$ torques. It is seen that $R\sim 0.25$ for this case.
However, when gas bombardment is taken into account, we found that R increases
to $R=0.75$ (see the lower panel  in Fig. \ref{f17}). In fact, this increase is
not unexpected, since for this case, grains are driven by gas bombardment to high attractor point, which increases the degree of alignment as discussed in
\S 8.

\section{Discussion}

\subsection{Central role of the AMO}

Our understanding of the RAT alignment has been improved considerably when a simple model of a helical grain (i.e. the AMO), was introduced in Paper I. With the assumption of perfect internal alignment, the AMO reproduced the alignment effects similar to irregular grains studied by DDSCAT (see Paper I). For instance, the AMO shows that for a given shape, size and radiation field, some grains can be spun up to suprathermal rotation and aligned at high-$J$ attractor points, while most of grains are driven to low-$J$ attractor points. Moreover, the AMO  demonstrated that whether RATs can spin-up grains to high J attractor points depends on the ratio of the torque components $Q_{e1}$ and $Q_{e2}$. 

The correspondence of the AMO to DDSCAT calculations of RATs for irregular grains in Paper I encouraged us to use the AMO for the case when thermal wobbling is taken into account.
In our study, we used the AMO assuming that its inertial properties are defined by a triaxial ellipsoid rather than a spheroid. This was very easy to accomplish,
as, within the AMO, torques acting upon either spheroidal, or ellipsoidal body 
induce only precession, while the actually important torques (i.e.
$Q_{e1}$ and $Q_{e2}$) arise from the mirror. 
We found a number of differences in the AMO dynamics for the two cases. We treat the AMO
with ellipsoid of inertia as our generic case. 

\subsection{Treatment of free wobbling}

In Paper I we showed that $Q_{e3}$ in the AMO can be negligible when considering the
alignment and spin-up torques. We have found again that the component $Q_{e3}$ for the AMO also
plays a marginal role in alignment, as well as, spinning up grains if we followed the algorithm in WD03 and averaged
RATs over a sufficient time-steps (e.g. about $N=10^{3}$; see \S 3.3). Using our new algorithm of averaging in which we solve the differential equation for the rotation angle $\zeta$ instead of assuming that this  angle ranges from $0$ to $2\pi$, we obtained similar results  when the averaging is perfomed over a much longer time (e.g. $P=10^{3} P_{\tau}$). It contributes
mainly to the precession about the radiation beam axis ${\bf k}$ (see Fig. 1). As a result, in most practical situations, it is sufficient to deal with only two RAT components $Q_{e1}$ and $Q_{e2}$.

We studied the alignment for two cases (1) when thermal relaxation
time-scale is much shorter than the crossover time (i.e., averaging RATs over
thermal fluctuations) and (2) when the thermal relaxation time scale is much longer than the crossover time. For the former case, grains are directly aligned at low-$J$ attractor points. In contrast, the grains undergo multiple crossovers in the second case.

The AMO enables us to average RATs analytically over thermal
fluctuations assuming a Gaussian distribution. The obtained expressions allowed
us to gain important insights into the role of thermal wobbling.

\subsection{Direction of grain alignment}

It has been shown in Paper I that the alignment by RATs occurs mainly with the
longer axis perpendicular to the magnetic field. However, there was still a narrow
range of radiation directions for which grains were aligned with longer axes parallel to
magnetic field  (i.e. ``wrong'' alignment). We have shown that the "wrong" alignment can no longer exist because of the fast wobbling of the grains. Indeed, in 
Paper I we found that ``wrong'' alignment takes place for a low-$J$
attractor point for a narrow range
of angles around the angle  between the light direction and magnetic field $\psi=\pi/2$ . This range is narrower than the range
of grain wobbling at the low-$J$ attractor point. The disappearance of the ``wrong'' alignment was, in fact, predicted in Paper I. 
A direct implication this effect is that allows a more reliable interpretation
of the observation data: grains are aligned with long axes perpendicular to magnetic field when they are aligned by RATs.

\subsection{Degree of grain alignment and critical size}

The degree of external alignment of angular momentum with the magnetic field can achieve unity. However, because of the weak coupling of the maximal inertia axis of grain with
the angular momentum at low-$J$ attractor points, the Rayleigh reduction factor $R$ is lower than unity. A new effect that we discuss in this
paper is the transition of grains from low-$J$ to high-$J$ attractor points,
in situations when such attractor points coexist. This increases 
the degree of alignment. Interestingly enough, the transition which makes grains better aligned is induced by {\it random bombardment} (see \S 9).

The degree of alignment depends on the angular momentum $J_{high}$ of the high attractor points. The value of $J_{high}$ is a function of RATs that increases with grain size. Because of the gas bombardment, only grains aligned with $J_{high}>J_{th}$ can maintain the stable alignment with the magnetic field. The minimal size corresponding to $J_{high,min}$ is named the critical size. For the irregular shape 1 and the ISRF, by studying the alignment for grains with size spanning from $0.025\mu m$ to $0.2\mu m$, we found that the critical size of aligned grains is about $0.05\mu m$, assuming that $J_{high,min}=3J_{th}$. Apparently, critical size depends on radiation field. Therefore, for attenuated field, e.g. dark molecular clouds, the critical size is shifted to larger value (i.e., only large grains can be aligned by RATs).

\subsection{Role of paramagnetic dissipation}

Paramagnetic dissipation does not play any role in the RAT alignment.
Although the predictions of RAT alignment coincides with the predictions by the Davis-Greenstein
mechanism (i.e. the alignment
 with long axes perpendicular to the magnetic field), RAT alignment occurs over shorter time scale compared to 
paramagnetic relaxation time $t_{DG}$, provided that grains are
paramagnetic. 

As a result, unless grains are superparamagnetic (see Jones \& Spitzer 1967;
Mathis 1986), paramagnetic relaxation is irrelevant. A common fallacy 
entranced in
the literature is that RAT alignment is a sort of paramagnetic
alignment of suprathermally rotating grains, i.e. kind of alignment of
fast rotating suggested by 
Purcell (1979), but with fast alignment produced by RATs. This way of thinking
of RAT alignment is erroneous, as, unlike the Purcell (1979) torques, RATs
induce their own alignment, which is faster than the paramagnetic one.  

\subsection{Effects of radiation field intensity}

When grains are aligned at high-$J$ attractor points, the variations
of radiation intensity induce the variation in $J$ value. However, a significant fraction of grains that align at low -$J$ attractor points at which $J$ does not vary with radiation intensity. Therefore, after
a certain threshold at which grains are aligned suprathermally (i.e.
with $\omega\gg \omega_{thermal}$) at high-$J$
 attractor point 
 the increase of radiation intensity does not increase
 the degree of alignment for grains aligned at such points.

When grains are aligned at low-$J$ attractor points,
a weaker alignment is expected, because of the poor coupling of the
maximal inertia axis with angular momentum for low $J$ (see Lazarian 1994; Lazarian \& Roberge 1997), irrespective of the intensity of radiation field. The increase of the radiation intensity does
not affect the position of lower-$J$ attractor points.

\subsection{Role of gas bombardment and $H_{2}$ torques}
This is the first study to take into account the effects of gaseous bombardment
and H$_2$ formation within the RAT alignment mechanism.
Random gas bombardment itself has always been considered as the
 cause of grain randomization. However, in the framework of the RAT alignment, its effect can be different.

As discussed in the paper, RATs align grains with respect to the magnetic field while driving grains to high-$J$ and low-$J$ attractor points.
Naturally, the later case is more sensitive to gas bombardment. This random process can increase the alignment degree by driving grains from the low-$J$ attractor points to high J attractor points. Such a process gives rise to the time-dependent alignment. The time over which the alignment can become  stable is about $10^{2} t_{gas}$. For high-$J$ attractor points, the alignment is marginally affected by the bombardment.

$H_2$ torques were believed to play a major role on the grain alignment. However, we found in the  present study that $H_{2}$ torques are indeed important for high attractor points for which they can spin grains up. For low attractor points with low angular momentum, the fast flipping of grain axis as a result of thermal fluctuations averages out any torques which are fixed in the grain body, e.g. $H_{2}$ torques.

\subsection{Time-scales of alignment}

If paramagnetic dissipation is neglected, one may guess that the only
characteristic time that can determine RAT alignment could be $t_{gas}$ (see
DW97). As pointed out in Paper I, strong radiation field can provide grain alignment over a time-scale shorter than
the gas damping time (i.e. fast alignment). The characteristic time of
this alignment can be $\sim 30 t_{phot}$, where $t_{phot}$ is the time over
which the amplitude value of radiative torques can deposit a grain
with the angular momentum of the order of its initial angular momentum. The
large coefficient in front of $t_{phot}$ reflects relative inefficiency of
radiative torques in the vicinity of low-$J$ attractor points.
The supernova shell or star forming regions are favorable medium for such 
a fast alignment.

Our present study shows that another time scale is appropriate, when the
phase trajectory map contains both low-$J$ and high-$J$ attractor points.
This time is about $\sim 30 t_{gas}$ and it corresponds to the time scale
over which the gaseous bombardment transfers grains from low-$J$ to high-$J$
attractor points.

\subsection{Role of initial $J$}

The role of initial $J$ for alignment has been studied in Paper I, where
we found that it determines the time of ``fast'' alignment $t_{phot}$. In addition, in the paper above, we have seen new
  low-$J$ attractor points present in the phase maps when thermal fluctuations
  are accounted for. However, if grains start from small $J$, the
grain trajectories may not enter the parameter space, that would
induce directing grains to high-$J$ attractor points. Instead, the trajectories
may, in the absence of gaseous bombardment, end up at the 
low-$J$ attractor points.

However, in practical terms, 
this particular difference is not crucial for the overall dynamics.
In the presence of gaseous bombardment, the grains will be moved to the high
attractor points, if such points exist. 

\subsection{Grain dynamics at $J\sim J_{th}$}

In Paper I, with the assumption of simplified dynamics, i.e., 
assuming that ${\bf J}$ is always aligned with the maximal 
inertia axis ${\bf a}_{1}$, we showed that  AMO induces a good alignment with respect to radiation direction or magnetic field. The assumption of ${\bf J}$ and $\ma_{1}$ coupling is correct only for $J \gg J_{th}$. In this paper, we use the analytical formulae for RATs and get the analytical expressions for torques components when thermal fluctuations are accounted for, and therefore the angle between ${\bf J}$ and $\ma_{1}$ fluctuates. Results show that torque components decrease substantially as thermal fluctuations become stronger (see Figs \ref{f41}), but the dynamics of grains does not change radically. This interesting finding testifies that thermal fluctuations as a result of internal relaxations do not play the key role in creating the low-$J$ attractor points as it was believed earlier, but they ``lift up'' the earlier existing attractor points. 

In order to see whether the irregularity of grain shape is important, we modify slightly AMO by replacing the spheroidal body by the ellipsoidal one. As earlier discussed, the ellipsoidal body does not affect on $Q_{e1}, Q_{e2}$ for AMO, 
but changes the dynamics of fluctuations and therefore the averaging.
As a result, the torques resulting from averaging over thermal wobbling are different, which induces the difference in the dynamics for irregular grains. Thus,
 the irregularity of shape and the presence of thermal fluctuations act
together when grain angular momentum is comparable with $J_{th}$.

\subsection{Crossovers}
We studied the crossover by following the variations of angular momentum and
the component parallel to the axis of major inertia in the case when thermal
fluctuations are negligible. 
Our detailed treatment of crossovers is in agreement with the earlier
studies by Spitzer \& McGlynn (1979) and Lazarian \& Draine (1997),
but is different from the treatment of crossovers in DW97. 

Whereas in Paper I we followed the regular phase trajectories of grains approaching
to the crossover, in the present study we take into account thermal wobbling
that becomes important as the grain angular momentum becomes comparable with the 
thermal one. Unlike the approach in WD03, which is applicable only to small 
grains, our treatment is also applicable to large
grains for which initially the time
of thermal wobbling (which is of the order of the internal relaxation time) 
is longer than that of crossovers. We find that both large and small grains
eventually get to a similar state, at which the amplitude
of the angular momentum gets of the order of its {\it thermal} value. This
corresponds to low-$J$ attractor point. 

Our work confirms the claim in Paper I that the low-$J$ attractor points 
are not metamorphosis of crossover points after which grains spin up again
as in DW97 (see also WD03), but a modification of zero-$J$ attractor points, that exists
in the absence of thermal fluctuations.

\subsection{Helicity of grains}

Helicity of grains as a factor of RAT processes was first mentioned in
Dolginov \& Mytrophanov (1976). However, there it was not identified what can
make grains helical. Moreover, it was claimed that some regions of space can
have non-helical grains, which implied that helicity of grains is something 
that requires rather special conditions to emerge\footnote{An additional
confusion in the aforementioned study stemmed from their claim that the
prolate grains should align with long axes perpendicular to the alignment
of the long axes of oblate grains. However, it was shown in Lazarian (1995) that both
types of grains align the same way, if internal relaxation is accounted for.}
. Later the RAT action was
demonstrated for rather arbitrarily shaped grains in DW96, but this and
subsequent studies, namely, DW97, WD03, did not use the concept of helicity.

Grain helicity, as an essential requirement for alignment was identified in 
Paper I, where by comparing results of numerical calculations for irregular
grains with AMO, it was established that grains can be classified as
 either of left or right helicity. Moreover, it was shown that irregular
grains demonstrate helicity and, therefore, alignment, not only subject to
radiation, but also to gaseous flows (see more in Lazarian \& Hoang 2007b). 

In Paper I, an irregular grain rotating about its axis corresponding to the maximal inertia axis was identified as a helical grain. Our present study demonstrates the
advantage of good internal relaxation of the maximal inertia axis and angular momentum. Indeed, we show that torques acting on a grain
decrease when the grain starts wobbling.

\subsection{Irregularity of shape and inertia}

Irregular grains, in general, have an irregular shape and have to be 
characterized by a triaxial ellipsoid of inertia, rather than a spheroid. The
irregularity of grain shape gives rise to grain helicity, while the irregularity
in terms of the moment of inertia modifies grain dynamics. One of the
interesting effects that we observed for grains, which 
can be characterized by a spheroid of inertia, is a transfer of grains from
low-$J$ attractor point (actually, zero-$J$ attractor point) to high-$J$ 
attractor point in the absence of gaseous bombardment. This effect disappears 
when thermal fluctuations are accounted for.

\subsection{Comparison with earlier studies}

The irregular torques arising from emission and absorption photons randomly depositing angular momentum to grains
were discussed in Harwit (1970). The marginal importance of the mechanism 
for the alignment was shown, however, in Purcell \& Spitzer (1971). Dolginov (1972) proposed the first model of RATs that act on chiral, e.g. quartz, grains.
Later, Dolginov \& Mytrophanov (1976) considered a more generic case of RATs,
i.e. RATs arising from ``twisted'', irregular grains. The idea of RAT alignment
was mostly ignored by the community (see Lazarian 1995, as an exception) until
numerical calculations of RATs got available in DW96, DW97. 
From the earlier studies, the most relevant to this work is WD03, where effects of thermal fluctuations within the RAT alignment were empirically studied.

This work extends our study in Paper I, which demonstrated the ability of the AMO to represent helical grains. The AMO allowed us to address the generic 
features of the RAT alignment. We note that our procedures of torque averaging and our 
results for angular momentum for low-$J$ attractor points differ from
those in WD03. Our interpretation of the origin of those points is also 
different.    

Extending AMO, we have proven
that thermal trapping of grains at the low $J$ attractor points
reported in WD03 is generic,
i.e. does not depend on the particular direction between the beam and the 
magnetic
field, or on the particular choice of the irregular grain or on the particular 
wavelength/radiation spectrum chosen.
Moreover, we have shown that the we expect the new low $J$ attractor points 
to appear close to $\cos\xi=1$, which makes grains aligned with {\bf B} at these new attractor
 points.
All the results we obtained with AMO were also tested with two irregular grain
shapes. In addition, we accounted for gaseous bombardment and H$_2$ formation.

\subsection{Accomplishments and limitations of the present study}

The study above clarified a number of outstanding issues in RAT alignment
theory. It relied on the guidance of the analytical studies of RATs and 
crossovers in Paper I and accounted or the proper averaging of RATs 
arising from thermal wobbling. It confirmed the predicted in Paper I the
gradual change of the position of lower attractor points on the trajectory maps of irregular grains from zero to the thermal value of angular momentum. 
In addition, our study accounted for the effects of gaseous bombardment and
H$_2$ formation on the RATs. This resulted in a discovery of a new important 
effect, namely, the transfer of grains from lower attractor points to high 
attractor points, when the latter points exist. This is a single most
important finding of our paper. It is also a practically important effect, as
the internal alignment is perfect for grains at high attractor points and 
reduced for grains at lower attractor points. All in all, the present paper 
substantially extends our analysis in Paper I. 

The limitations of the present study stem from the fact, that, while
addressing the issue of how the degree of alignment changes in different
circumstances, it does not calculate the degree of alignment exactly. This is
done in Hoang \& Lazarian (in preparation).       

\subsection{Towards modeling of grain alignment}

Present day modeling of grain alignment both in molecular clouds (CL05;
Pelkonen et al. 2007; Bethell et al. 2007) and accretion disks 
(Cho \& Lazarian 2007) uses rather heuristic recipes for determining whether
grains are aligned. In these studies the amplitude values of the RATs are
calculated and used to calculate the maximal angular velocities achievable for
a given damping of grain rotation. Such velocities parametrize the efficiencies
of grain alignment in the chosen environments. In the paper above, we have
confirmed the utility of such a parametrization and improved the criterion for
the alignment to be efficient. However, we still have to obtain better measures
of the expected alignment.

\section{Summary}

In the present paper, we studied the role of thermal fluctuations, thermal
flipping, efficiency of $H_{2}$ torques and influence of random collisions on
the RAT grain alignment for both the AMO and irregular grains. Our main results are summarized below:

1. We have used an AMO with inertia defined by a triaxial ellipsoid and averaged numerically RATs over thermal fluctuation. For this {\it ellipsoidal} AMO, we have found the increase of angular momentum of low-$J$ attractor points from $J=0$ to $J \sim J_{th}$. We also found the similar effect for irregular grains in which RATs are calculated using DDSCAT and for the entire spectrum of the ISRF.

2. We have proved that ``wrong'' alignment corresponding to low angular momentum for a narrow range of radiation direction reported in Paper I is eliminated when thermal fluctuations and thermal flipping are considered.

3. We have studied the crossover for the AMO and irregular grains and found that grains experience multiple crossovers before being trapped at low-$J$ attractor points in the presence of thermal fluctuations.

4. We have found that random collisions of atomic gas increase the degree of
alignment when grains are aligned by RATs with both low-$J$ and high-$J$ attractor points by driving grains from low-$J$ to high-$J$ attractor points. When the angular momentum of the high-$J$ attractor point $J_{high}(\psi)$ greater than $3 J_{th,gas}$, a significant degree of alignment can be achieved. 

5. We have studied the influence of $H_{2}$ formation torques in the frame of
the RAT alignment for different resurfacing times, showing that it can, in particular circumstances, enhance the degree of alignment
with respect to the magnetic field.

\section*{Acknowledgments}
We thank  Bruce Draine for clarifying to us some points in their DDSCAT code. TH
acknowledges Joseph Weingartner for clarifications provided at the initial stage of this work. We thank Wayne Roberge and Ford for sharing with us their results from Roberge \& Ford (2000). The support by the NSF Center for Magnetic Self-Organization in Laboratory and Astrophysical
Plasmas and NSF grant AST 0507164 is acknowledged.

\appendix
\section{Relation of angles $\Theta, \beta, \Phi$ and $\xi, \phi, \psi$}\label{apen1}
\begin{align}
\mc\Theta&=\ma_{1}.\me_{1},\\
\Phi&=2{\mbox tan}^{-1}\left(\frac{\ms\Theta-\ma_{1}.\me_{2}}{\ma_{1}.\me_{3}}\right),\\
\beta&=2{\mbox tan}^{-1} \left(\frac{\ms\Theta+\ma_{2}.\me_{2}}{\ms\Theta(\ma_{2}.\me_{3}\mc\Phi-\ma_{2}.\me_{2}\ms\Phi)}\right),
\end{align}
where $\ma_{1}, \ma_{2}, \ma_{3}$ are unit vectors in grain reference, and
$\me_{1}, \me_{2}, \me_{3}$ are unit vectors in the lab system.
Dot products in above equations can be obtained from following expressions
\begin{align}
\ma_{i}.\me_{1}&=\mc\psi\mz_{B}.\ma_{i}-\ms\psi\mx_{B}.\ma_{i},\\
\ma_{i}.\me_{2}&=\mc\psi\mx_{B}.\ma_{i}-\ms\psi\mz_{B}.\ma_{i},\\
\ma_{i}.\me_{3}&=\my_{B}.\ma_{i},
\end{align}
where
\begin{align}
\mx_{B}.\ma_{1}&=\mx\xi\mc\phi\ms\zeta\ms\gamma+\ms\phi\mc\zeta\ms\gamma\nonumber\\
&+\ms\xi\mc\phi\mc\gamma,\\
\mz_{B}.\ma_{1}&=\mc\xi\mc\gamma-\ms\xi\ms\zeta\ms\gamma,\\
\my_{B}.\ma_{1}&=\mc\xi\ms\phi\ms\zeta\ms\gamma-\mc\phi\mc\zeta\ms\gamma\nonumber\\
&+\ms\xi\ms\phi\mc\gamma,\\
\mx_{B}.\ma_{2}&=\mc\xi\mc\phi(\mc\alpha\mc\zeta-\ms\alpha\ms\zeta\mc\gamma)\nonumber\\
&-\ms\phi(\mc\alpha\ms\zeta+\ms\alpha\mc\zeta\mc\gamma)\nonumber\\
&+\ms\xi\mc\phi\ms\alpha\ms\gamma,\\
\mz_{B}.\ma_{2}&=-\ms\xi(\mc\alpha\mc\zeta-\ms\alpha\ms\zeta\mc\gamma)\nonumber\\&+\mc\xi\ms\alpha\ms\gamma,\\
\my_{B}.\ma_{2}&=\mc\xi\ms\phi(\mc\alpha\mc\zeta-\ms\alpha\ms\zeta\mc\gamma)\nonumber\\
&+\mc\phi(\mc\alpha\ms\zeta+\ms\alpha\mc\zeta\mc\gamma)\nonumber\\
&+\ms\xi\ms\phi\ms\alpha\ms\gamma,\\
\end{align}
with $\mx_{B}, \my_{B}, \mz_{B}$ are unit vectors of magnetic field coordinate
system in which $\mz_{B}\| {\bf B}$. Here $\alpha, \gamma, \zeta$ are Euler
angles shown in Fig. \ref{fap2}.
\begin{figure}
\includegraphics[width=0.49\textwidth]{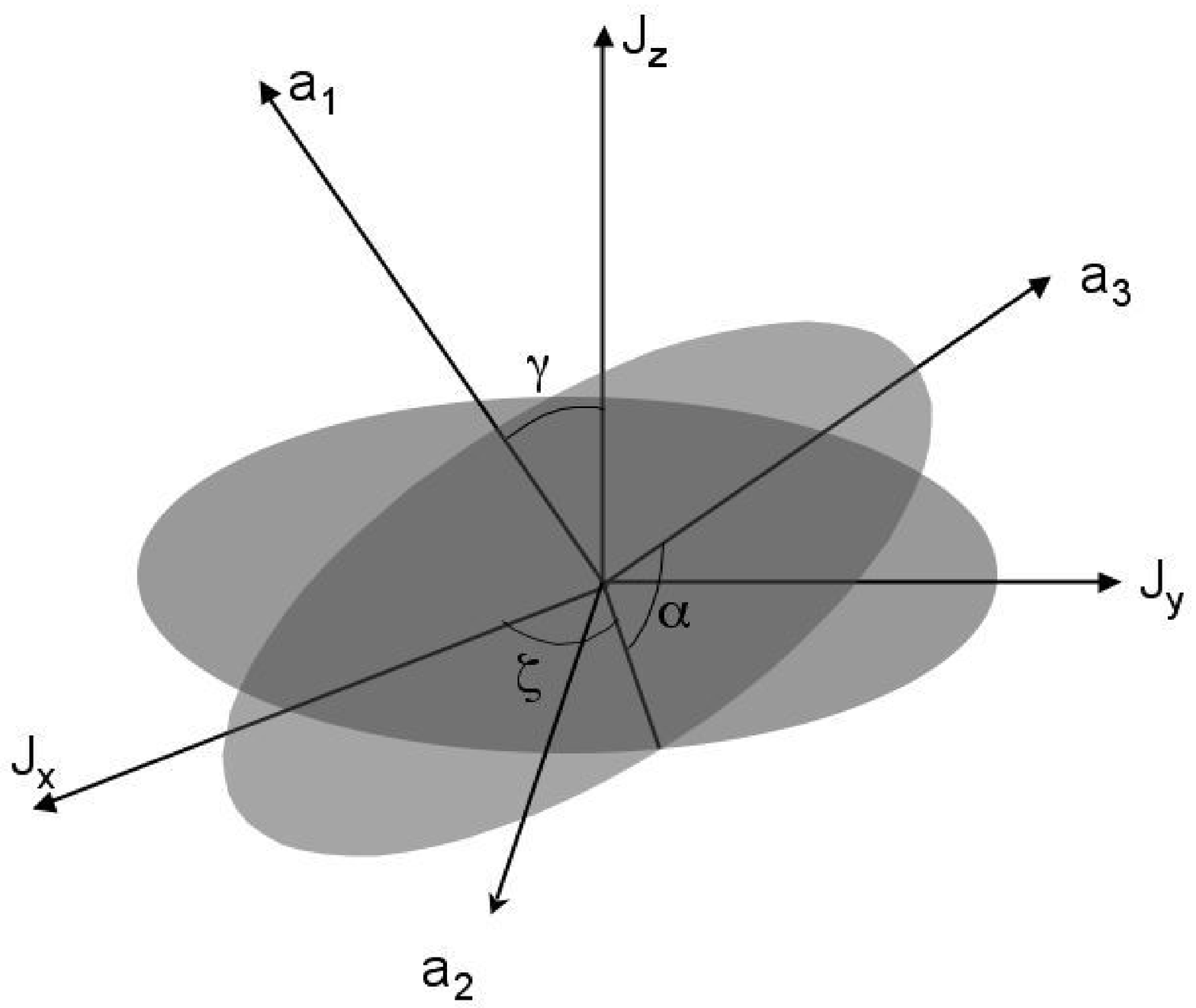} 
\caption{Orientation of principal axes in the angular coordinate system are described by Euler angles $\gamma, \alpha$ and $\zeta$.}
\label{fap2}
\end{figure}
\section{RATs}
Similar to Paper I, in order to make an easy relation of our results to those in
earlier works, wherever it is possible, we preserve notations adopted in DW97.
Mean radiative torque efficiency over wavelengths, $\overline{\bQ(\Theta,\beta,\Phi)}$ is defined as
\bea
\overline{\bQ}=\frac{\int \bQ_{\lambda}u_{\lambda}d\lambda}{\int 
  u_{\lambda}d\lambda} \label{eq6}, 
\ena 
where $u_{\lambda}$ is the energy density (see Mathis, Mezger \& Panagia 1983), and $\bQ_{\lambda}$ is the RAT efficiency corresponding to wavelength
$\lambda$.
RAT from the anisotropic component of radiation is defined by 
\bea 
\bG_{rad}=\frac{\overline{u}_{rad}a_{eff}^{2}\overline{\lambda}}{2}\gamma\overline{\bQ}.\label{eq7}
\ena 
where $\gamma$ is the anisotropy degree of radiation, $a_{eff}$ is the
effective size of the grain (see DW96; Paper I), $\overline{\lambda},
\overline{u}_{rad}$ are mean wavelength and mean energy density of radiation
field, which are respectively given by
\begin{align}
\overline{\lambda}&=\frac{\int \lambda d\lambda}{\int d\lambda},\\
\overline{u}_{rad}&=\frac{\int u_{\lambda}\lambda d\lambda}{\int \lambda d\lambda}.
\end{align}

\section{Averaging over free-torque motion}\label{apen3}

Consider an ellipsoid with three principal axes ${\bf a}_{1}, {\bf a}_{2},
{\bf a}_{3}$ and moment of inertia $I_{1}> I_{2}> I_{3}$, respectively. 
Dynamics of such an ellipsoid is clearly presented in the textbook of Landau
\& Liftshift (1972). 
Let define a dimensionless quantity
\begin{align}
k^{2}=\frac{(I_{2}-I_{3})(q-1)}{(I_{1}-I_{2})(1-I_{3}q/I_{1})},
\end{align}
where $q=\frac{2I_{1}E}{J^{2}}$ is the ratio of total kinetic energy to the rotational energy along
the maximal inertia axis.

For $q<I_{1}/I_{2}, k^{2}<1$, the solution of Euler equations is
\begin{align}
\omega_{1}&=\pm\frac{J}{I_{1}}(\frac{I_{1}-I_{3}q}{I_{1}-I_{3}})^{1/2} dn(\tau),\\
\omega_{2}&=-\frac{J}{I_{2}}(\frac{I_{2}(q-1)}{I_{1}-I_{2}})^{1/2} sn(\tau),\\
\omega_{3}&=\pm\frac{J}{I_{3}}(\frac{I_{3}(q-1)}{I_{1}-I_{3}})^{1/2} cn(\tau)
\end{align}
where cn, sn, tan are  hyperbolic trigonometric functions, and $\tau$ is
given by
\begin{align}
\tau\equiv tJ\left[\frac{(I_{1}-I_{2})(1-I_{3}q/I_{1})}{I_{1}I_{2}I_{3}}\right]^{1/2}.
\end{align}

The rotation period around the maximal inertia axis is
\begin{align}
P_{\tau}=4F(\pi/2,k^{2}),
\end{align}
where $F$ is the elliptic integral defined by
\begin{align}
F(\epsilon, m)=\int_{0}^{\epsilon} d\theta (1-m \mss\theta)^{-1/2}.
\end{align}

For $q> I_{1}/I_{2}$, angular velocities are given by
\begin{align}
\omega_{1}&=\pm\frac{J}{I_{1}}(\frac{I_{1}-I_{3}q}{I_{1}-I_{3}})^{1/2} cn(\tau),\\
\omega_{2}&=-\frac{J}{I_{2}}(\frac{I_{2}(1-I_{3}q)}{I_{2}-I_{3}})^{1/2} sn(\tau),\\
\omega_{3}&=\pm\frac{J}{I_{3}}(\frac{I_{3}(q-1)}{I_{1}-I_{3}})^{1/2} dn(\tau),
\end{align}
where 
\begin{align}
\tau\equiv tJ\left[\frac{(I_{2}-I_{3})(q-1)}{I_{1}I_{2}I_{3}}\right]^{1/2}.
\end{align}
Rotation period for this case is given by
\begin{align}
P_{\tau}=4F(\pi/2,k^{2}).
\end{align}
In above equations, $\pm$ stand for the positive and negative flipping states.

Euler angles $\alpha, \gamma, \zeta$ (see Fig. \ref{fap2})  can be deduced from angular velocity as followings
\begin{align}
\ms\alpha\ms\gamma&=\frac{I_{2}\omega_{2}}{J},\\
\mc\alpha\ms\gamma&=\frac{I_{3}\omega_{3}}{J},\\
\mc\gamma&=\frac{I_{1}\omega_{1}}{J}.
\end{align}
WD04 averaged RATs over free torque motion as following,
\begin{align}
\overline{A}(q)&=\frac{1}{P_{\tau}}\int_{0}^{P_{\tau}} d\tau
\frac{1}{2\pi}\int_{0}^{2\pi}A(\alpha, \zeta, \gamma).
\end{align}
Here they assume that the rotation of the grain about the angular momentum is much faster than the rotation about the maximal inertia axis (i.e. at each a moment $t$ with angles $\gamma$ and $\alpha$, the angle $\zeta$ can vary from $0$ to $2\pi$). Clearly, this assumption is not correct because these processes are comparable. We have implemented an algorithm to average RATs over free torque motion in which the angle $\zeta$ is obtained by solving an Euler differential equation. We found that RATs obtained by our method are in good agreement with what obtained using equation (C16) when they are averaged over a time scale greater than $10^{3} P_{\tau}$.

\section{Averaging over thermal fluctuations}\label{apen4}
Average of a quantity $A$ over thermal fluctuations is defined by
\begin{align}
\langle A \rangle&=\frac{\int_{0}^{1} ds  A(q,J)\mbox{exp}[-q(s)J^{2}/2I_{1}kT_{d}]}{\int
    ds \mbox{exp}[-qJ^{2}/2I_{1}kT_{d}]}.\label{ap2}
\end{align}
Number of state $s$ in equation (\ref{ap2}) is given by
\begin{align}
s\equiv 1-\frac{2}{\pi}\int_{0}^{\alpha_{1}}d\alpha \left[\frac{I_{3}(I_{1}-I_{2}q)+I_{1}(I_{2}-I_{3})\mcs\alpha}{I_{3}(I_{1}-I_{2})+I_{1}(I_{2}-I_{3})\mcs\alpha}\right]^{1/2},
\end{align} 
with $\alpha_{1}$ is given by
\begin{align}
\alpha_{1}&=\mc^{-1}\left[\frac{I_{3}(I_{2}q-I_{1})}{I_{1}(I_{2}-I_{3})}\right]^{1/2},
\end{align}
for $q>I_{1}/I_{2}$, and $\alpha_{1}=\pi/2$ for $q\leq I_{1}/I_{2}$ (see WD03).

\end{document}